\documentclass[%
 reprint,
nofootinbib,
 amsmath,amssymb,
 aps,
]{revtex4-2}

\usepackage{graphicx}
\usepackage{dcolumn}
\usepackage{bm}

\usepackage{mathtools}

\usepackage{array}
\newcolumntype{P}[1]{>{\centering\arraybackslash}p{#1}}
\newcolumntype{M}[1]{>{\centering\arraybackslash}m{#1}}
\usepackage{lineno}
\usepackage[dvipsnames]{xcolor}
\usepackage{longtable}
\usepackage{url}
\usepackage{verbatim}
\usepackage[utf8]{inputenc}
\usepackage{makecell}
\usepackage{colortbl}
\usepackage{comment}
\usepackage{cleveref}

\usepackage[T1]{fontenc} 
\graphicspath{ {Figures/} }

\defcitealias{CoronadoBlazquez2019a}{Paper~I}
\defcitealias{CoronadoBlazquez2019b}{Paper~II}

\begin{document}

\title{Spatial extension of dark subhalos as seen by \textit{Fermi}-LAT\\ and implications for WIMP constraints}

\author{Javier Coronado-Bl\'azquez}
\author{Miguel A. S\'anchez-Conde}
 \email{miguel.sanchezconde@uam.es}
\author{Judit P\'erez-Romero}
\affiliation{Departamento de F\'{i}sica Te\'{o}rica, Universidad Aut\'{o}noma de Madrid, Madrid, Spain $\&$  Instituto de F\'{i}sica Te\'{o}rica, UAM-CSIC, E-28049 Madrid, Spain}

\collaboration{\textit{Fermi}-LAT Collaboration}

\author{Alejandra Aguirre-Santaella}
\affiliation{Departamento de F\'{i}sica Te\'{o}rica, Universidad Aut\'{o}noma de Madrid, Madrid, Spain $\&$  Instituto de F\'{i}sica Te\'{o}rica, UAM-CSIC, E-28049 Madrid, Spain}

\date{March 2022}

\begin{abstract}
Spatial extension has been hailed as a ``smoking gun'' in the gamma-ray search of dark galactic subhalos, which would appear as unidentified sources for gamma-ray telescopes. {  In this work, we study the sensitivity} of the \textit{Fermi}-LAT to extended subhalos using simulated data based on a realistic sky model. We simulate spatial templates for a set of representative subhalos, whose parameters were derived from our previous work with N-body cosmological simulation data. We find that detecting an extended subhalo and finding an unequivocal signal of angular extension requires, respectively, a flux 2 to 10 times larger than in the case of a point-like source. By studying a large grid of models, where parameters such as the WIMP mass, annihilation channel or subhalo model are varied significantly, we obtain the response of the LAT as a function of the product of annihilation cross section times the J-factor. Indeed, we show that spatial extension can be used as an additional “filter” to reject subhalos candidates among the pool of unidentified LAT sources, as well as a ``smoking gun'' for positive identification. {For instance, typical angular extensions of a few tenths of degree are expected for the considered scenarios.} Finally, we also study the impact of the obtained LAT sensitivity to such extended subhalos on the achievable dark matter constraints, which are a few times less constraining than comparable point-source limits.
\end{abstract}

\maketitle


\section{Introduction}
\label{sec:intro}
The nature of dark matter (DM), a non-baryonic component of the Universe accounting for $\sim$25\% of its matter-energy density \cite{Aghanim2020}, is one of the most pressing questions in modern physics. Many candidates have been proposed to explain its ultimate nature, such as the \textit{Weakly Interacting Massive Particles} (WIMPs) \cite{Bertone2010}. These are well-motivated particles that decouple from the thermal plasma through a freeze-out mechanism, and are able to reproduce the measured cosmological abundance with an interaction cross section coincident with the electroweak scale, known as the ``WIMP miracle'' \cite{Bertone+05}. In the last few years, different strategies have been employed in the quest for WIMP DM detection \cite{GarrettDuda09}. From direct detection experiments, aiming to observe the scattering of DM-nuclei interactions; and DM production in particle accelerators such as the \textit{Large Hadron Collider} (LHC), to indirect searches, looking for the annihilation (or decay) products of DM. These approaches are in many cases complementary, which allow different models of DM to be probed, subject to different uncertainties \cite{Schumann2019, Kahlhoefer2017, Gaskins2016}.

In the case of indirect detection, gamma rays have been considered a ``golden channel'', as they are not deflected by magnetic fields (as opposed to antimatter produced in DM annihilation or decays), and are easier to detect than neutrinos \cite{Conrad2017}. With gamma-ray telescopes, many different targets have been observed. Among the most promising sources are the Galactic center (GC), e.g., \cite{fermi_gc_paper16,fermi_gc_paper17, Abdallah2016, CTA_GC_2020}, dwarf spheroidal galaxies (dSphs) \cite{Ackermann2015, Oakes2019, MAGIC_fermi_dsphs16} and galaxy clusters \cite{fermi_cluster_paper, Ackermann_2010, Acciari_2018}, as well as the M31 galaxy, where a gamma-ray signal has been detected both from the inside and towards the outer halo \cite{Ackermann_2017, Karwin_2019, Karwin_2021}. The GC is expected to be the brightest source of gamma rays produced by DM annihilation, yet the astrophysical gamma-ray background in the area, in the form of unresolved point sources (PS) or diffuse Galactic emission, is dominant and still poorly understood \cite{fermi_gc_pulsar_paper,Bartels2017}. On the other hand, while the astrophysical background in dSphs is expected to be negligible \cite{2016ApJ...832L...6W}, their predicted DM annihilation signal is thought to be comparatively fainter \cite{2017ApJ...834..110A, Pace_2018, 2020ARNPS..70..455M}.

An interesting and complementary target for gamma-ray searches are the so-called dark subhalos (also known as dark satellites). Within the standard $\Lambda$--Cold Dark Matter ($\Lambda$CDM) cosmological paradigm, the DM halos that host galaxies should contain a large amount of substructure (subhalos, i.e., halos within halos) as a consequence of the bottom-up hierarchical structure formation scenario \cite{Madau2008}. The largest members of this subhalo population (above $\sim10^{7}$ M\textsubscript{\(\odot\)}) will host dSphs, yet most of them are not massive enough to retain baryonic content, i.e., gas, and form stars, and therefore should remain completely dark \cite{Walker13,Sawala_2015}. These objects are predicted to be compact and very concentrated and so, if near enough and composed of WIMPs, they would yield a significant flux of gamma rays that could be detected by gamma-ray telescopes such as the \textit{Fermi}-Large Area Telescope (LAT) on board the NASA Fermi satellite~\cite{fermi_instrument_paper}.

The main advantages of these subhalos compared to other competitive DM targets are i) they are not expected to host any non-DM astrophysical gamma-ray emitters, as they do not retain baryonic content and, thus, no astrophysical background is expected; ii) their large number density in the Galaxy; and iii) their high concentrations of DM, as inferred from DM-only N-body sims within $\Lambda$CDM \cite{Diemand2008,Springel2008_nature,Moline+17}. The main disadvantages come from uncertainties related to the lack of observations/identification of these objects in the sky, such as their precise structural properties, abundance (particularly at the smallest predicted scales) and radial distribution within the Galaxy. Indeed, there is an absolute lack of knowledge on their exact position in the sky. Dark subhalos are expected to be almost isotropically distributed from the Earth's point of view; yet, in general terms, only those located very close to us may yield annihilation fluxes large enough to be detected \cite{CoronadoBlazquez2019a,hutten19b}.

A large number of \textit{Fermi}-LAT sources lack clear astrophysical association and have no multi-wavelength counterparts, remaining as unidentified sources (unIDs) at present. These are perfect candidates to perform a search for Galactic dark subhalos, as WIMPs may be annihilating within them, appearing as unIDs in the LAT sky. Many works have already used these targets for indirect DM detection \cite{Bertoni+15, Bertoni+16, Calore+17, Schoonenberg+16,Hooper_2017,BerlinHooper14, Zechlin+12, ZechlinHorns12, Belikov2012, BuckleyHooper10, fermi_dm_satellites_paper, CoronadoBlazquez2020, coronadoblazquez2021sensitivity}. {  In our previous works \citep{CoronadoBlazquez2019a,CoronadoBlazquez2019b}, which we will refer to as 
\citetalias{CoronadoBlazquez2019a} and \citetalias{CoronadoBlazquez2019b} respectively,} starting from a total of 1235 unIDs in the latest LAT point-source catalogs, we applied a series of filters according to the expected emission from dark subhalo annihilation. Once the number was greatly reduced to a pool of just 44 unIDs compatible with DM annihilation in \citetalias{CoronadoBlazquez2019a}, a careful, dedicated spectral analysis was performed in \citetalias{CoronadoBlazquez2019b}, to enforce or reject the DM hypothesis. Only a handful of them, with marginal statistical significance, were more compatible with a DM origin over a traditional astrophysical explanation. Therefore, we proceeded to set upper limits on the $\langle\sigma v\rangle-m_{\chi}$ parameter space. As it was shown in \citetalias{CoronadoBlazquez2019a}, the fewer the sources compatible with DM among the pool of unIDs, the stronger the DM constraints. To set constraints, the catalog filtering results were combined with expectations from state-of-the-art N-body simulations, that were conveniently repopulated with low-mass subhalos below the original mass resolution limit using well-motivated $\Lambda$CDM recipes. In parallel to this N-body simulation work, it was also necessary to perform a full characterization of the \textit{Fermi}-LAT point-source sensitivity to DM annihilation for different annihilation channels and sky positions. The obtained constraints were competitive, ruling out canonical thermal WIMPs up to $\sim$20 GeV in the case of $\tau^+\tau^-$ annihilation, this way being also complementary to constraints obtained by other methods and targets.

In \citetalias{CoronadoBlazquez2019b}, source spatial extension was pointed out to be a ``smoking gun'' for this kind of indirect DM search, as already noted by many authors \cite{Bertoni+15,Bertoni+16,Hooper_2017,Xia2017,Facchinetti2020,Mauro2020}. {  Indeed, in \citetalias{CoronadoBlazquez2019b} we quantified the expected extension of the brightest (and therefore, easiest to detect) subhalos according to N-body simulations. Very large angular sizes, $\mathcal{O}(5^\circ)$ were found when integrating the annihilation signal up to the scale radius, this being in line with previous estimates \citep{hutten16,Mauro2020}.}

Aside from known extended sources (e.g., \cite{Ackermann_2018}), all sources in official LAT catalogs \cite{2015ApJS..218...23A, 2016ApJS..222....5A, 2017ApJS..232...18A, 2020ApJS..247...33A} are modeled as point sources, and no additional spatial analysis is performed. Therefore, this allows for a dedicated spatial analysis on the candidates surviving the DM spectral analysis. Yet, none of them shown any preference for a spatially extended signal over a point-like model in \citetalias{CoronadoBlazquez2019b}. This would in principle rule out all remaining candidates, or at least put some tension between simulations and data. Nevertheless, it is important to note that a cuspy DM annihilation profile can appear point-like if the extended emission is below the detection threshold, as is likely the case for faint unIDs. For this reason, we could not discard our candidates straight away in \citetalias{CoronadoBlazquez2019b}.

{  In \cite{Mauro2020} the spatial extension of subhalos as it could be actually detected by the \textit{Fermi}-LAT was assessed with simulated gamma-ray data and a semi-analytical subhalo population model.} In this work, we address it quantitatively using a realistic sky model based on LAT data, where the subhalo is injected in real data by adopting both a specific input WIMP spectrum and spatial annihilation profile. The latter is chosen according to results from our own N-body simulation suite, which takes into account low-mass subhalos.

In particular, we first characterize the properties of Milky Way (MW) subhalos as well as their expected annihilation fluxes in Section \ref{sec:DM_modeling}. More precisely, in Section \ref{sec:models} we select {  representative cases of the subhalo population in the Galaxy according to masses and distances.} These are studied in detail in Section \ref{sec:j-factors}, where also their expected annihilation fluxes are computed. Indeed, we build DM profiles for these objects and compute the expected DM annihilation signal as two-dimensional templates accounting for the spatial morphology. Then, we compute the LAT sensitivity (i.e., the threshold flux) to extended dark subhalos in Section \ref{sec:analysis_bigsection}, and study how this sensitivity is degraded with respect to the point-like case computed in \citetalias{CoronadoBlazquez2019a}. 

We consider a grid of possible scenarios, varying different analysis parameters such as the annihilation channel or the WIMP mass. For these scenarios, we look for any hint of angular extension and characterize it in case of detection. After performing this exercise, we are also able to determine if spatial extension is, indeed, a ``smoking gun'' for dark subhalo detection and, if so, in which cases and under which configurations. In Section \ref{sec:implications} we  discuss and quantify the impact of this subhalo extension study on the DM constraints, namely those presented in \citetalias{CoronadoBlazquez2019a} and \citetalias{CoronadoBlazquez2019b}, where point-like subhalos were assumed. Finally, we conclude in Section \ref{sec:conclusions}.

\section{Gamma-ray signal from Milky Way subhalos}
\label{sec:DM_modeling}

\subsection{Building of the subhalo DM density profile}\label{sec:models}

As said before, we can distinguish between dSphs, which have a baryonic counterpart, and dark satellites, which do not have it. Most Galactic DM subhalos are expected to be dark satellites, lacking the baryonic content of dSphs. We apply the same conservative \footnote{Conservative in the sense that this will reject some subhalos with large J-factors in the simulation. Objects above $10^7 \rm M_{\odot}$ may not host baryonic content and be dark subhalos, yet only below this value it is completely safe to rule out dSphs formation \cite{Sawala_2015}.} upper mass limit of $M_{sub}<10^7 \rm M_{\odot}$ for these objects, as in \citetalias{CoronadoBlazquez2019a} and \citetalias{CoronadoBlazquez2019b}, thus we will stick to this bound here as well.

Formerly, in \citetalias{CoronadoBlazquez2019a}, we analyzed how useful state-of-the-art N-body cosmological simulations of MW-like halos are to learn about the properties of the subhalo population existing at the present time. In both \citetalias{CoronadoBlazquez2019a} and \citetalias{CoronadoBlazquez2019b}, we made use of the outcome of one of them, namely the Via Lactea II DM-only simulation  (VL-II,~\cite{Diemand2008}), which resolves subhalo masses as light as $10^4 \mathrm{M_\odot}$ and is complete in terms of subhalo abundance down to $\sim 10^6 \mathrm{M_\odot}$.
In particular, in our previous work we repopulated the original VL-II to also include  subhalos less massive than those in the parent simulation, as some of these small subhalos could still be relevant for DM searches. We did so using theoretically-motivated recipes arising within the $\Lambda$CDM structure formation scenario. Once this was done, we also obtained subhalo annihilation fluxes. 
Furthermore, in \citetalias{CoronadoBlazquez2019b} we computed the angular extension of the brightest subhalos in our repopulated simulation, using their scale radii (defined as the radius at which the slope of the density profile is $-2$) as a proxy for subhalo extension. From this study, we found large angular extensions of $\sim1-10^\circ$, and concluded that the majority of them should be actually seen as extended by \textit{Fermi}-LAT, given the instrument's point-spread function (PSF), i.e., its angular resolution (see Figure 9 from \citetalias{CoronadoBlazquez2019b}).

\begin{figure}[!ht]
\centering
\includegraphics[width=1\linewidth]{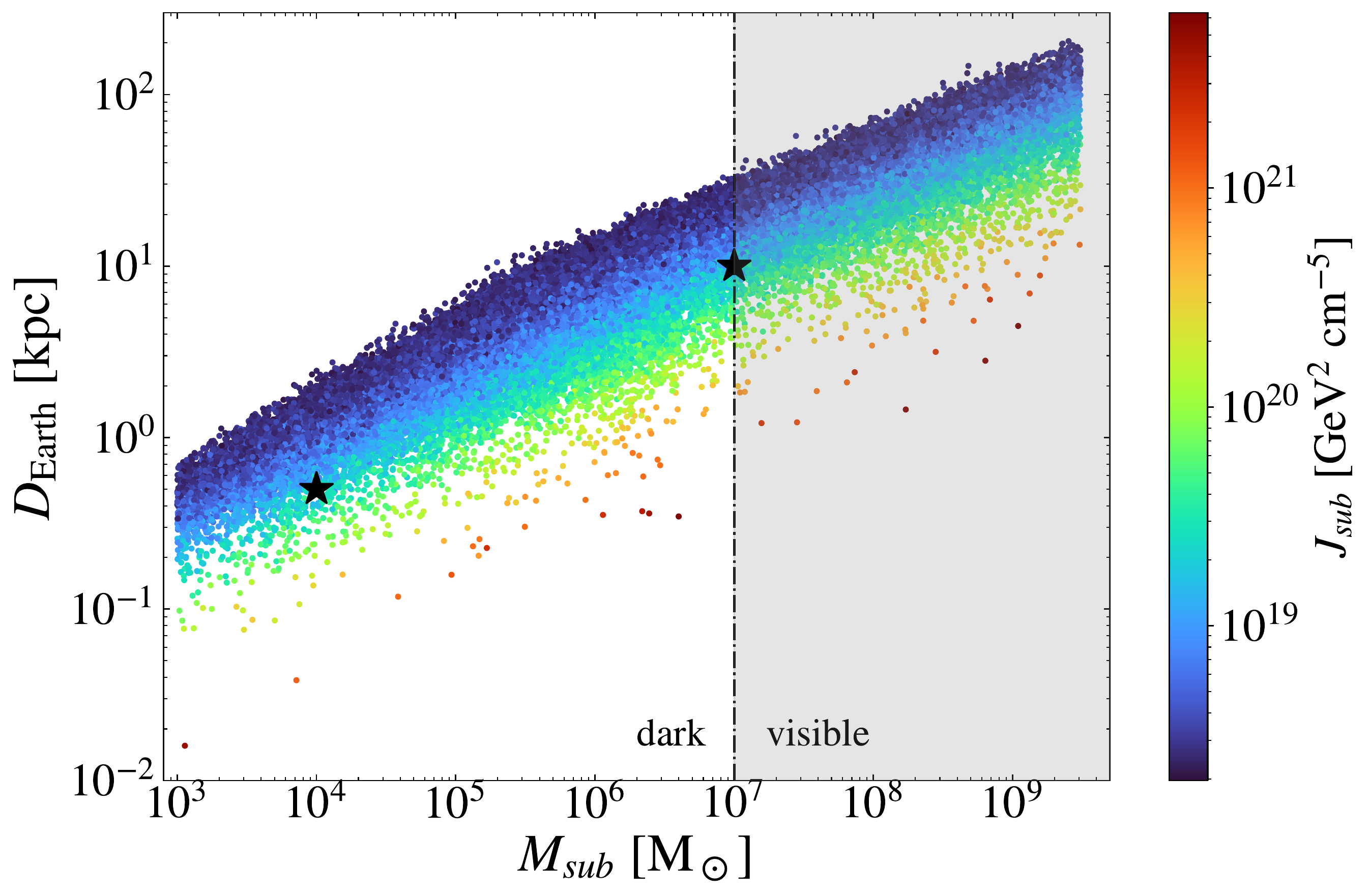}
\caption{Distance to the Earth versus the subhalo mass for the 100 brightest subhalos in 1000 repopulations of the VL-II N-body cosmological simulation. The colored z-axis represents the J-factor. The vertical, dark dash-dotted line separates between dSphs and dark subhalos, and the black stars show the mass and distance of the two subhalos that were chosen as representative to perform the \textit{Fermi}-LAT analysis in the next sections.}
\label{fig:vliirep}
\end{figure}

We use this study as the starting point to perform the analysis of detecting a possible extension of these objects with \textit{Fermi}-LAT. To quantitatively measure the expected annihilation flux, we define the so-called J-factor for a subhalo as:

\begin{equation}\label{eq:j-factor}
J_{sub}(\Delta\Omega, l.o.s) = \int_0^{\Delta\Omega}d\Omega\int_{l.o.s}\rho^2(r)dl,
\end{equation}

\noindent where $\Delta\Omega=2\pi(1-\cos\alpha_{int})$, with $\alpha_{int}$ the integration angle, the integration is calculated along the line of sight ($l.o.s.$) and $\rho(r)$ is the DM density profile. We show in Figure~\ref{fig:vliirep} the J-factor as a function of mass and distance to the Earth for the brightest subhalos in the repopulated VL-II. We note that subhalos with the highest expected annihilation fluxes, i.e., at least $J_{sub}\gtrsim 10^{19}$ GeV$^2$ cm$^{-5}$, are the most interesting for our purposes. Given the large number of subhalos above this value, as well as the large spread in their masses and distances, it becomes unfeasible to perform a \textit{Fermi}-LAT analysis simulation for each of them. In view of this, {  we decided to select two representative scenarios among the brightest subhalos, in terms of subhalo mass and its corresponding typical distance.} As already stated before, we want to restrict ourselves to dark subhalos. This implies that the highest mass we can reach would be $10^7\, \rm M_{\odot}$, following our previous work and results found, e.g., in~\cite{Sawala_2015}.  With this in mind, and taken into account what is shown in the top panel of Figure~\ref{fig:vliirep}:

\begin{itemize}
    \item[(i)] the first chosen scenario is a subhalo with a small mass of $M_{sub} = 10^4\, \rm M_{\odot}$ located near Earth ($D_{\rm Earth} = 0.5$ kpc);
    \item[(ii)] the second scenario is a subhalo with a mass in the limit of $M_{sub} = 10^7\, \rm M_{\odot}$ and situated far from Earth ($D_{\rm Earth} = 10$ kpc).
\end{itemize}

With this selection of representative cases, we aim at covering the range in masses, distances and expected angular extensions of the fraction of the subhalo population expected to be relevant for our purposes. In Section \ref{sec:j-factors}, we will study in further detail the expected DM-induced emission from these two scenarios as well as their similarities and differences.

For each of these two representative subhalos, we construct a consistent DM density profile following the results found in N-body, DM-only cosmological simulations in a $\Lambda$CDM universe~\cite{Diemand2008, 2008MNRAS.391.1685S, 2016MNRAS.457.3492H, 2020arXiv200714720I}, which points to a universal DM density profile. More precisely, we adopt the Navarro-Frenk-White (NFW) profile \cite{Navarro1997}. We note though that deviations from this universal profile are particularly expected for subhalos, which are subject to multiple and complex processes within their hosts (e.g. tidal stripping, dynamical friction, interaction with baryonic material, etc.). The magnitude and impact of these effects on the subhalo population is still a matter of debate, e.g.~\cite{stref17,2018MNRAS.474.3043V, 2018MNRAS.475.4066V, 2019MNRAS.485..189O}. These deviations are expected to be particularly relevant in the outer regions of the subhalos, where the mass loss is more significant.

Yet, in such regions the annihilation flux has already decreased very significantly compared to the one originated in the inner regions. For this reason, for the sake of our work we will proceed with an NFW profile to model the inner structure of subhalos\footnote{We note that, since we are interested in the innermost region of the subhalo, i.e. the one within its scale radius where most of the annihilation is produced, any other realistic DM density profile such as Einasto \cite{1965TrAlm...5...87E} or a truncated NFW \cite{Kazantzidis_2004} is expected to render similar results. Only core profiles would potentially yield significantly different results, however these profiles are not expected for these small objects, as they are already devoid of baryons, which are the ones that drive the formation of cores \cite{Pontzen_2012,Cintio_2014}.}. The NFW profile can be written as
\begin{equation}\label{eqn:NFW}
 \rho_{\mathrm{NFW}}(r)=\frac{\rho_0}{\left(\frac{r}{r_s}\right)\left(1+\frac{r}{r_s}\right)^{2}},
\end{equation}
where $r_s$ and $\rho_0$ are the scale radius and the normalization density, respectively.

One key ingredient to build this NFW profile is to define a mass for an enclosed overdensity of $\Delta_{200} = 200$ within the virial radius of the object, with respect to the critical density $\rho_{crit}=135.99$ M$_{\odot} \rm kpc^{-3}$, $M_{200}$. However, for subhalos we recall that the meaning of this quantity becomes unclear. After repeatedly suffering tidal stripping, which removes part of the mass, the profile gets modified, especially in the outermost regions. This causes the standard definition of virialized objects to lack a physical meaning in the subhalo case, motivating new definitions \cite{Moline+17}. Thus, in order to construct an NFW profile from the value of $M_{200}$, a specific concentration-mass relation for the subhalos must be adopted. In \cite{Moline+17}, the authors propose a parametrization of this relation specifically for subhalos, that involves the dependence of the concentration not only with mass but also with subhalo position within the host halo. With this new parametrization, the problem of the definition of $M_{200}$ in subhalos is accounted for by properly correcting their concentrations. This makes this parametrization ideal for our purposes. This relation is expressed as: 

\begin{equation}
\begin{split}
    c_{200}(M_{sub},x_{sub})=\\=c_0 \left[ 1 + \sum_{i=0}^3 \left[a_i \ln\left(\frac{M_{sub}}{10^8 h^{-1} \rm M_\odot}\right)\right]^i\right]\times [1+b\ln(x_{sub})],
    \label{eq:csub}
\end{split}
\end{equation}

where $h = 0.7$ is the reduced Hubble constant and $c_0,\ a_i$ and $b$ are obtained from fits to N-body cosmological simulations at different subhalo mass scales. The concentration is written as a function of the masses of the subhalos $M_{sub}$ and as a function of their positions within the host halo ($D_{sub}$) through $x_{sub} \equiv D_{sub}/R_{host}$. In order to compute $x_{sub}$ for subhalos, we assume $R_{host} = 402$ kpc, i.e. the virial radius of the Milky Way in the VL-II simulation. For our two representative subhalos, $D_{sub}$ is estimated to simply be $D_{sub}= D_{\rm Earth} + d_{\odot}$, with $d_{\odot}=8.5$ kpc the galactocentric distance from the Sun. This provides a conservative estimate of the annihilation flux (as the concentration decreases with increasing distance to the host halo center). Having computed the subhalo concentration via the Eq. \ref{eq:csub}, we can then obtain $r_s$: 

\begin{equation}\label{eqn:scale radius}
r_s = R_{sub}/c_{200},
\end{equation}

\noindent for which we previously computed $R_{sub}$ via the definition of the overdensity $\Delta_{200}$:

\begin{equation}\label{eqn:vitual_virial_radius}
R_{sub}=\left(\frac{3 M_{sub}}{4\pi \Delta_{200}\rho_{crit}}\right)^{1/3}.
\end{equation}

We stress that the purpose of obtaining $R_{sub}$ is only to build a coherent NFW profile for subhalos, since by adopting the $c_{200}(M_{sub})$ relation in \cite{Moline+17} we mostly overcome the problem to describe subhalos with standard, non-truncated and less concentrated NFW profiles. Indeed, the variable that we will use to estimate the actual spatial extension of the subhalos is the tidal radius $R_{tidal}$, obtained by applying the Roche criterium~\cite{1962AJ.....67..471K}:

\begin{equation}
R_{tidal}= D_{GC} \left( \frac{M_{sub}}{M (<D_{GC})} \right)^{1/3},
\label{eqn:roche_crit}
\end{equation}

\noindent where $D_{GC}$ is the distance to the GC. The last parameter that is needed to build the subhalos DM density profiles is the normalization density, $\rho_0$. We impose the normalization condition $M_{sub}=\int_0^{R_{sub}}\rho_{NFW}(r)r^2drd\Omega$, and rewriting this in terms of the concentration parameters, we can finally compute $\rho_0$ as

\begin{equation}\label{eqn:density_norm}
\rho_0 = \frac{2~\Delta_{200}~\rho_{crit}~c_{200}}{3~F(c_{200})},
\end{equation}

\noindent where $F(c_{200})=\frac{2}{c_{200}^2}\left(\ln{(1+c_{200})}-\frac{c_{200}}{1+c_{200}}\right)$. The obtained profile parameters for each of the two representative subhalo scenarios are given in Table \ref{tab:parameters_nfw}. 

\begin{table}[ht!]
\caption{Parameters needed to build the NFW DM profiles of our two subhalo scenarios, each of them representatives of the fraction of the Galactic dark subhalo population that is relevant in terms of expected annihilation flux. See Section \ref{sec:models} for details on the derivation of these parameters.}
\centering
\begin{tabular}{  l | c | c | c | c | c | c }
\hline
\hline
Name & $D_{\rm Earth}$ & $M_{sub}$ & $c_{200}$ & $r_s$ & $R_{tidal}$ & $\log_{10}\rho_0$ \\
 & (kpc) & $\rm (M_{\odot})$ &  & (kpc) & (kpc) & $\rm (M_{\odot}/kpc^3)$ \\
\hline
Subhalo 1 & 0.5 & $10^4$ & 71.36 & 0.006 & 0.036 & 9.00 \\
Subhalo 2 & 10 & $10^7$ & 42.31 & 0.107 & 0.516 & 2.95 \\ 
\hline
\hline
\end{tabular}
\label{tab:parameters_nfw}
\end{table}

\subsection{Gamma-ray flux from subhalos}\label{sec:j-factors}
The DM modeling performed in the previous section is the starting point for obtaining the expected DM-induced gamma-ray flux for the two representative subhalo cases. Assuming that the DM is made of Majorana WIMPs, the annihilation flux can be computed as: 

\begin{equation}\label{eq:dm-flux}
\begin{split}
\frac{d\phi_{\gamma}}{dE}(E, \Delta\Omega, l.o.s.)=\\=\frac{1}{8\pi}\frac{\langle\sigma v\rangle}{ m^2_{DM}}\frac{dN_{\gamma}}{dE}(E)\times J(\Delta\Omega, l.o.s.),
\end{split}
\end{equation}

\noindent where $\frac{d\phi_{\gamma}}{dE}$ is the differential gamma-ray flux, $\langle\sigma v\rangle$ is the velocity-averaged annihilation cross-section, $m_{DM}$ is the DM mass, $\frac{dN_{\gamma}}{dE}$ is the photon spectrum and $J$ is the astrophysical factor or J-factor defined as in Eq.~\ref{eq:j-factor}. We notice that two main dependencies can be identified in the flux. First, an energy dependence that appears only in the particle physics term, carrying all the information of the mass of the DM candidate and the annihilation channels. Secondly, the spatial dependence appearing as the J-factor. This allows us to factorize these terms independently, and to model the spatial distribution of the DM independently of energy. The photon spectrum from DM annihilation strongly depends on the mass of the dark matter particle and on the annihilation channel, which indeed leads to very different spectra depending on which combinations are selected. We will explore the impact on the gamma-ray flux of these dependencies in Section \ref{sec:LAT_analysis}.



The computation of the J-factors is performed using the \texttt{CLUMPY} code \cite{Charbonnier:2012gf, Bonnivard:2015pia, Hutten:2018aix}. \texttt{CLUMPY} is a very flexible code written in C++, which allows us to model the DM profiles of the subhalos as described above and to compute J-factors, annihilation fluxes, etc. The obtained values of the integrated J-factors for each considered subhalo model are shown in Table \ref{tab:j-factors} and also highlighted in Figure~\ref{fig:vliirep}. We note that, since we do not expect subhalos to host a significant population of sub-subhalos, we do not add further levels of subhalo substructure to the computation of the J-factors, as this effect has been shown to be very subdominant \cite{Moline+17}. 

\begin{table}[ht!]
\caption{Total, integrated J-factors for the two representative subhalo scenarios, in the first column integrated up to the whole object ($R_{sub}$), and up to $r_s$ in the second. }
\centering
\begin{tabular}{  l  c  c  }
\hline
\hline
Name & $\log_{10}J_{sub}$ $\rm (GeV^2 cm^{-5})$ & $\log_{10}J_{s}$ $\rm (GeV^2 cm^{-5})$ \\
\hline
Subhalo 1 & 19.20 & 19.17 \\
Subhalo 2 & 19.14 & 19.10 \\ 
\hline
\hline
\end{tabular}
\label{tab:j-factors}
\end{table}

\begin{figure}[!ht]
\centering
\includegraphics[width=0.8\linewidth]{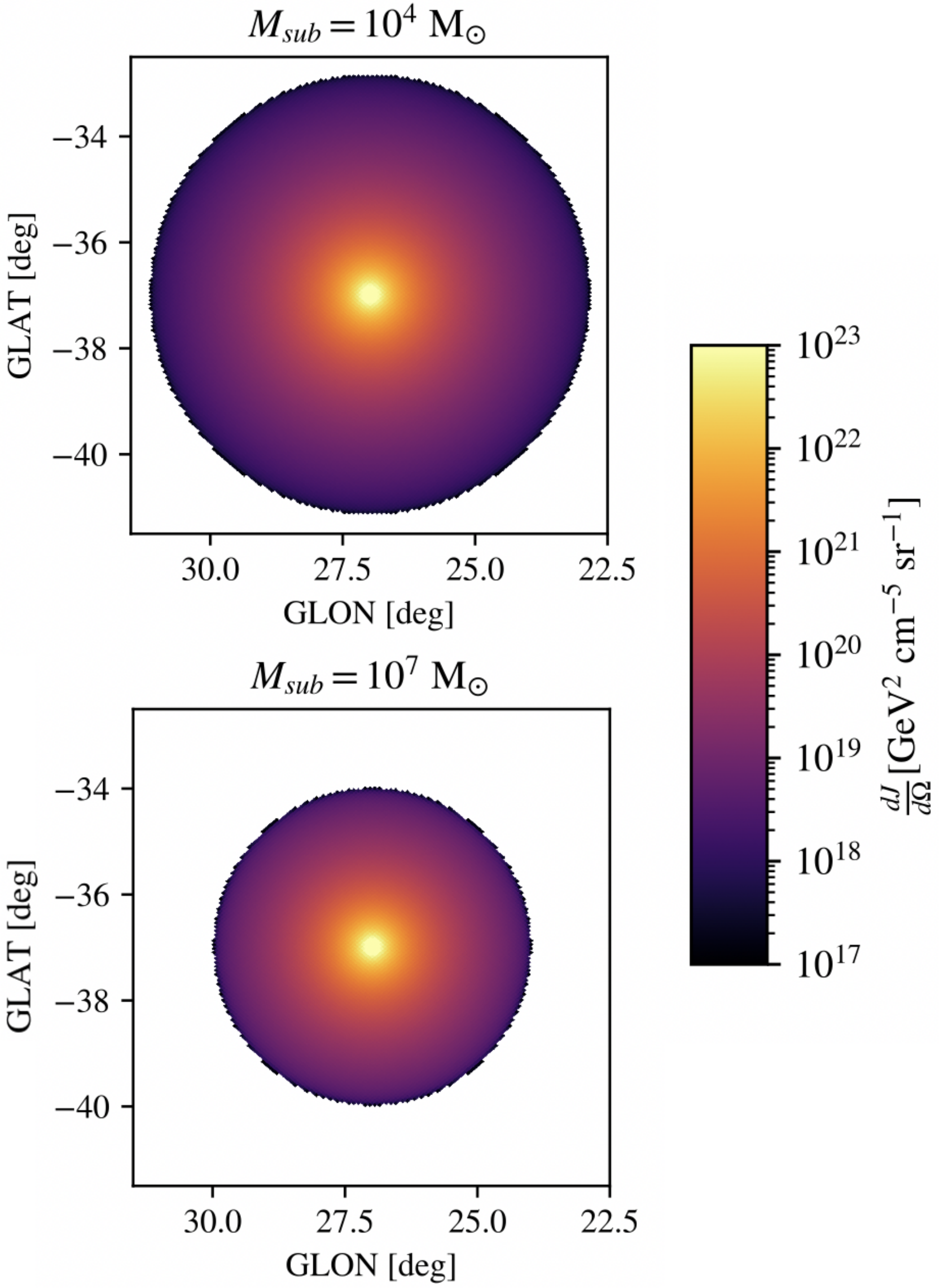}
\caption{Two-dimensional templates of the DM-induced gamma-ray spatial morphology for the two representative subhalo scenarios defined in Section \ref{sec:models}. The z axis represents the differential J-factor $\left(\frac{dJ}{d\Omega}\right)$ computed according the DM modeling explained in Section \ref{sec:models}.}
\label{fig:clumpy_maps}
\end{figure}

The results shown in Table \ref{tab:j-factors} show an emerging degeneracy between the two considered physical scenarios chosen as representative of the population of subhalos in the MW relevant in terms of annihilation flux. Indeed, both scenarios will be nearly indistinguishable in terms of the expected annihilation flux, despite the fact that both physical scenarios were defined independently in Section \ref{sec:models}. In order to take a closer look at this degeneracy, we can also use \texttt{CLUMPY} to create two-dimensional templates, reproducing the spatial morphology of the expected DM annihilation signal. These maps are shown in Figure \ref{fig:clumpy_maps} for the two representative subhalos, and represent, indeed, the main \texttt{CLUMPY} result that will be later used for the LAT analysis. The figure also exhibits the same degeneracy between the two scenarios that appeared before in Table \ref{tab:j-factors} for the integrated J-factors. To better characterize both the spatial morphology and this apparent degeneracy, we define two angular quantities. The first one is the angular extension in the sky associated to the scale radius:

\begin{equation}\label{eqn:theta_s}
\theta_{s} = \arctan\left(\frac{r_{s}}{D_{\rm Earth}}\right).
\end{equation}
Also, it is convenient to define the angular extension associated to the angle that contains 68\% of the DM annihilation flux, $\theta_{68}$, i.e.:
\begin{equation}\label{eqn:theta_68}
J(\theta_{68}) = 68\%\times J_{sub}.
\end{equation}

\begin{figure}[!ht]
\centering
\includegraphics[width=1\linewidth]{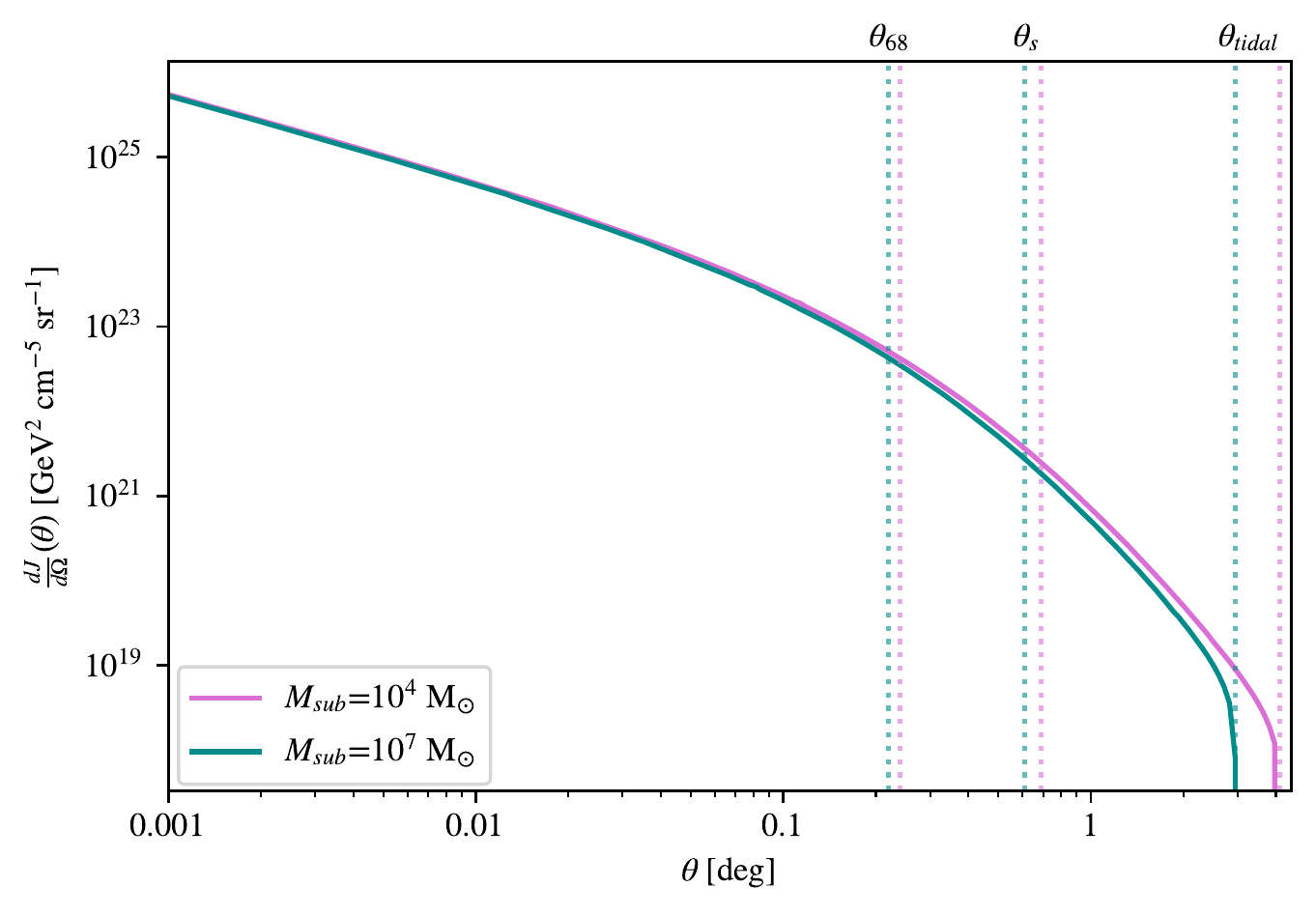}
\caption{Differential J-factor versus the radial angle for the two representative subhalo scenarios defined in Section \ref{sec:models}. Here, we also show $\theta_{tidal}$, which represents the angle subtended by $R_{tidal}$, $\theta_s$ as defined in Eq.~\ref{eqn:theta_s} and $\theta_{68}$ as defined in Eq.~\ref{eqn:theta_68}, all depicted with a different color in each subhalo case. The spatial distributions of the J-factor are almost identical until about 1$^\circ$, emphasizing the degeneracy discussed in the main text.}
\label{fig:degeneracy}
\end{figure}

These two quantities as well as the annihilation flux profile within the two representative subhalos are illustrated in Figure~\ref{fig:degeneracy}. From this Figure, we can observe that the two scenarios exhibit very similar flux profiles, with only small differences arising in the outermost regions, for angles above 1$^\circ$. 
As it will be shown later on, the fact that the selected scenarios are mostly indistinguishable from the spatial point of view will impact the \textit{Fermi}-LAT analysis of these objects, where we expect this degeneracy to persist.

\section{LAT sensitivity to extended dark matter subhalos}
\label{sec:analysis_bigsection}
In this section, we will compute the actual sensitivity of the LAT to an extended DM subhalo in the two subhalo mass modeling scenarios described in Section \ref{sec:DM_modeling}. For each subhalo model, we will consider two benchmark annihilation channels: $b\bar{b}$, representative of hadronic final states, and $\tau^+\tau^-$, representative of leptonic ones. We first compute the minimum flux required to detect an extended subhalo in Section \ref{sec:fmin}, and compare it with the case of point-like sources. Then, in Section \ref{sec:LAT_analysis} we extend the computation to a grid of different setups to study the LAT response to an extended signal. 

\subsection{Threshold flux for extended subhalo detection}
\label{sec:fmin}

First, we compute the minimum flux that is needed in order to detect an extended subhalo with the \textit{Fermi}-LAT. To do so, we simulate the LAT sensitivity by placing a putative, extended DM subhalo in the sky. Then, we vary the normalization of the source until it reaches detection -- at that very moment the integrated flux will be the minimum detection flux.

To perform such computation, we rely on \textit{fermipy} \cite{Wood2017}. We use 11 years of LAT data, with an energy range between 500 MeV and 2 TeV, Pass 8 events \cite{Atwood2013}, with the \texttt{P8R3\_SOURCE\_V2} instrumental response functions (IRFs, \cite{Bruel2018}), and utilize all available photons (FRONT+BACK), excluding those arriving with zenith angles greater than $105^\circ$.

As we are interested in high-latitude sources, like in \citetalias{CoronadoBlazquez2019a} and \citetalias{CoronadoBlazquez2019b}, we first performed some tests placing the subhalo at different, high Galactic latitudes; yet we found the corresponding variations in our results to be negligible when compared to other uncertainties that will be presented in this work. Nevertheless, we assume the subhalo to lie in a relatively clean patch of the sky, with no sources nearby ($\leq 1^\circ$).\footnote{We note that, in case that the subhalo is close enough to another gamma-ray emitter due to projection effects, it is likely that Fermi will identify it as an extended source when in reality there are two distinct objects, e.g., \cite{2018PDU....21....1C}. This could also alter the spectral properties due to the spill-over of photons in the ROI. Further tests were performed with no near sources closer than $2^\circ$ with no difference in the results.} Because of that, the results in this section refer only to a specific region of interest (ROI) of $15^\circ \times 15^\circ$ centered at $(l,b)=(27^\circ,-37^\circ)$. To model point sources lying within the ROI, we use the second release of the latest LAT point-source catalog, 4FGL--DR2 \cite{Abdollahi2020, Ballet2020}, version \texttt{gll\_psc\_v27.fit}. In the analysis, the \texttt{gll\_iem\_v07.fits} and \texttt{iso\_P8R3\_SOURCE\_V2.txt} templates\footnote{\url{https://fermi.gsfc.nasa.gov/ssc/data/access/lat/BackgroundModels.html}} are used to model the Galactic and isotropic diffuse emission, respectively. Table \ref{tab:setup_summary} summarizes the analysis setup, with an spatial extension given by the templates shown in Figure \ref{fig:clumpy_maps}.

\begin{table}[h!]
\caption{Summary of the \textit{fermipy} analysis setup.}
\centering
\begin{tabular}{l c}
\hline
\hline
Coordinates (l,b) & $(27^\circ,-37^\circ)$\\
Time domain (ISO 8601) & 2008-08-04 to 2019-08-03\\
Time domain (MET) & 239557417 to 586490000\\
Energy range & 500 MeV -- 2 TeV\\
IRFs & \texttt{P8R3\_SOURCE\_V2}\\
Event type & FRONT+BACK\\
Point-source catalog & 4FGL--DR2\\
ROI size & $15^\circ \times 15^\circ$\\
Max zenith angle & $105^\circ$\\
Galactic diffuse model & \texttt{gll\_iem\_v07.fits}\\
Isotropic diffuse model & \texttt{iso\_P8R3\_SOURCE\_V2.txt}\\
\hline
\hline
\end{tabular}
\label{tab:setup_summary}
\end{table}

The subhalo is injected with \texttt{gta.add\_source()} and, adopting a specific input WIMP spectrum (defined by its mass and annihilation channel) as obtained by means of DMFit \cite{Jeltema2008}, we run \texttt{$\mathrm{\texttt{gta.simulate\_roi()}}$} to simulate the full ROI with the photon events\footnote{This procedure assumes a good description of the background components. Yet, being our interest subhalos well outside the Galactic plane with no nearby sources, large deviations from this assumption are not expected. \label{foot:injection}}. Then, we free the subhalo flux normalization, defined as,

\begin{equation}
\label{eq:normalization}
\mathcal{N}=J \times \langle\sigma v\rangle~\left[\mathrm{GeV^2\cdot cm^{-2}\cdot s^{-1}}\right],
\end{equation}

\noindent where $J$ is the J-factor and $\langle\sigma v\rangle$ is the velocity-averaged annihilation cross section as both introduced in Eq.~\ref{eq:j-factor}. Indeed, the spectrum normalization is completely degenerate between these two quantities, and therefore we will discuss the results based on this overall normalization from now on. In practice, given a certain value of J-factor or $\langle\sigma v\rangle$, it will be trivial to obtain the complementary quantity as a function of $\mathcal{N}$.

To compute the detection significance of the source, we adopt the Test Statistic \cite{Rico_2020},

\begin{equation}
\label{eq:TS_det}
\mathrm{TS_{det}}=2\cdot\textrm{log}\left[\frac{\mathcal{L}(H_1)}{\mathcal{L}(H_0)}\right],
\end{equation}

\noindent where $\mathcal{L}(H_0)$ and $\mathcal{L}(H_1)$ are, respectively, the likelihoods under the null (no source) and alternative (existing DM source) hypotheses. In a similar fashion, one can define the extension significance of a source,

\begin{equation}
\label{eq:TS_ext}
\mathrm{TS_{ext}}=2\cdot\textrm{log}\left[\frac{\mathcal{L}(EXT)}{\mathcal{L}(PS)}\right]
\end{equation}

\noindent where $\mathcal{L}(EXT)$ is the likelihood of the extended hypothesis and $\mathcal{L}(PS)$ is the one of the point source. To ensure either a robust source detection or an unequivocal extended source, we will search for the minimum flux value ($F_{min}$) that would yield at least $\mathrm{TS_{det}}=25$ or $\mathrm{TS_{ext}}=25$, respectively.\footnote{This threshold is approximately a 5-$\sigma$ significance in the case of detection, as $\textrm{TS}\sim\sigma^2$.} Note that, as it will be shown later in this same section, it is not possible to simultaneously have $\mathrm{TS_{det}=TS_{ext}}=25$, as for a source to appear unequivocally as extended in the analysis, ($\mathrm{TS_{ext}}=25$), the required signal must be properly characterized, thus exhibiting $\mathrm{TS_{det}}\gg25$.

We use the \texttt{gta.optimize} and \texttt{gta.fit} \textit{fermipy} subroutines to simultaneously fit all the sources in the ROI and to compute the spectral properties of the subhalo, including its $\mathrm{TS_{det}}$. We then run the code iteratively, increasing or decreasing the normalization, until the source exhibits just $\mathrm{TS_{det}}=25$.\footnote{Throughout this section, we will assume a $\pm3$ numerical tolerance in all TS computations in order to ensure convergence.} At this point, the obtained integrated flux is $F_{min}^{det}$.

In order to compute $F_{min}^{ext}$, i.e., the minimum flux required to have $\mathrm{TS_{ext}}=25$, we will follow a similar procedure: once the source is detected, a fit to the spatial extension is performed via \texttt{gta.extension()}. There are two available 2D fitting templates in \textit{fermipy}, a 2D uniform disk and a 2D Gaussian profile. We adopt the latter from now on, as no differences were found when using the 2D disk. Although neither the Gaussian nor the disk profile are a perfect model of the expected DM source spatial profile, an important advantage of using these generic spatial templates is that they are agnostic to the real, underlying annihilation profile. Indeed, should a DM subhalo be detected with the LAT, the specific spatial template would be unknown and, therefore, the computations reported in this work would be actually closer to reality. This time, as we are only interested in the preference for extension with respect to the PS hypothesis, $\mathrm{TS_{ext}}$ (Eq. \ref{eq:TS_ext}), we vary iteratively the normalization until we reach the required $\mathrm{TS_{ext}}=25$. At that point, we save the integrated flux of the source, $F_{min}^{ext}$.\footnote{We note that the convergence of the code in this case is slower, as the analysis of extension does not behave as smoothly as the detection, due to the photon spill over, which makes the relation between  $\mathrm{TS_{ext}}$ and normalization more unstable (see Figure \ref{fig:signal_morphology_comparison} in Appendix \ref{app:signal_normalization}).}

\begin{figure*}[!ht]
\centering
\includegraphics[width=0.45\linewidth]{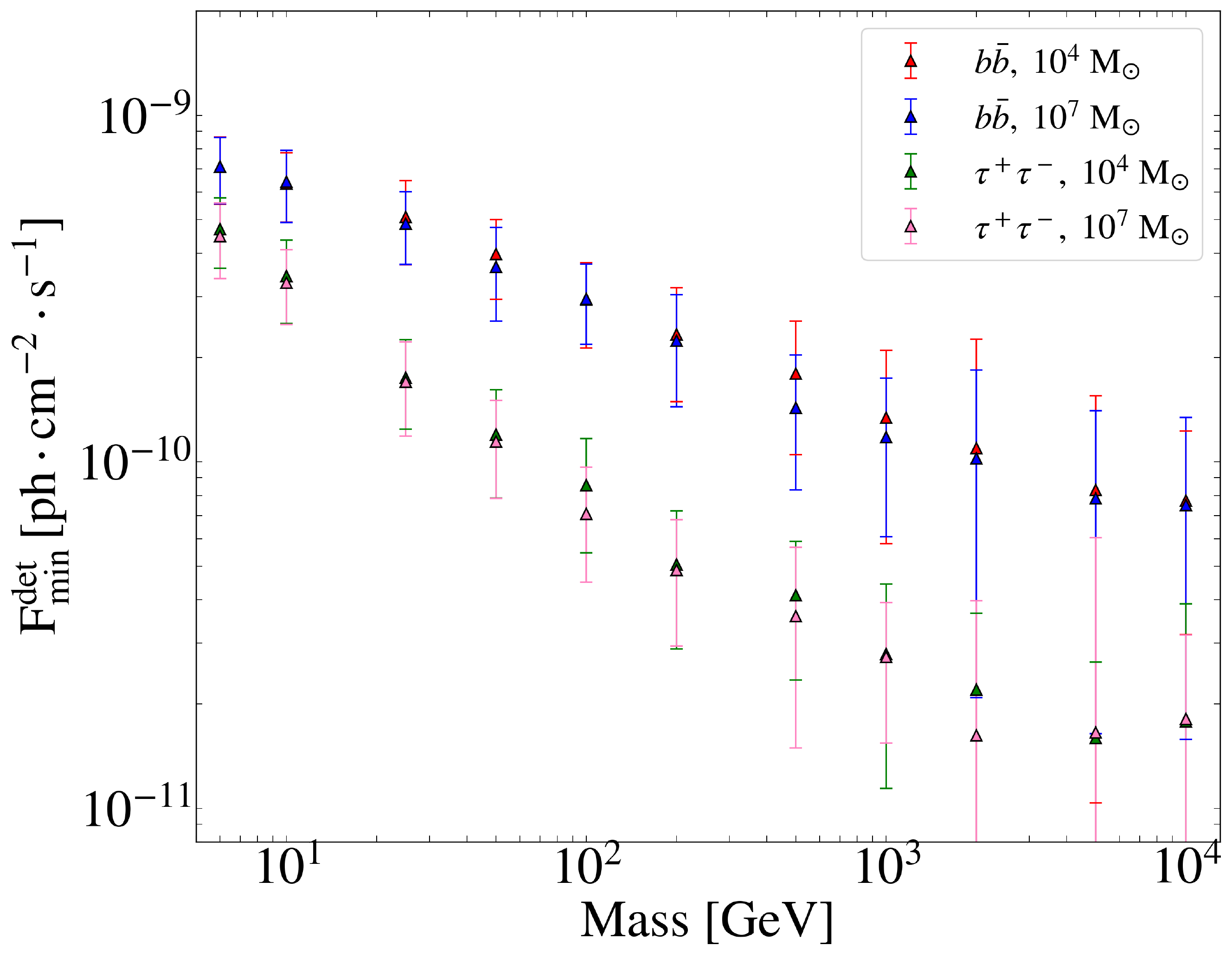}
\includegraphics[width=0.45\linewidth]{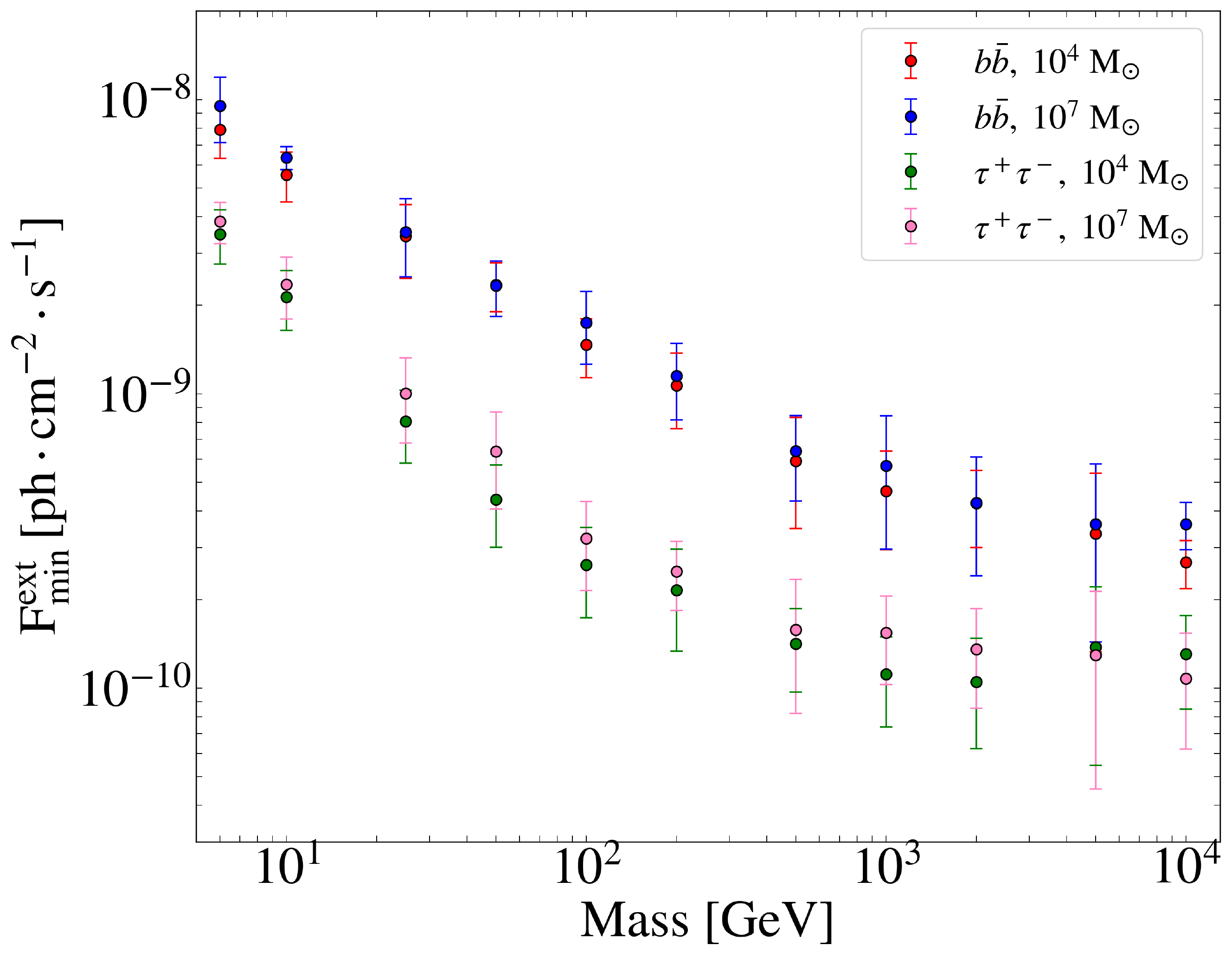}
\caption{Comparison between minimum fluxes for extended subhalos {as a function of the mass of the DM candidate}. {Left panel:} Minimum flux for detection ($\mathrm{TS_{det}=25}$), for the two considered annihilation channels and subhalo models. {Right panel:} The same for the case of minimum flux for extension ($\mathrm{TS_{ext}=25}$). The uncertainty bars come from averaging across ten realization {of the same simulation setup}. Note the different flux scale in each panel.}
\label{fig:fmin_det_fmin_ext_all}
\end{figure*}

Finally, in order to have proper statistics, we repeat each configuration of subhalo modeling, WIMP mass, and annihilation channel 10 times for both our evaluation of subhalo detection and detection of extension. The results of both $F_{min}^{det}$ and $F_{min}^{ext}$ for every channel and subhalo model are shown in Figure \ref{fig:fmin_det_fmin_ext_all}, where the threshold flux is lowered for larger masses in every case. As can be seen, the two considered subhalo models are completely degenerate from the point of view of detection and extension analysis. As a result, in the remaining of this section we will focus on only one of them, the conclusions being identical for the other.

\begin{figure*}[!ht]
\centering
\includegraphics[width=0.45\linewidth]{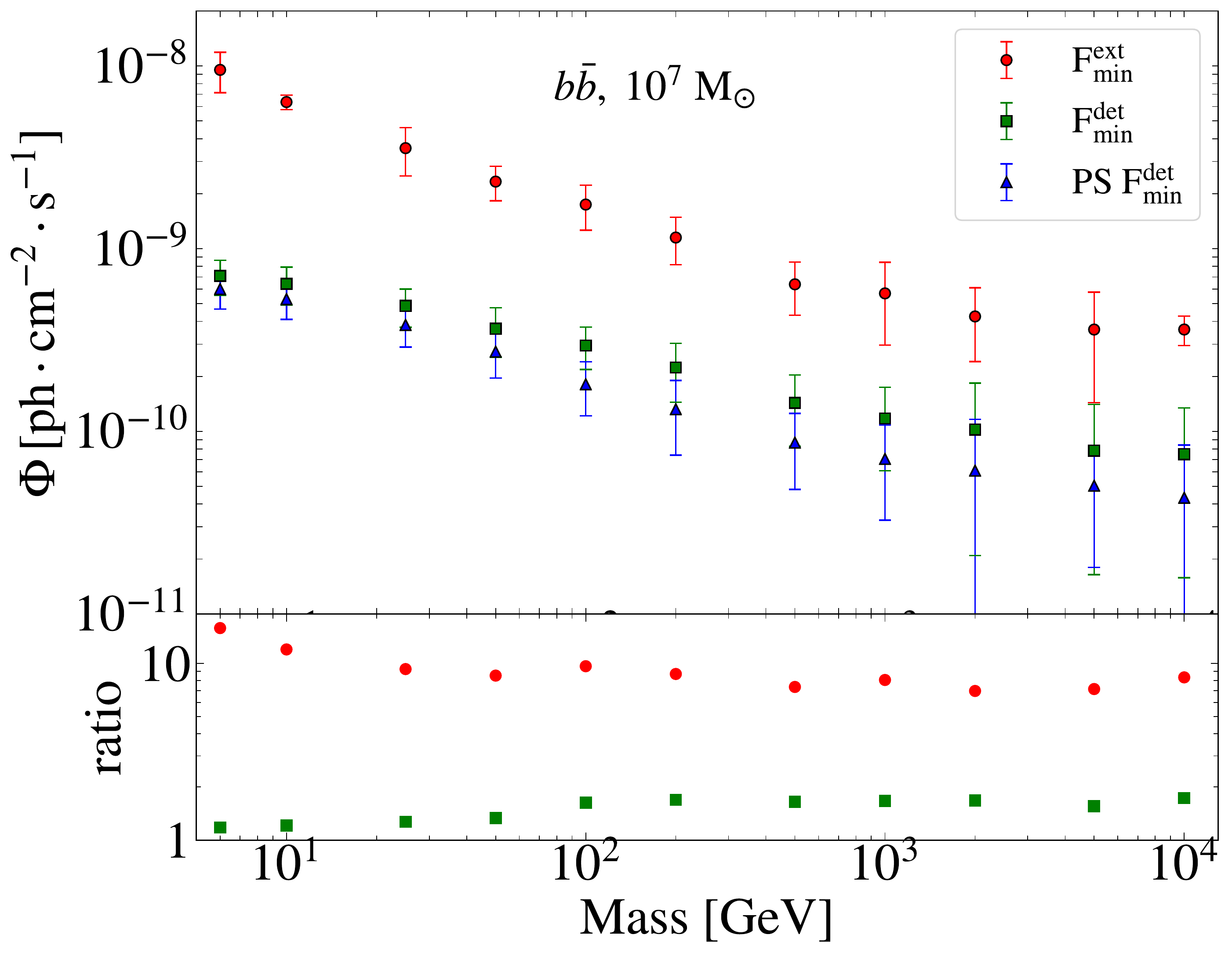}
\includegraphics[width=0.45\linewidth]{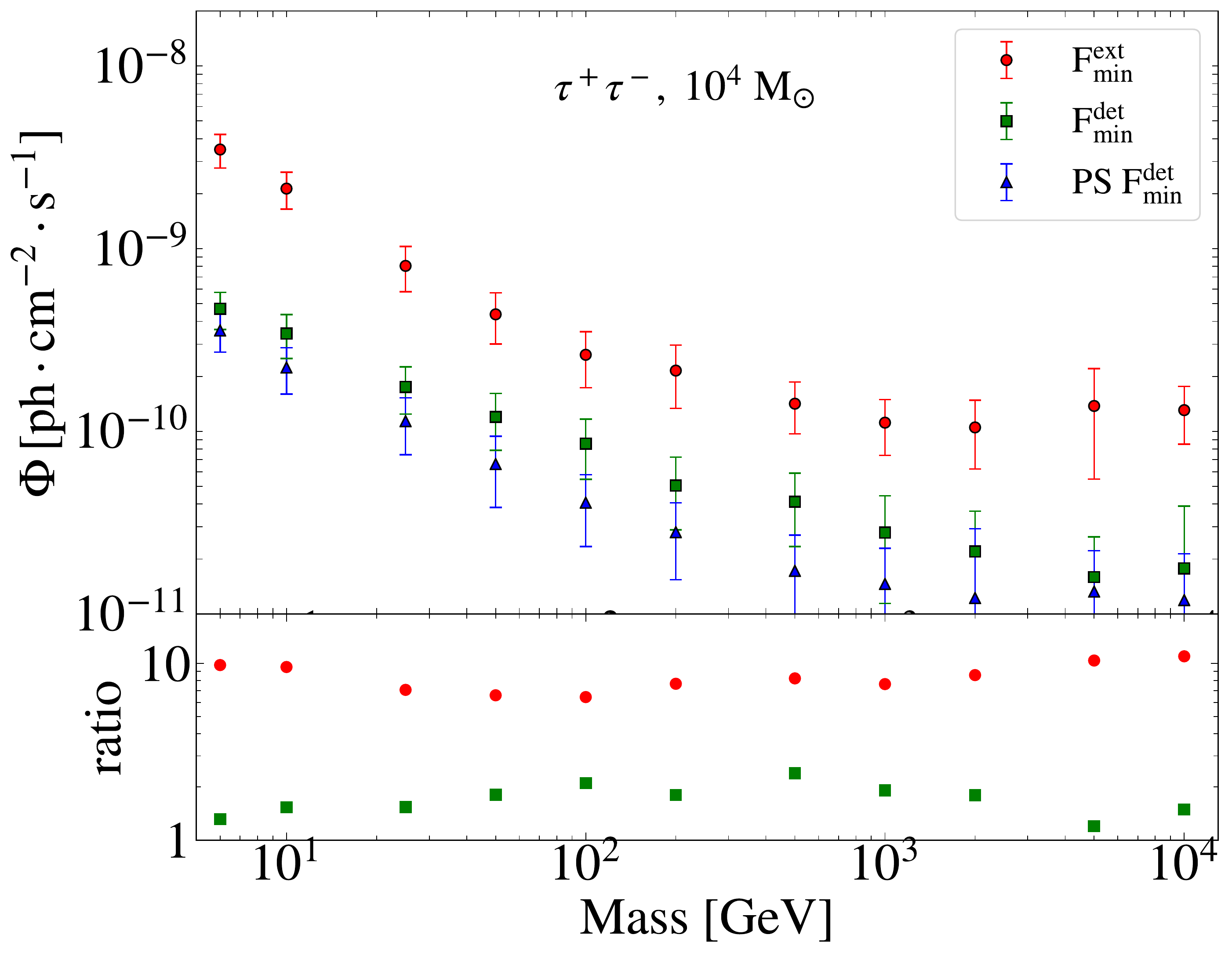}
\caption{  Minimum fluxes for point-source detection ($\mathrm{TS_{det}}=25$; PS $F_{min}^{det}$), extended subhalo detection ($\mathrm{TS_{det}}=25$; $F_{min}^{det}$), and detection of extension ($\mathrm{TS_{ext}}=25$; $F_{min}^{ext}$), for $b\bar{b}$ (left panel) and $\tau^+\tau^-$ (right panel) annihilation channels. Lower panels show the ratio between $F_{min}^{det}$ (green squares) and $F_{min}^{ext}$ (red circles) when compared to the point-source detection threshold (PS $F_{min}^{det}$).}
\label{fig:fmins_bb_tautau}
\end{figure*}

We can also compare the detection and extension fluxes with respect to the PS case, i.e., with no spatial template. This is shown in Figure \ref{fig:fmins_bb_tautau}. In both Figure \ref{fig:fmin_det_fmin_ext_all} and \ref{fig:fmins_bb_tautau} we note that annihilation into $\tau^+\tau^-$, yielding a harder spectrum than the $b\bar{b}$ annihilation channel, is easier to detect (i.e., requires a lower flux). There is a consistent trend of obtaining lower fluxes for larger WIMP masses in every case. Moreover, it is interesting to compare the ratios between the three of them; when analyzing detection ($\mathrm{TS_{det}=25}$), we can distinguish between the PS and extended cases. The latter turns out to be a factor $1.5-2$ larger (meaning more difficult to detect), as the photons are spatially spread, and therefore the sensitivity is slightly decreased. Nevertheless, being that the DM annihilation profile is very cuspy (see Section \ref{sec:DM_modeling}), most of the signal comes from the innermost part of the profile, and there is little difference with respect to the PS case. This is no longer true for the case of detection of extension ($\mathrm{TS_{ext}}=25$), where the required fluxes are a factor $8-10$ larger than in the PS case (and therefore a factor $4-6$ larger than the $F_{min}^{det}$ for the extended template). This trend is observed in every channel and subhalo model.

\begin{figure*}[!ht]
\centering
\includegraphics[width=0.45\linewidth]{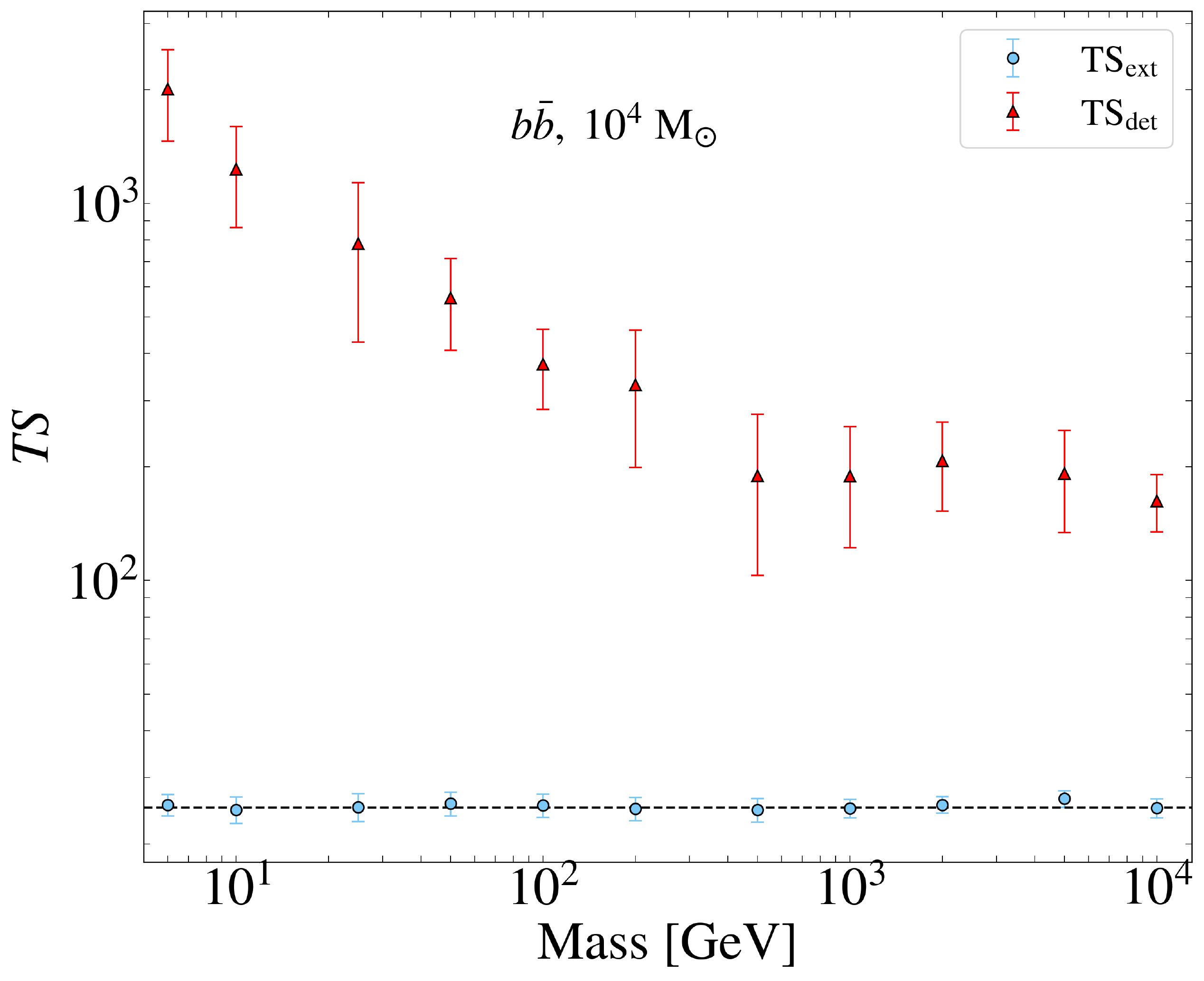}
\includegraphics[width=0.45\linewidth]{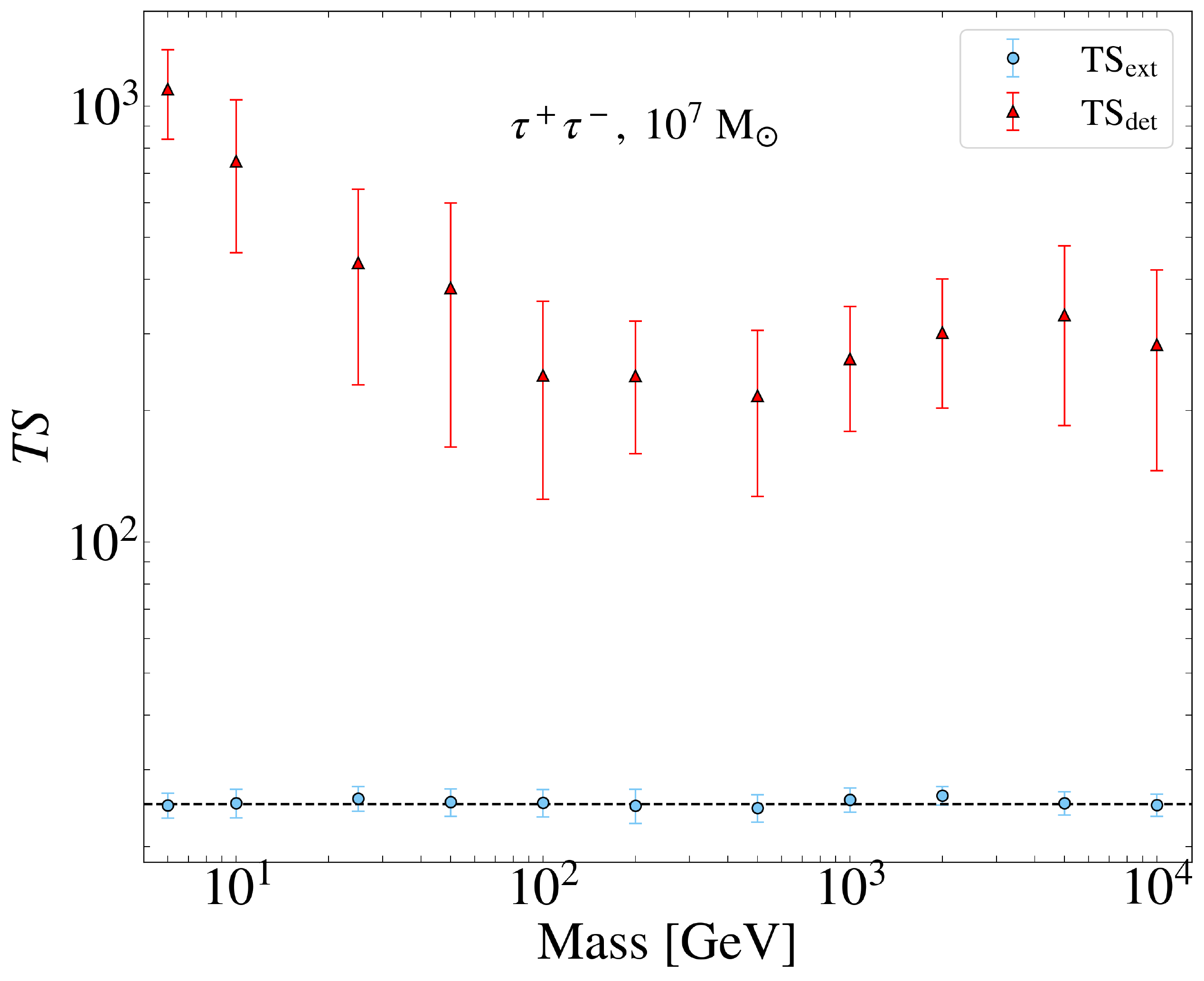}
\caption{Comparison between the TS of detection ($\mathrm{TS_{det}}$) and TS of extension ($\mathrm{TS_{ext}}$) for the case of achieving $F_{min}^{ext}$, i.e., $\mathrm{TS_{ext}}=25$ (we use a tolerance of $\pm3$), {  as a function of the mass of the DM candidate}. Left panel shows the results for $b\bar{b}$ and right panel for $\tau^+\tau^-$ annihilation channel.}
\label{fig:fmin_tsdet_tsext}
\end{figure*}

Overall, the different behaviors found for both annihilation channels can be explained due to the convolution of the LAT sensitivity and the peak energy of the DM spectrum: the LAT is more sensitive to energies $\sim1-10$ GeV, while the DM spectrum peaks at energies $\sim m_{\chi}/30$ in the case of $b\bar{b}$ and $\sim m_{\chi}/3$ in the case of $\tau^+\tau^-$. Therefore, depending on the annihilation channel, increasing the WIMP mass will either improve the instrument's performance or reach a sensitivity plateau. In the case of $b\bar{b}$, the spectrum peaks at the LAT maximum sensitivity for WIMP masses $m_{\chi}\sim50-500$ GeV, while in the case of $\tau
^+\tau^-$, this maximum sensitivity is reached for masses $m_{\chi}\sim5-50$ GeV. Thus, for masses larger than these values, the corresponding $\mathrm{TS_{det}}$ required to have $\mathrm{TS_{ext}=25\pm3}$ is approximately constant. These trends are observed in Figure \ref{fig:fmin_tsdet_tsext}.

As a general conclusion of this Section, our results allow for a quick evaluation of a potential DM subhalo candidate found among the LAT unIDs -- depending on the value of $\mathrm{(TS_{det}, TS_{ext})}$, one can evaluate how likely it is that the emission is coming from a DM subhalo, according to the previous figures. {  Indeed, for subhalo candidates whose best-fit DM masses and annihilation cross sections were not yet discarded by the best DM constraints to date \cite{2017ApJ...834..110A,Di_Mauro_2021}, the relation between their values of $\mathrm{TS_{det}}$ and $\mathrm{TS_{ext}}$ should agree with the results in this section. This way spatial extension information would be used as an additional DM `filter' to spot potential dark subhalos.}

This is especially interesting when considering the most robust LAT DM constraints \cite{2017ApJ...834..110A} (see also \cite{Di_Mauro_2021}), which do not constrain canonical thermal WIMPs for masses larger than $\sim50$ GeV and $\sim100$ GeV in the case of $b\bar{b}$ and $\tau^+\tau^-$ final states, respectively. A putative candidate, not discarded by these constraints, should be evaluated according to the results in this Section, taking the information on expected spatial extension as an additional DM ``filter'' to spot potential dark subhalos.

\subsection{Recovery of spatial information from dark subhalos}
\label{sec:LAT_analysis}

Once the LAT threshold sensitivity to extended dark subhalos has been properly characterized, we can study a more general performance of the instrument in a wider range of subhalo configurations, varying the same parameters as in Section \ref{sec:fmin} (different subhalo models, WIMP masses, annihilation channels) and including new normalization values -- not searching for a specific value of detection/extension significance but rather blindly studying the output with different normalizations.

To do so, we will first define the grid of models to be studied. As in Section \ref{sec:fmin}, we consider two subhalo models, ($\mathrm{10^4~M_{\odot}}$, 0.5 kpc) and ($\mathrm{10^7~M_{\odot}}$, 10 kpc), that were fully described in Section \ref{sec:DM_modeling}. We will consider 11 WIMP masses that span over the WIMP range, $m_\chi=\left[6, 10, 25, 50, 100, 200, 500, 1000, 2000, 5000, 10000\right]$ GeV, and the same two annihilation channels as before are selected as benchmark models, $b\bar{b}$ and $\tau^+\tau^-$. Finally, we choose an $\mathcal{N}$ grid which will be representative of viable DM models (even if extreme in some cases), with 15 values logarithmically spaced between $\mathcal{N}=10^{-6}-10^{-3}$ (i.e., 5 values per order of magnitude). The lower limit would be representative of, e.g., a $\langle\sigma v\rangle=10^{-26}~\mathrm{cm^3\cdot s^{-1}}$ and $J=10^{20}~\mathrm{GeV^2\cdot cm^{-5}}$, this is, a (quasi)canonical thermal WIMP with a J-factor similar to that of the best dSphs. The upper limit would correspond to, e.g., a $\langle\sigma v\rangle=10^{-24}~\mathrm{cm^3\cdot s^{-1}}$ and $J=10^{21}~\mathrm{GeV^2\cdot cm^{-5}}$ model, i.e., an annihilation cross section roughly 2 orders of magnitude above the thermal value, and an extreme J-factor value according to our results from N-body simulations (see Figure \ref{fig:vliirep}), therefore an extreme yet possible DM scenario.

With this grid definition and set of models, we  proceed similarly to Section \ref{sec:fmin}, this time directly saving the output of each setup rather than forcing the analysis to converge to some predefined TS values. This is, for each 10 repetitions of the ($m_\chi$, $\mathcal{N}$, channel, subhalo model) combination, we perform the analysis of the ROI and, provided the source is detected ($\mathrm{TS_{det}\geq25}$), we search for spatial extension and perform a spectral fit to DM. A discussion on the precision of \textit{fermipy} to recover the spectral information is deferred to Appendix~\ref{app:precision_dm_spectrum}. This way we obtain a full characterization of the source, including its detection significance, integral flux, extension significance, angular extension and best-fit DM parameters.

\begin{figure*}[!ht]
\centering
\includegraphics[height=6.2cm]{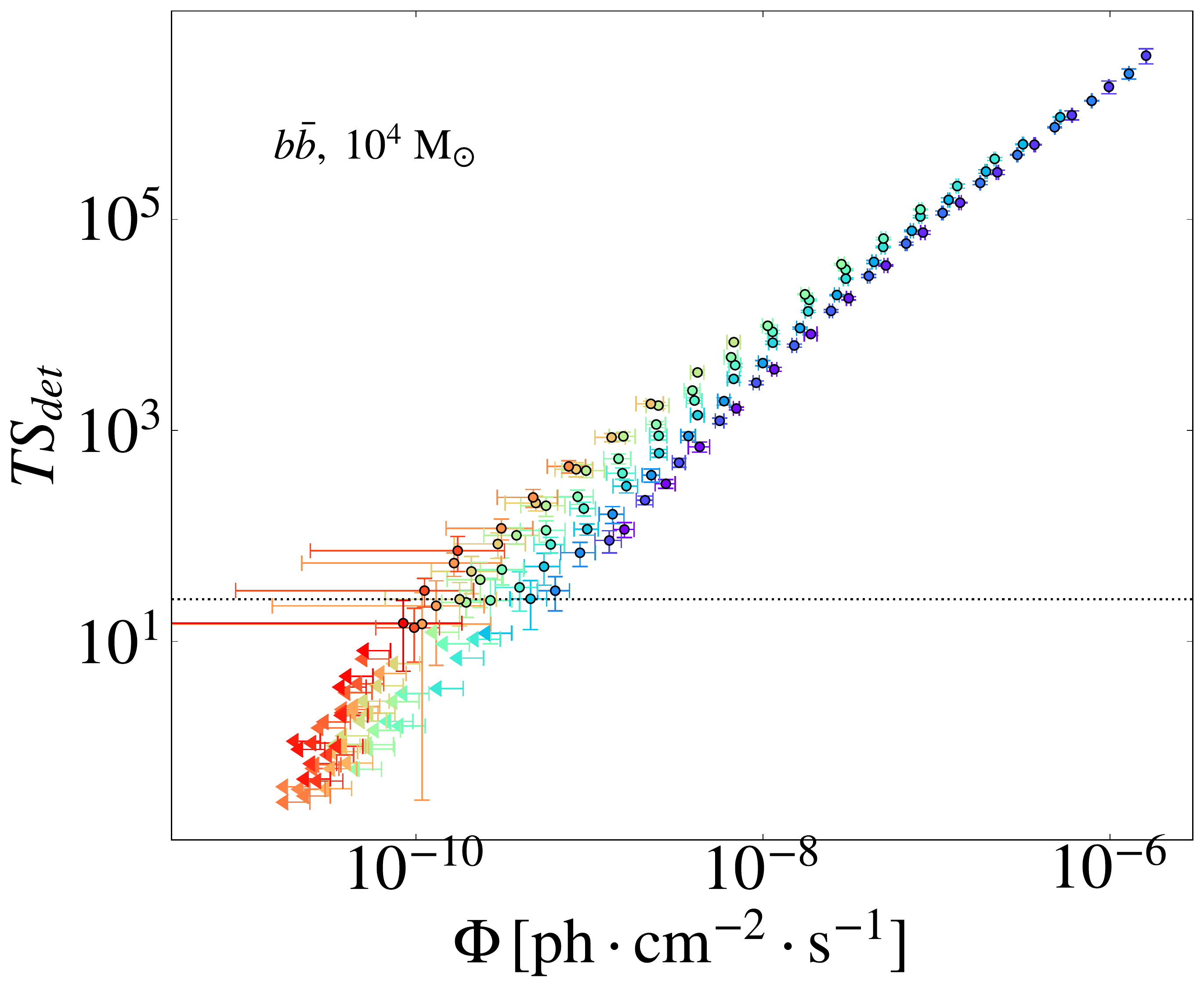}
\includegraphics[height=6.2cm]{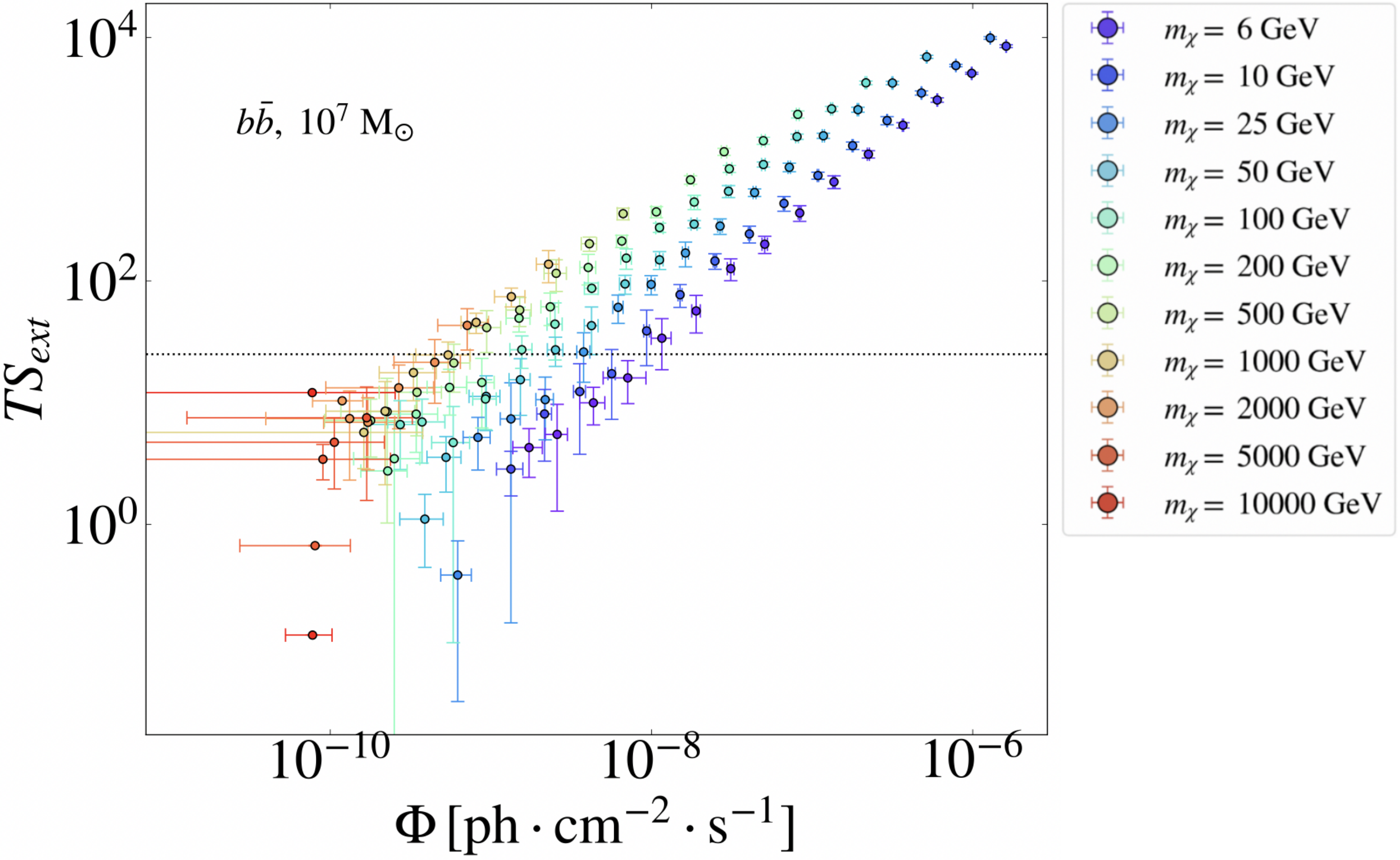}
\caption{Comparison of the integrated source flux with the corresponding detection ($\mathrm{TS_{det}}$; left panel) and extension significances ($\mathrm{TS_{ext}}$; right panel), for the two subhalo models and $b\bar{b}$ annihilation channel, {  as a function of the source flux.} Each point is averaged across the 10 different realizations of the same setup. As in Section \ref{sec:fmin}, the two subhalo models that we consider turn out to be completely indistinguishable within errors, and therefore we show only one of them in each case.}
\label{fig:flux_tsdet_tsext}
\end{figure*}

Figure \ref{fig:flux_tsdet_tsext} shows the photon flux versus both the detection and extension significances. The relations are remarkably linear (in log-space), meaning that the more flux emitted by the subhalo, the easier will be to have a signal detection and the more preference for extension will be found, both as expected. Additionally, the uncertainty in the flux among the 10 realizations gets smaller for larger $\mathcal{N}$, as the source can be better characterized for higher fluxes. 

Interestingly, low WIMP masses are easier to detect and to characterize. Indeed, as the input $\mathcal{N}$ values are all the same for every WIMP mass in Figure~\ref{fig:flux_tsdet_tsext}, we can conclude that the lowest WIMP masses considered in our analysis are always detected and spatially resolved, while this is not true for heavier WIMPs. 
The reason is mainly twofold: i) for a fixed $\mathcal{N}$, the DM spectrum per annihilation is dimmer as we go towards higher energies, and ii) the sensitivity of the LAT peaks at roughly 1--5 GeV, while the energy peak of a DM spectrum is $E_{peak}^{b\bar{b}}\sim m_{\chi}/30$ for $b\bar{b}$ and $E_{peak}^{\tau^+\tau^-}\sim m_{\chi}/3$ for $\tau^+\tau^-$. Therefore, masses larger than $m_{\chi}\sim 150~(50)$ GeV for $b\bar{b}~(\tau^+\tau^-)$ will be increasingly difficult to detect, and therefore progressively more difficult to be spatially resolved as well.

\begin{figure*}[!ht]
\centering
\includegraphics[height=6.2cm]{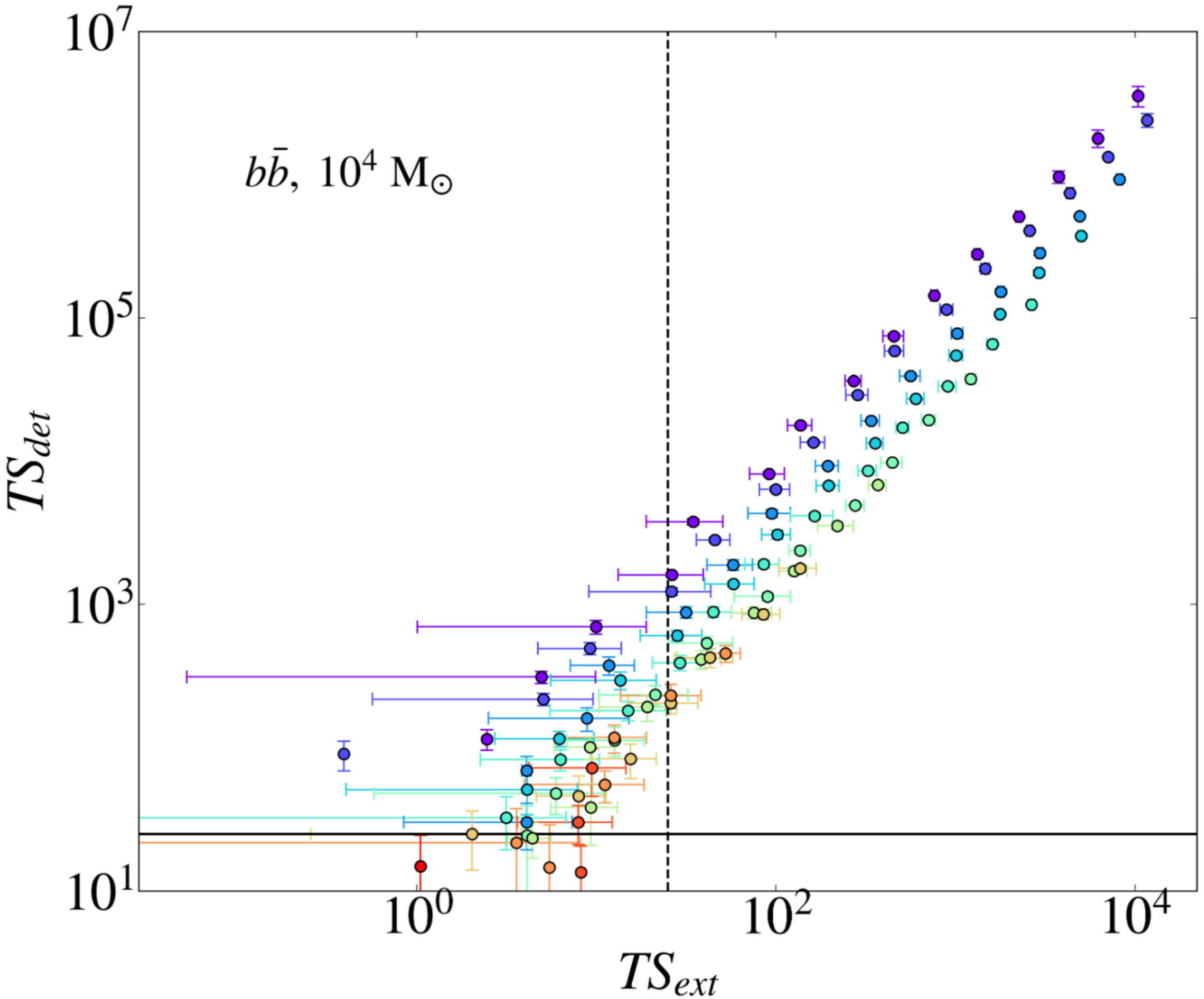}
\hfill
\includegraphics[height=6.2cm]{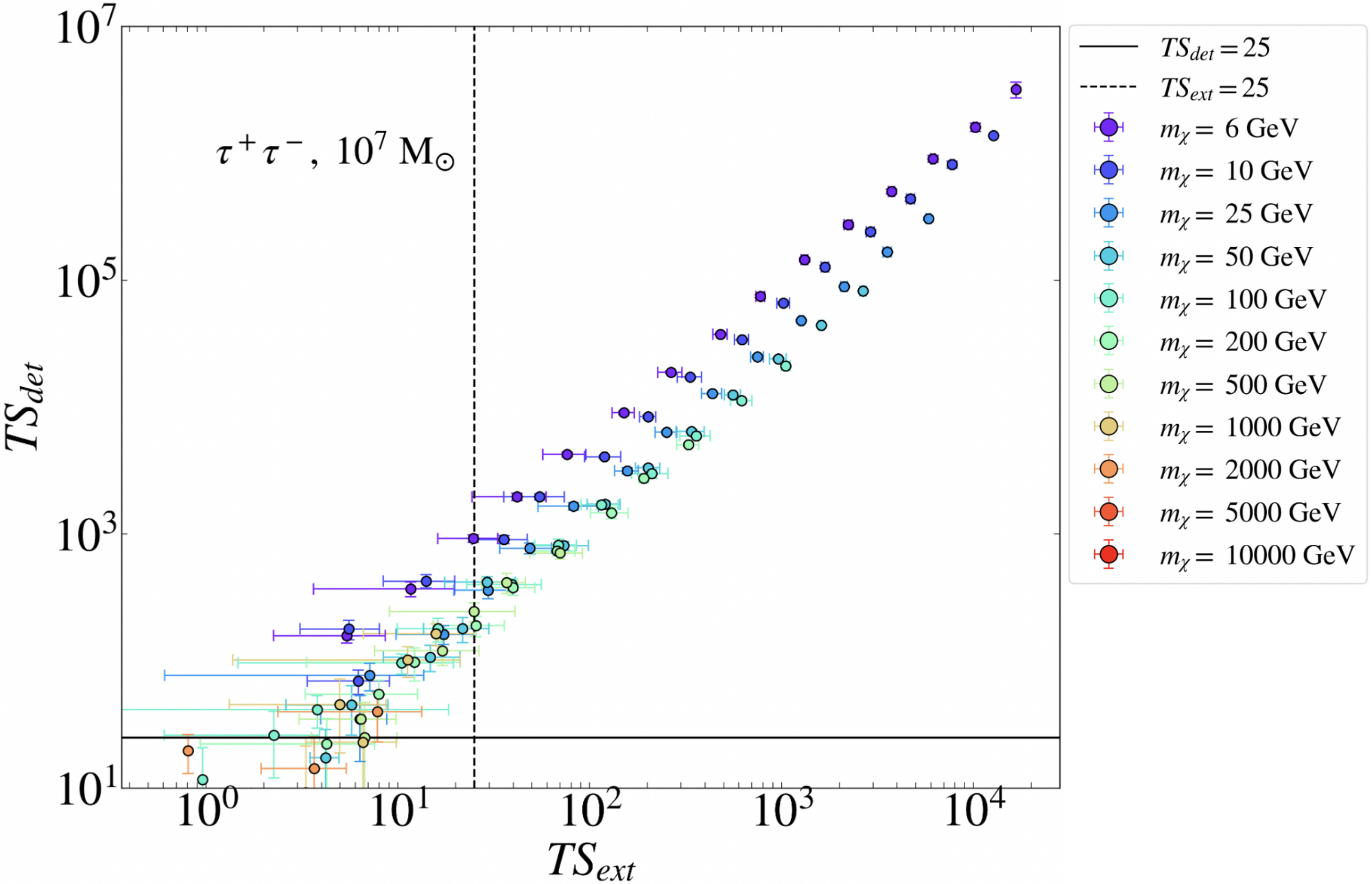}
\caption{{  Detection $\mathrm{TS_{det}}$ as a function of extension detection} $\mathrm{TS_{ext}}$ significances for the ($b\bar{b}$, $\mathrm{10^4~M_{\odot}}$) and the ($\tau^+\tau^-$, $\mathrm{10^7~M_{\odot}}$) models (left and right panels, respectively). Vertical and horizontal lines mark, respectively, $\mathrm{TS_{ext}}=25$ and $\mathrm{TS_{det}}=25$. Each point is averaged across the 10 different realizations of the same setup.}
\label{fig:tsext_vs_tsdet}
\end{figure*}

\begin{figure*}[!ht]
\centering
\includegraphics[height=5.8cm]{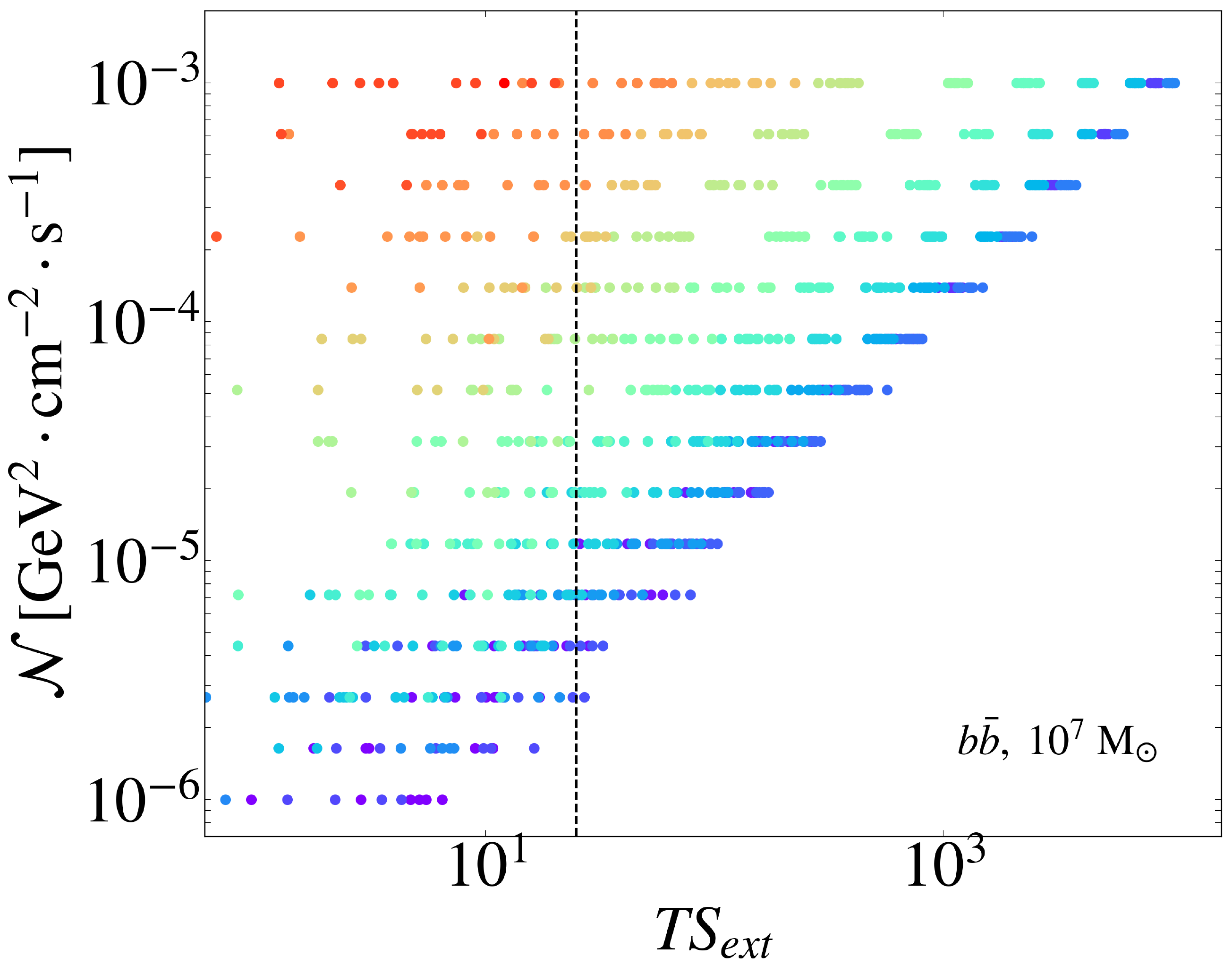}
\hfill
\includegraphics[height=5.8cm]{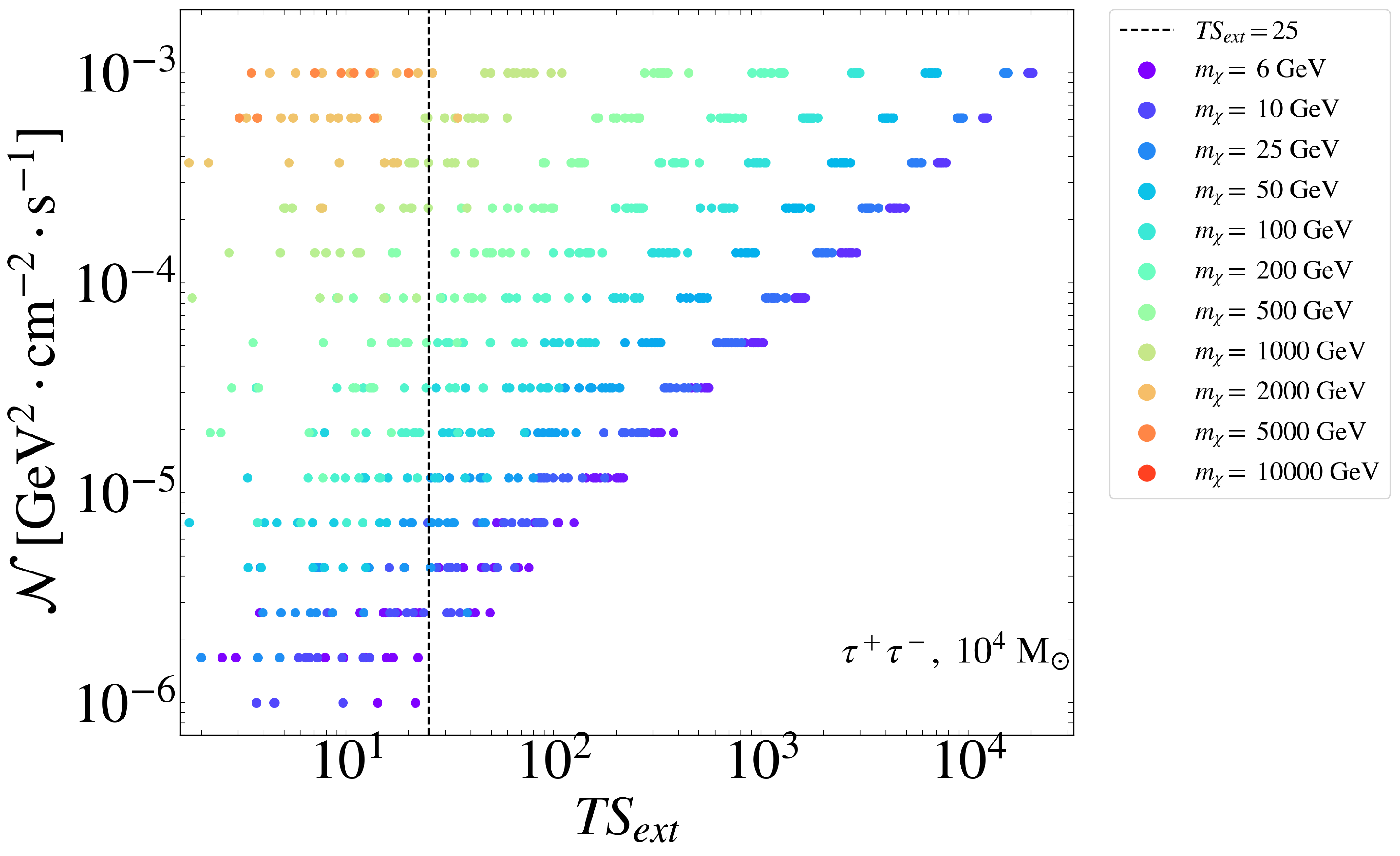}
\caption{Normalization of the source (J-factor $\times~\langle\sigma v\rangle$) versus the extension significance $\mathrm{TS_{ext}}$, for the ($b\bar{b}$, $\mathrm{10^7~M_{\odot}}$) and the ($\tau^+\tau^-$, $\mathrm{10^4~M_{\odot}}$) models, in all realizations. Vertical, black line marks $\mathrm{TS_{ext}}=25$.}
\label{fig:tsext_vs_norm}
\end{figure*}

In Figure \ref{fig:tsext_vs_tsdet}, we show the relation, again linear, between detection and extension significances for both $b\bar{b}$ and $\tau^+\tau^-$, a complementary version of the previous figure. The data points, averaged over the 10 different realizations for each analysis setup, also exhibit a trend of decreasing uncertainty\footnote{We define our uncertainty as quadratic combination of individual error plus standard deviation among realizations.} with increasing $\mathcal{N}$ values, as expected. Note that the uncertainties associated to $\mathrm{TS_{ext}}$ are consistently larger than the ones for $\mathrm{TS_{det}}$ (see Appendix \ref{app:signal_normalization}). This can also be seen in Figure \ref{fig:tsext_vs_norm}, where $\mathcal{N}$ is shown against $\mathrm{TS_{det}}$ and $\mathrm{TS_{ext}}$, in this case for every single realization of the analysis. For small normalization values, only low masses yield a detection, while for the largest values heavy WIMPs also show up.

In our analysis, we find that larger normalization values consistently yield smaller uncertainties, in both individual and averaged realizations. Indeed, this is what one would expect, as the signal is better characterized and the uncertainties in the parameters are reduced. Nevertheless, \cite{Mauro2020} finds the opposite trend, with relative errors up to 50\% for the largest cross sections. We stress that our work is based on results from N-body simulations and real LAT data, while \cite{Mauro2020} used a semi-analytical subhalo population model and mock gamma-ray data.

\begin{figure*}[!ht]
\centering
\includegraphics[width=0.9\linewidth]{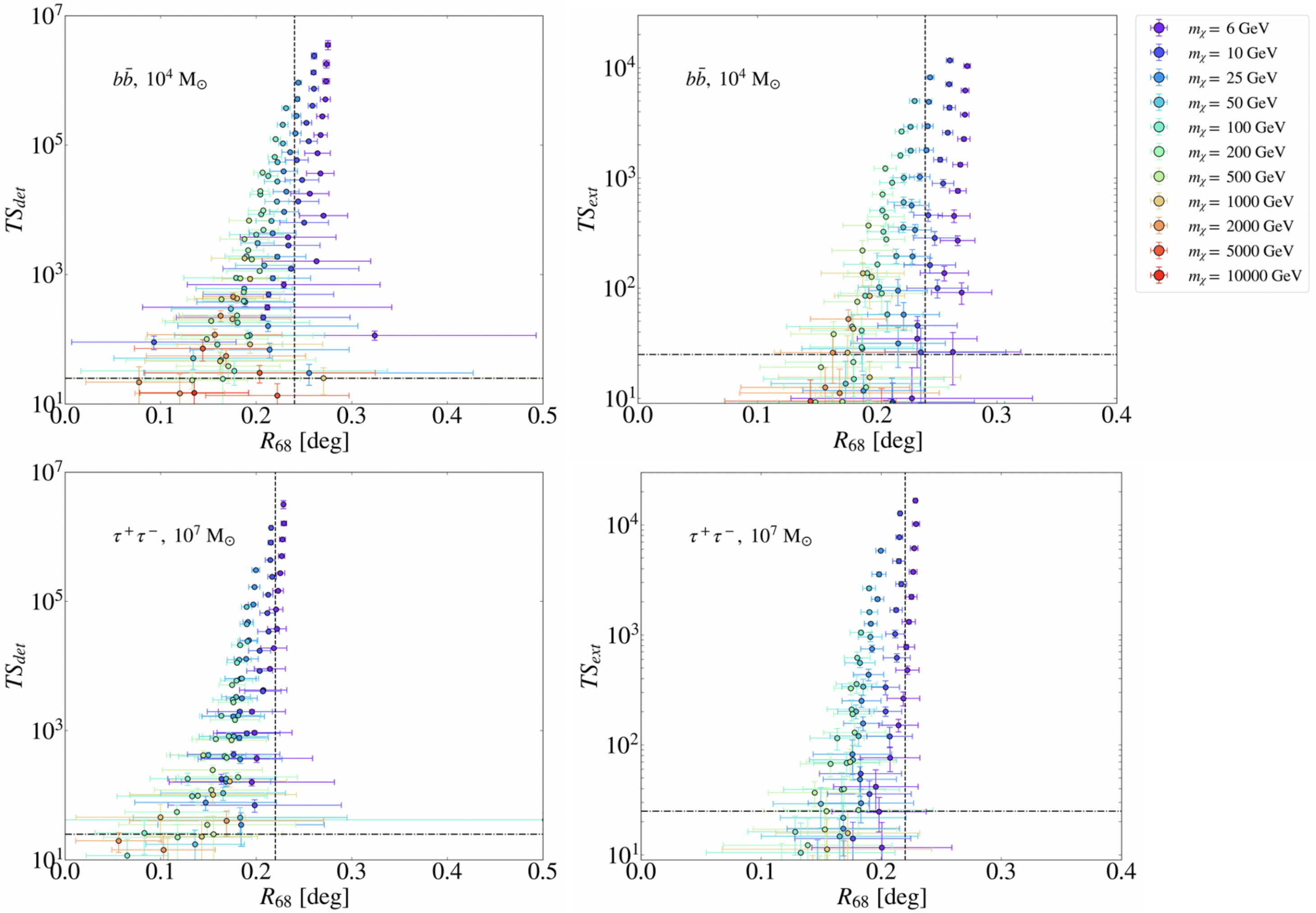}
\caption{Best-fit extension $R_{68}$ versus the extension ($\mathrm{TS_{ext}}$) and detection ($\mathrm{TS_{det}}$) significance, for the ($b\bar{b}$, $\mathrm{10^4~M_{\odot}}$) and the ($\tau^+\tau^-$, $\mathrm{10^7~M_{\odot}}$) models. Each $R_{68}$ is the average value of the 10 realizations, while their uncertainties are computed via a quadratic sum of the individual realization uncertainties and the $1\sigma$ spread across realizations.}
\label{fig:r68_tsdet_tsext}
\end{figure*}

Figures \ref{fig:flux_tsdet_tsext} to \ref{fig:tsext_vs_norm} were mainly focused on the detection/extension significances, yet it is worth also asking what is the actual spatial extension we can recover in case of having an extended signal. This is shown in Figure \ref{fig:r68_tsdet_tsext}, where the best-fit angular extension $R_{68}$ (containing 68\% of the DM-induced emission) is given versus both $\mathrm{TS_{det}}$ and $\mathrm{TS_{ext}}$. We recall here that the predicted $\theta_{68}$ value for the considered subhalo models is $0.22^{\circ}$.

As found in previous figures, one can see that the lowest WIMP masses are the ones better resolved for every $\mathcal{N}$ value, while larger masses have associated larger error bars, and these increase for lower $\mathcal{N}$ values as naturally expected. Additionally, we note that, for a fixed WIMP mass, the central value of $R_{68}$ is stable, regardless of $\mathcal{N}$, while $R_{68}$ will shift to lower values as we increase the WIMP mass. Focusing e.g., on the $b\bar{b}$ case, the lowest WIMP mass predicts $R_{68}\sim0.28^{\circ}$, larger than the theoretical value. For $m_{\chi}\sim 25$ GeV they coincide, while it shifts down to $R_{68}\sim0.18^{\circ}$ for $m_{\chi}\sim 1$ TeV. This is possibly due to i) the fact that we are modeling the spatial extension with a Gaussian profile, which represents only a fair approximation of the input spatial DM profile, as previously discussed; ii) the DM spectrum peaking at higher energies for heavier WIMPs, which, convoluted with the IRFs and PSF, may also partially produce this shift; or iii) most likely, a combination of both effects.

The angular sizes and their associated uncertainties as a function of $\mathrm{TS_{det}}$ and $\mathrm{TS_{ext}}$ may allow us to define an additional ``filter'' that unIDs should pass should they be viable DM subhalo candidates (see \citetalias{CoronadoBlazquez2019a}). Indeed, our results in this Section allow to predict the expected angular extension and significance of the subhalo candidate provided that the unID is bright enough and is well fitted to a DM spectrum. In the most optimistic case, a bright unID source with a spectrum compatible with DM but appearing as strictly point-like could in principle be ruled out as a viable DM candidate. However, in practice most point-like sources cannot be safely ruled out as DM, except for sufficiently large detection significances that would unequivocally predict angular extension. This is so given the typical uncertainties present in the case of low $\mathcal{N}$ values (i.e., faint sources, which is the case for most unIDs). Likewise, if an unID turns out to be a potential subhalo candidate, i.e., it exhibits both a significant spatial extension and prefers a spectral fit to DM, one could start from its $\mathrm{TS_{det}}$, $\mathrm{TS_{ext}}$, $R_{68}$ and $m_{\chi}$ and see where it falls in Figure~\ref{fig:r68_tsdet_tsext}. This exercise will be performed elsewhere on particular dark subhalo candidates among the LAT unIDs.

\section{Implications for WIMP dark matter constraints}
\label{sec:implications}
Our study on the characterization of the spatial extension of a potential dark subhalo signal in LAT data could have relevant implications for the case of setting DM constraints with unIDs. Following the methodology in \citetalias{CoronadoBlazquez2019a}, the constraints on the WIMP mass vs. annihilation cross section parameter space we can set are given by,

\begin{equation}
\label{eq:master_formula}
\langle\sigma v\rangle_{UL}=\frac{8\pi\cdot F_{min}^{det}}{J\cdot N_{\gamma}}~m_{\chi}^2,
\end{equation}

\noindent where $F_{min}^{det}$ is the minimum detection flux (see Section \ref{sec:fmin}), $J$ the J-factor (see Section \ref{sec:DM_modeling}) and $N_{\gamma}$ the integrated DM spectrum per annihilation, also known as the ``particle physics factor'', which is obtained from \cite{Cirelli2011} (see Section \ref{sec:DM_modeling} and Eq. \ref{eq:dm-flux}). The DM spectrum is integrated from 500 MeV, to match the $F_{min}$ computations through this paper. As in our previous papers, the J-factor is defined as the one above which 95\% of the J-factor distribution across different realizations of the Galactic subhalo population is contained, thus allowing us to derive 95\% CL upper limits (see \citetalias{CoronadoBlazquez2019a} for more details). Throughout this Section, we will assume this same definition when discussing specific J-factor values.

There are two main differences with respect to our previous constraints in \citetalias{CoronadoBlazquez2019a} and \citetalias{CoronadoBlazquez2019b}, namely:

\begin{itemize}
    \item {Point sources:} The constraints were set assuming point-like sources, i.e., using a detection threshold flux $F_{min}^{PS}$, while in Section \ref{sec:fmin} we computed how the actual flux for extended source detection is degraded by a factor $\sim1.5$ (see Figure \ref{fig:fmin_tsdet_tsext}).
    
    \item {J-factor:} The J-factor was computed by integrating the signal up to the scale radius ($J_s$), as we expected this to be representative of the emission that may be observed by the LAT. However, as seen in Figure \ref{fig:degeneracy}, $\theta_s$, i.e., the radius subtended by the scale radius, is a factor $\sim3$ larger than the one given by $\theta_{68}$, containing 68\% of the emission -- the theoretical equivalent of the \textit{fermipy} $R_{68}$.
\end{itemize}

Our work in the previous sections allows us to finally quantify the expected spatial signal of dark subhalos, whose angular size is far from being $\theta_s$. In Figure \ref{fig:r68_tsdet_tsext} it can be seen that the {\it recovered} angular extension is centered around $0.3^\circ$, and can take larger or lower values depending on the WIMP mass (this is also true for the considered subhalo models). \footnote{Ideally, one would need to analyze further subhalo models; yet, each model is extremely expensive from the computational point of view, as it requires thousands of individual analysis runs. It is a remarkable coincidence that two different subhalo models present almost identical spatial properties, and are also coincident with the theoretical containment angle. Therefore, this value is motivated as a conservative choice to ensure that 68\% of the subhalo emission is contained, yet is not integrating too much signal.} Therefore, we now study the impact in the DM constraints of integrating the J-factor just up to the innermost $0.3^\circ$ of subhalos (we refer to it as $J_{03}$ instead of up to $\theta_s$ as before in \citetalias{CoronadoBlazquez2019a} and \citetalias{CoronadoBlazquez2019b}).

\begin{figure}[!ht]
\centering
\includegraphics[width=1\linewidth]{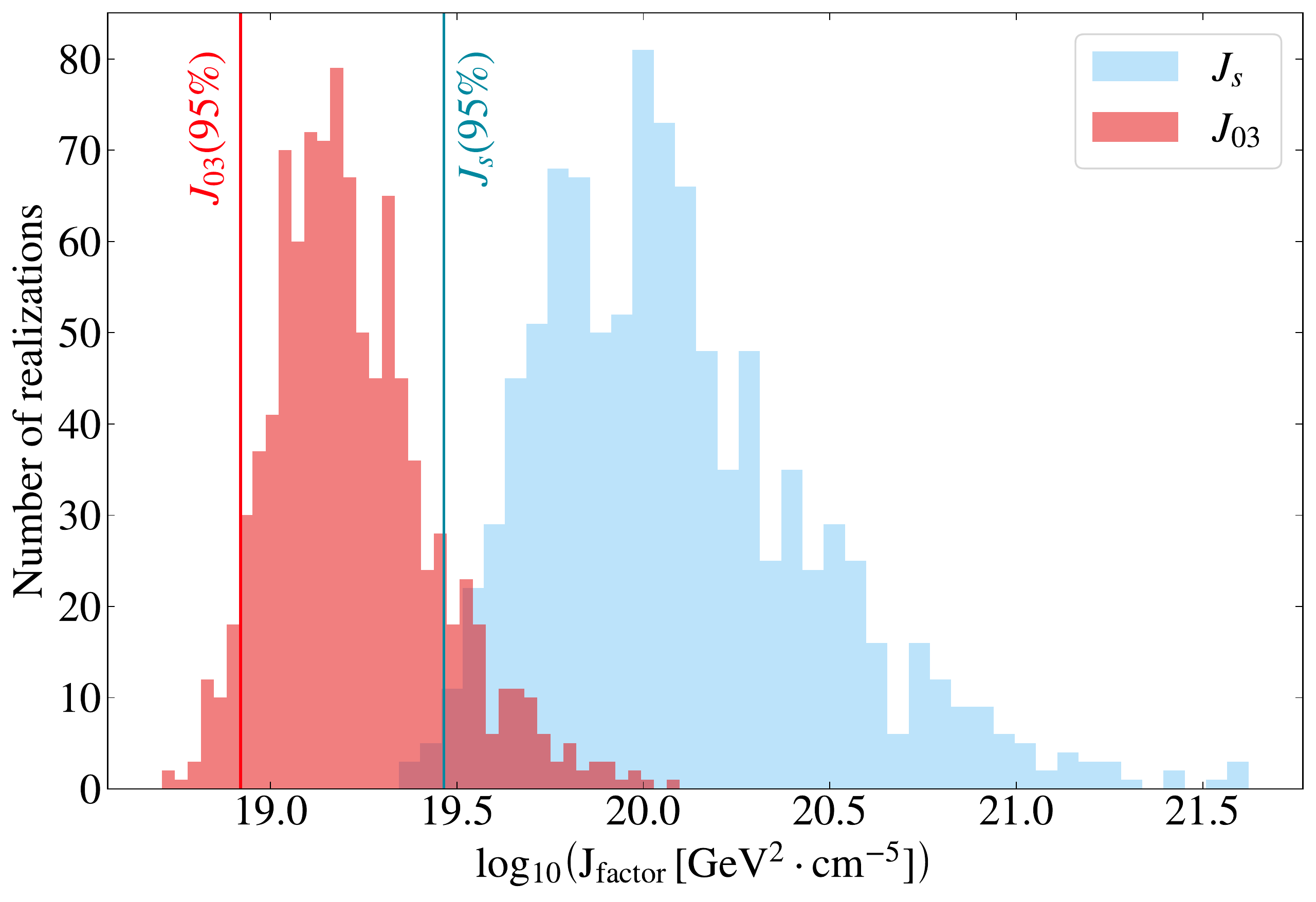}
\caption{Distribution of the brightest subhalo J-factor, across 1000 realizations of the N-body simulation (see \citetalias{CoronadoBlazquez2019a} for more details), integrating the signal either up to the scale radius (blue) or the innermost $0.3^\circ$ (red). Vertical lines mark the value at which 95\% of the distribution is contained, i.e., the J-factor used to set the DM constraints.}
\label{fig:jfact_dist_rs_03}
\end{figure}

We can also estimate how the resulting J-factor will degrade compared to the one used in \citetalias{CoronadoBlazquez2019a} and \citetalias{CoronadoBlazquez2019b}. In Figure \ref{fig:jfact_dist_rs_03}, we show how the distribution of the 95\% C.L. J-factor changes when integrating the signal either up to the scale radius or to the innermost $0.3^\circ$, for the brightest subhalo\footnote{Indeed, this would coincide with our result in \citetalias{CoronadoBlazquez2019b} for the $\tau^+\tau^-$ annihilation channel, where only one source was found to be compatible with a DM hypothesis.}. As in the previous papers, we draw the distribution filtered by mass ($\mathrm{M\leq10^7~M_{\odot}}$) and Galactic latitude ($\mathrm{|b|>10^\circ}$). The J-factor shifts towards lower values, up to a factor $\sim4$, which will directly impact the achievable constraints.

\begin{figure}[!ht]
\centering
\includegraphics[width=1\linewidth]{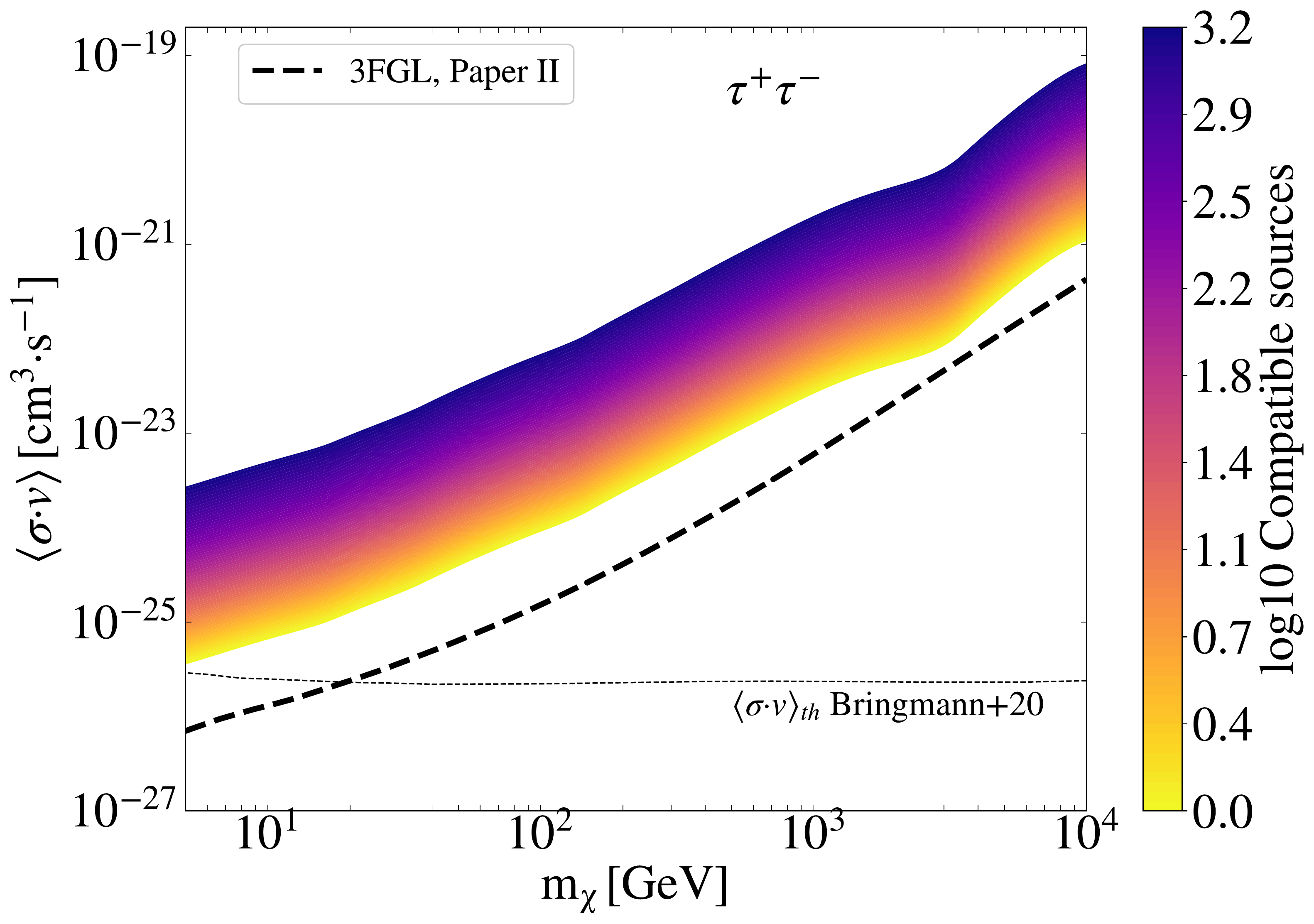}
\includegraphics[width=1\linewidth]{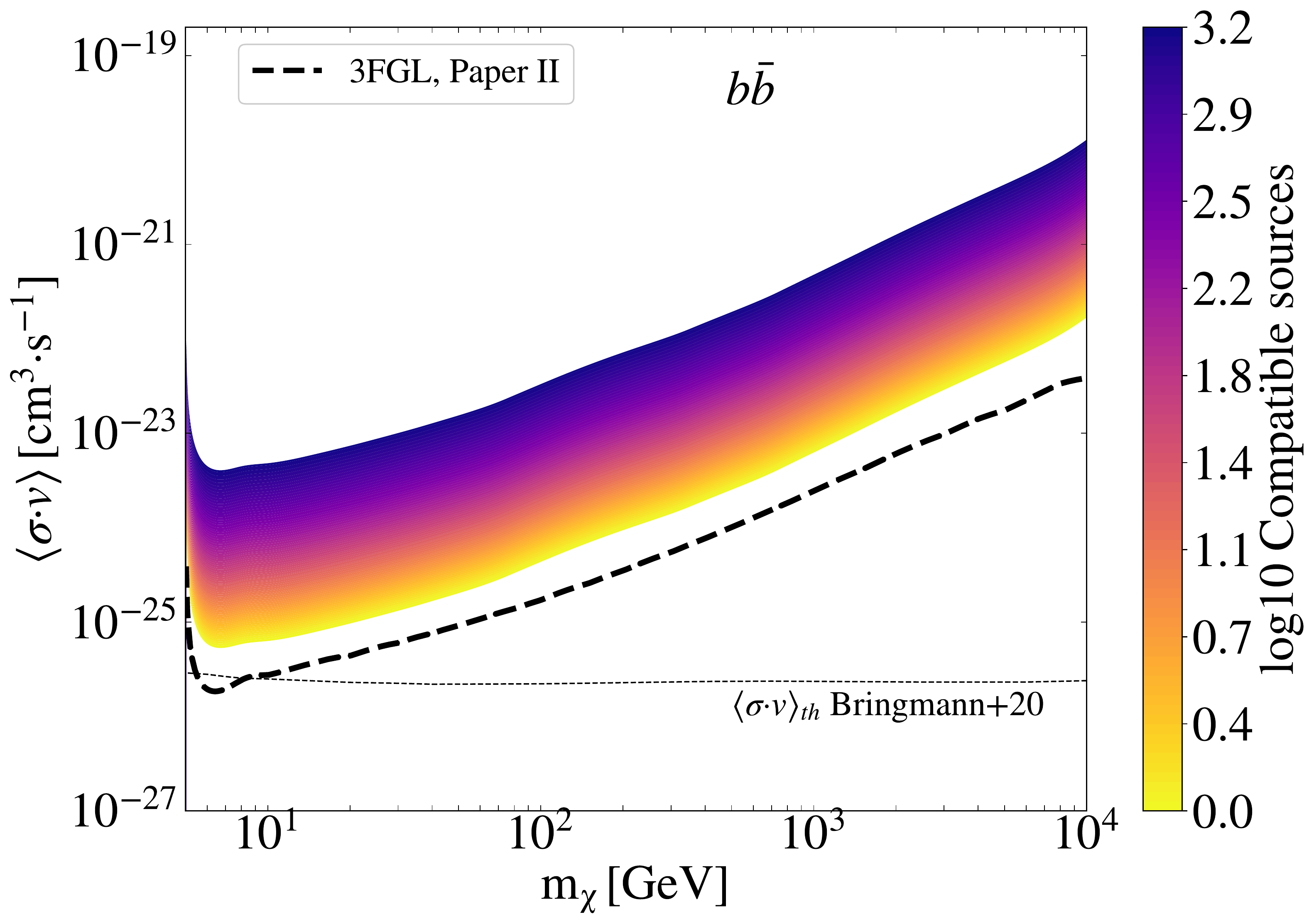}
\caption{95\% C.L. sensitivity predictions for DM annihilating to $\tau^+\tau^-$ (top panel) and $b\bar{b}$ (bottom panel), as a function of the number of 4FGL--DR2 unIDs compatible with DM. Black, dashed line is the constraint obtained in \citetalias{CoronadoBlazquez2019b} using the previous 3FGL catalog \cite{Acero2015} with 1 (5) compatible sources in the case of $\tau^+\tau^-$ ($b\bar{b}$) channel, starting the analysis at 100 MeV, assuming point-like sources and integrating the J-factor up to the scale radius. The horizontal dashed line is the thermal relic cross section \cite{Bringmann2020}.}
\label{fig:constraints_contour}
\end{figure}

In Figure \ref{fig:constraints_contour}, we show {sensitivity predictions} for both the $\tau^+\tau^-$ and $b\bar{b}$ annihilation channels as a function of the number of sources (in log-scale) compatible with DM. {  We recall that this type of DM constraint can be set by comparing the number of unIDs that are compatible with being subhalo candidates with results from numerical simulations (i.e., the J-factors), in such a way that the fewer the sources that are compatible with DM among the pool of unIDs, the stronger the DM constraints. This is so because the subhalo J-factor value used in the derivation of the limits (see Eq.~\ref{eq:master_formula}) is larger as the number of putative subhalos decrease.} As can be seen in Figure \ref{fig:constraints_contour}, even in the sensitivity reach case (only one source left), we do not achieve the constraints we set in \citetalias{CoronadoBlazquez2019b} by means of the 3FGL catalog \cite{Acero2015}. We remind that in the case of \citetalias{CoronadoBlazquez2019b}, the constraints for the $\tau^+\tau^-$ channel were already at the sensitivity reach level, which are now a factor $\sim$8 away for low WIMP masses and a factor $\sim$5 away for the heaviest WIMPs. The difference in this factor at small and large masses is probably due to the fact of having started this data analysis at 500 MeV instead of 100 MeV as in \citetalias{CoronadoBlazquez2019b}. Indeed, this is expected to particularly impact the constraints at low WIMP masses, as observed. Overall, in the sensitivity reach case, our new constraints are right at the thermal relic cross section value for the lowest considered WIMP masses and a factor $\sim10$ away at 100 GeV in the case of $\tau^+\tau^-$ annihilation channel, while similar conclusions can be drawn for $b\bar{b}$, yet a factor $\sim4$ far from the thermal value at best.

As already pointed out in \citetalias{CoronadoBlazquez2019a}, Figure \ref{fig:constraints_contour} shows that the achievable constraints would only marginally improve if we filtered less than $\sim80\%$ of the unID sources in the catalog, while the improvement will be more than two orders of magnitude from this number of sources down to zero. 

Finally, we also note that in this work we assumed two different subhalo models, representative of the subhalo population that would be most relevant from the point of view of annihilation flux (see top panel in Figure \ref{fig:vliirep}). Yet, when setting constraints we adopt the J-factor above which 95\% of the realizations are contained. This is conservative, as using e.g. the median value of the subhalo J-factor instead would improve our limits by a factor $\sim2$.

\section{Discussion and conclusions}
\label{sec:conclusions}
In this work, we have presented a quantitative assessment of the role of spatial extension in DM subhalo detection with the \textit{Fermi}-LAT using actual data. In the first place, 
we have chosen two representative physical scenarios for the subhalos, a massive and far subhalo, and a lighter and closer one, covering all interesting ranges. 
Relevant properties of such representative subhalos, namely angular extension and annihilation flux, were chosen following our previous results from the VL-II N-body cosmological simulation of a MW size halo~\citep{CoronadoBlazquez2019a}. 
In Section \ref{sec:DM_modeling} we built coherent DM density profiles taking into account the N-body simulation work and taking into account processes like tidal disruption \citep{Moline+17}. 
Both the computed J-factors and the spatial distribution of the corresponding putative DM signals revealed a degeneracy between our selected representative subhalos, as noted in Figure \ref{fig:vliirep}. This effect can be understood by analysing the dependencies of the J-factor computation for NFW DM density profiles, which can be described as:

\begin{equation}\label{eq:j-propto}
J_{sub}\propto\frac{M_{sub}c_{200}^3}{D_{\rm Earth}^2}\rho_{crit}.
\end{equation}

Indeed, the combination of the involved masses, distances and obtained concentrations for the selected subhalos resulted in similar values of their J-factors, as shown in Table \ref{tab:j-factors}.

Using \texttt{CLUMPY}, we built spatial templates for the annihilation signal (see Figure \ref{fig:clumpy_maps}), that were later used as inputs for the LAT data analysis. We performed two types of analysis. First, we computed the threshold flux to detect an unequivocal extension, $F_{min}^{ext}$, i.e., the minimum flux to have an extension TS of 25, equivalent to $\sim 5\sigma$. We also computed the detection minimum flux, $F_{min}^{det}$, and compared it to the point-source case. We found that, in order to detect an extended subhalo, the LAT requires a factor $\sim2$ more flux than the one needed to detect a point-like source, and a factor $\sim10$ more flux to properly characterize the spatial extension of the subhalo, as reported in Figure \ref{fig:fmins_bb_tautau}. The two subhalo models, as guessed from the \texttt{CLUMPY} templates, are indistinguishable also from the point of view of the data analysis.

In a second stage of the analysis, we performed ``blind'' computations to characterize the LAT sensitivity to extended subhalos, by analysing a larger grid of different parameters such as the WIMP mass, annihilation channel, subhalo model and the normalization of the DM spectrum. We defined the latter as the product of the cross section and the J-factor, $\langle\sigma v\rangle \times J$. This provides a unified framework in which, given one of the two magnitudes, the other can be easily estimated. The explored normalizations were chosen to lie within a range of extreme, yet possible DM scenarios, such as cross sections two orders of magnitude larger than the thermal relic value or ``outlier'' J-factors according to the N-body simulation results. From our full grid analysis computations, we found that the LAT should be capable of resolving the angular extension for most WIMP masses (Figures \ref{fig:tsext_vs_tsdet}--\ref{fig:r68_tsdet_tsext}). Indeed, for low WIMP masses, whose spectra peak around the maximum sensitivity of the LAT, even low normalization values provide a good signal characterization, both spectrally and spatially. Nevertheless, as we shift towards large WIMP masses, the annihilation spectrum yields less photons; this, together with the decreased sensitivity at high energies, provides no detection for most considered normalization values, as well as a poor (if any) spatial characterization.

{  We note a number of differences in our results with respect to the previous work of \cite{Mauro2020} on extended subhalo detection with the LAT. To model the Galactic subhalo population, we use state-of-the-art N-body simulation data, including low-mass subhalos below the original resolution limit according to the recipe in \citetalias{CoronadoBlazquez2019a}, while \cite{Mauro2020} used semi-analytical models. Also, we use a realistic sky model based on real LAT data, instead of simulated gamma-ray data. This allows us to account for putative observational biases that would be present in a real blind subhalo search. 
We investigate two extreme subhalo models that are representative of the entire subhalo population according to N-body simulations, 
while \cite{Mauro2020} adopted only the brightest subhalo predicted by their model. We extend the computations for a much finer grid of different models (660, with 10 iterations each). The behaviour of the uncertainties as a function of the input parameters is also quantified, finding that these are reduced for large values of J-factor/cross sections. In contrast, \cite{Mauro2020} found the opposite trend, i.e., bright subhalos being poorly characterized while faint sources present smaller uncertainties. Ref.~\cite{Mauro2020} studied the response of the LAT to different parameters such as WIMP mass and DM cross sections, while we also study the threshold sensitivity of the instrument, both for detection and extension.
    

Finally, we explore the impact of this subhalo spatial extension study on the DM constraints, an issue not explored in Ref.~\cite{Mauro2020}.} In our previous works \citetalias{CoronadoBlazquez2019a,CoronadoBlazquez2019b} and \cite{CoronadoBlazquez2020}, DM constraints were always derived assuming point-source subhalos, namely adopting a J-factor equal to the integration of the annihilation signal up to the subhalo scale radius. In our work, we found that a better estimate of the subhalo J-factor, as it could be actually observed by the LAT, would be the one integrated up to the innermost $0.3^\circ$ instead (see Figure \ref{fig:jfact_dist_rs_03}). Thus, we now updated the DM constraints with this new information at hand, shown in Figure \ref{fig:constraints_contour} as a function of the number of sources compatible with a DM origin.

Our analysis allows us to define, in a similar fashion to the work done in \citetalias{CoronadoBlazquez2019a}, an additional ``filter'' to identify DM subhalo candidates among the pool of unidentified LAT sources. Indeed, we conclude that, in most cases -- especially for low WIMP masses -- the LAT should be able to detect at least a hint of spatial extension if the source (subhalo) is bright enough\footnote{Yet, as mentioned in Section \ref{sec:analysis_bigsection}, the LAT may detect as extended source two nearby, individual point-like objects \cite{2018PDU....21....1C}.}. An analysis of spatial extension in the latest gamma-ray catalogs is already ongoing and will be presented elsewhere. This additional DM filter, now robustly defined, together with new generation instruments such as the Cherenkov Telescope Array \cite{coronadoblazquez2021sensitivity} or AMEGO \cite{mcenery2019allsky}, will be able to explore the whole WIMP range. CTA will be specially sensitive to large, TeV WIMP masses and will also possess a superb angular resolution of about $0.03^\circ$ in the TeV energy range, while AMEGO will be sensitive to lower energies than the LAT with 5 times better angular resolution, therefore ideal for this extension search as well.

\acknowledgments

The authors would like to thank N\'estor Mirabal and Chris Karwin for their comments on the manuscript, and Eric Charles for his valuable help with \textit{fermipy}. J.C.-B. and M.A.S.-C. are supported by the {\it Atracci\'on de Talento} contract no. 2016-T1/TIC-1542 and 2020-5A/TIC-19725 granted by the Comunidad de Madrid in Spain. {  J.P.-R.'s work is supported by grant SEV-2016-0597-17-2 funded by MCIN/AEI/10.13039/501100011033 and ``ESF Investing in your future''. The work of J.C.-B., M.A.S.-C., J.P.-R. and A.A.-S. was additionally supported by the grants PGC2018-095161-B-I00 and CEX2020-001007-S, both funded by MCIN/AEI/10.13039/501100011033 and by ``ERDF A way of making Europe''. }The work of A.A.-S. was also supported by the Spanish Ministry of Science and Innovation through the grant FPI-UAM 2018.

The \textit{Fermi} LAT Collaboration acknowledges generous ongoing support from a number of agencies and institutes that have supported both the development and the operation of the LAT as well as scientific data analysis. These include the National Aeronautics and Space Administration and the Department of Energy in the United States, the Commissariat `a l’Energie Atomique and the Centre National de la Recherche Scientifique / Institut National de Physique Nucl\'eaire et de Physique des Particules in France, the Agenzia Spaziale Italiana and the Istituto Nazionale di Fisica Nucleare in Italy, the Ministry of Education, Culture, Sports, Science and Technology (MEXT), High Energy Accelerator Research Organization (KEK) and Japan Aerospace Exploration Agency (JAXA) in Japan, and the K. A. Wallenberg Foundation, the Swedish Research Council and the Swedish National Space Board in Sweden. Additional support for science analysis during the operations phase is gratefully acknowledged from the Istituto Nazionale di Astrofisica in Italy and the Centre National d'Etudes Spatiales in France.

This research made use of Python, along with community-developed or maintained software packages, including IPython \cite{Perez2007}, Matplotlib \cite{Hunter2007} and NumPy \cite{Walt2011}. This work made use of NASA’s Astrophysics Data System for bibliographic information.

\appendix

\section{Precision of recovered signal}
In this Appendix, we discuss how well \textit{fermipy} can reconstruct an injected DM signal. This reconstruction check will be divided in two parts: first, we will study how precise are the recovered mass and normalization values, i.e., the spectral parameters. In the second part, we will focus on the evolution of the signal as we increase the normalization, both spectral and spatially. 

\subsection{DM spectrum}
\label{app:precision_dm_spectrum}
\begin{figure*}[!ht]
\centering
\includegraphics[width=0.45\linewidth]{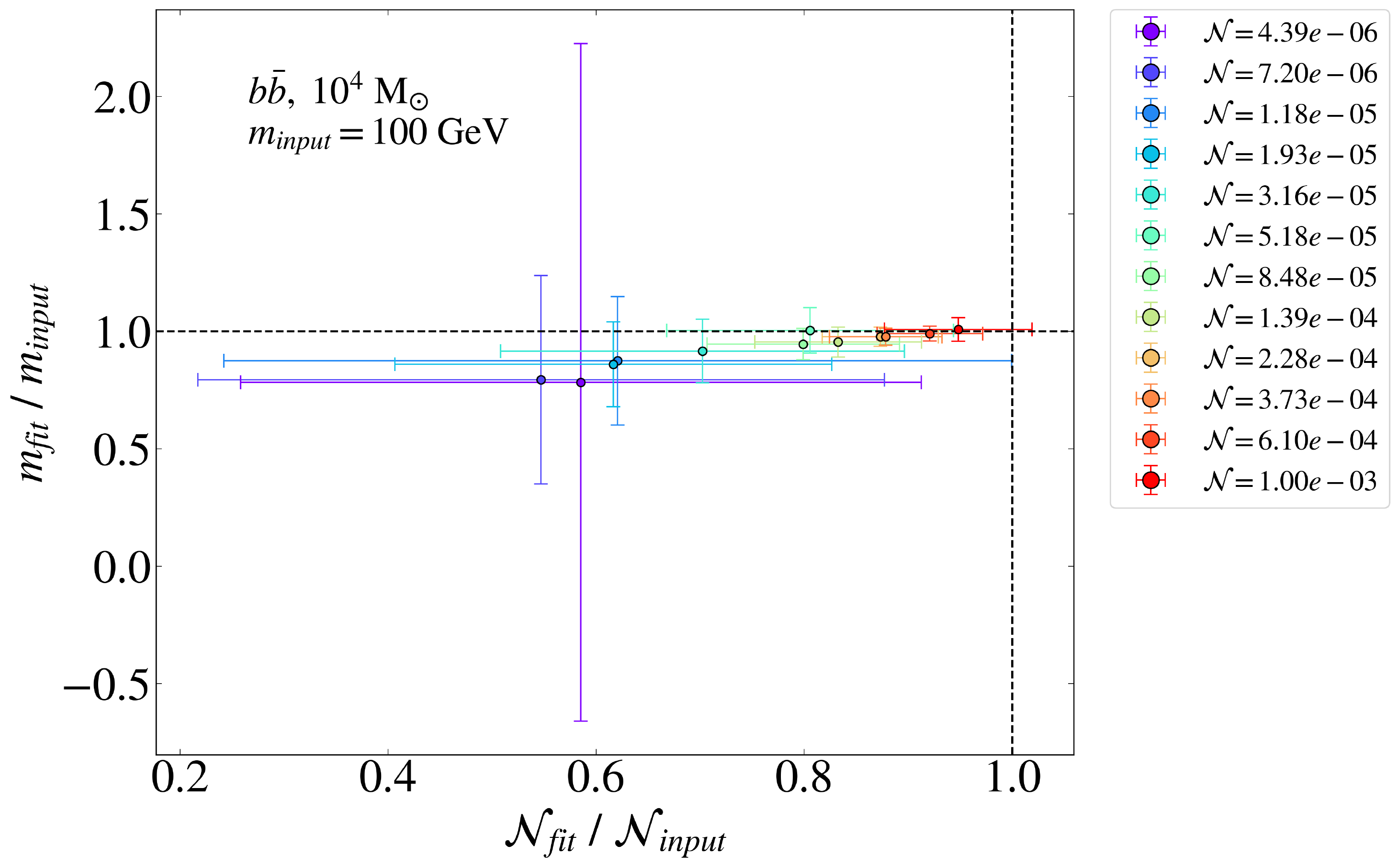}
\includegraphics[width=0.45\linewidth]{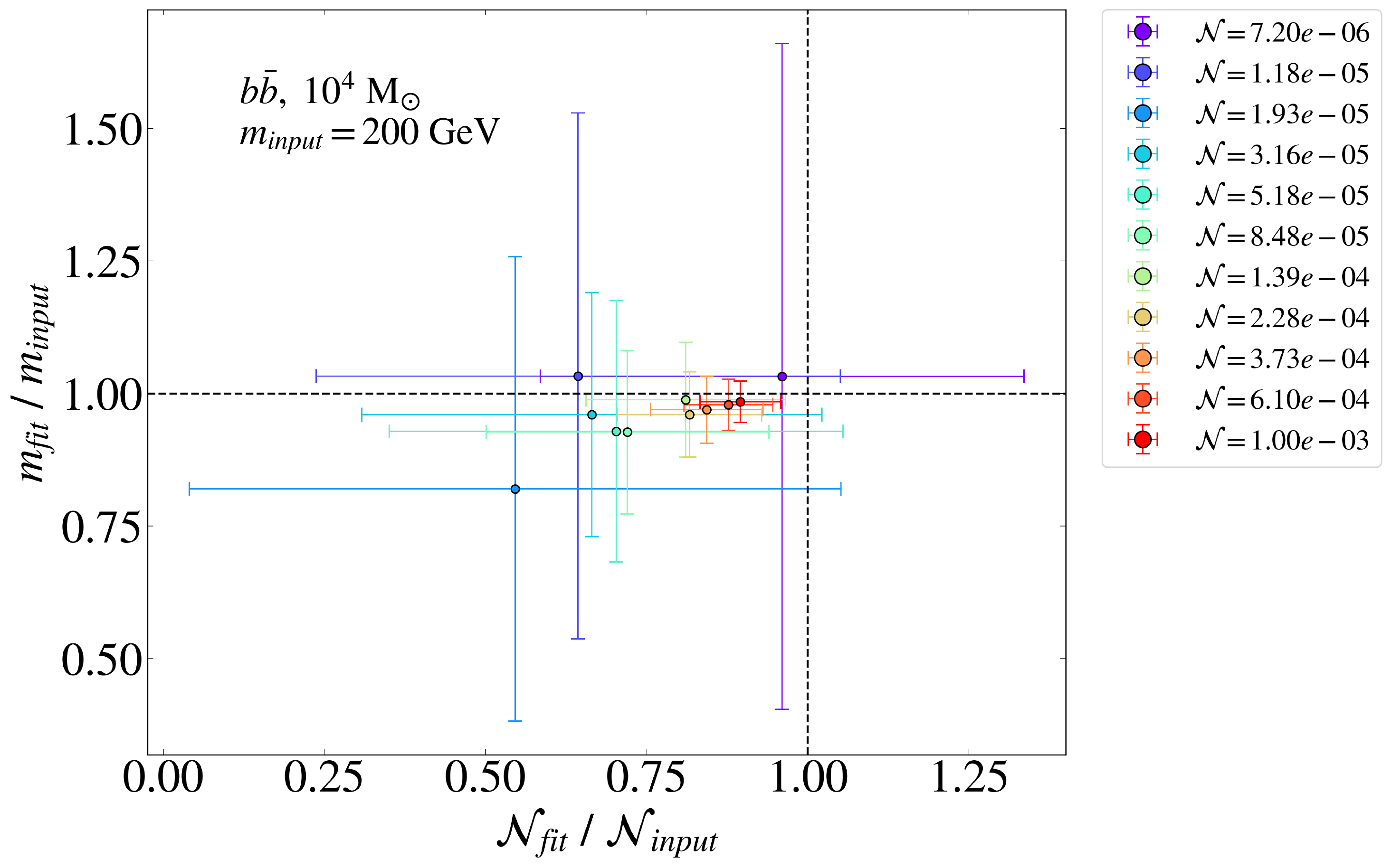}
\includegraphics[width=0.45\linewidth]{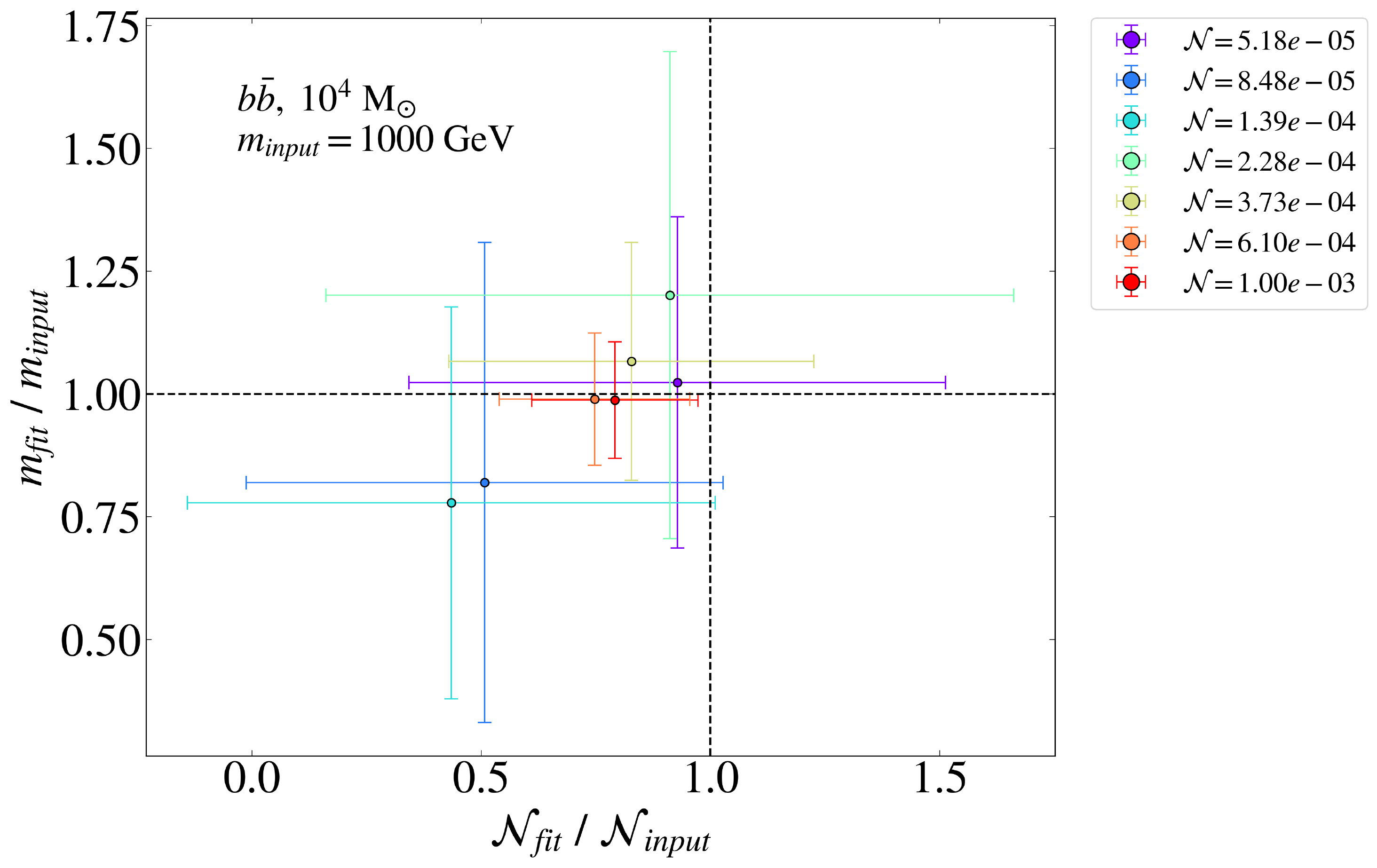}
\includegraphics[width=0.45\linewidth]{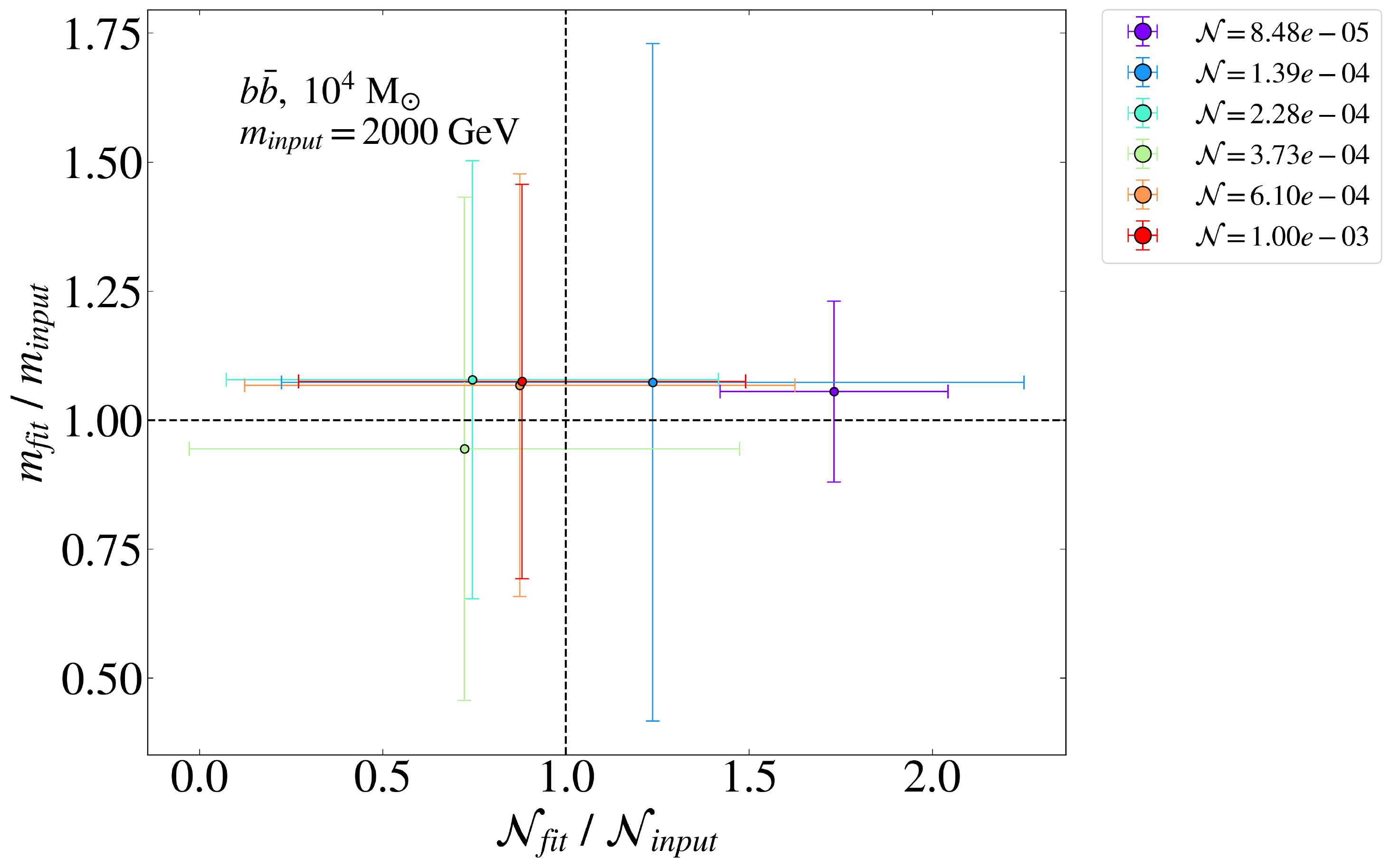}
\caption{Comparison between the injected DM signal and the recovered one, in terms of recovered WIMP mass and normalization (J-factor $\times$ $\langle\sigma v\rangle$, with units $\mathrm{GeV^2\cdot cm^{-2}\cdot s^{-1}}$), for the $b\bar{b}$ annihilation channel and the $10^4~\mathrm{M_\odot}$ subhalo model. Note that, as the input mass is increased, less points are present, as the lowest normalization values do not provide detection, indeed only the largest normalizations allowing for a proper characterization of the DM spectrum. Values and error bars refer to average and standard deviation as obtained from 10 realizations. Black, dashed line marks the true values of WIMP mass and normalization (unity ratios).}
\label{fig:mass_norm_bb_model1}
\end{figure*}

\begin{figure*}[!ht]
\centering
\includegraphics[width=0.45\linewidth]{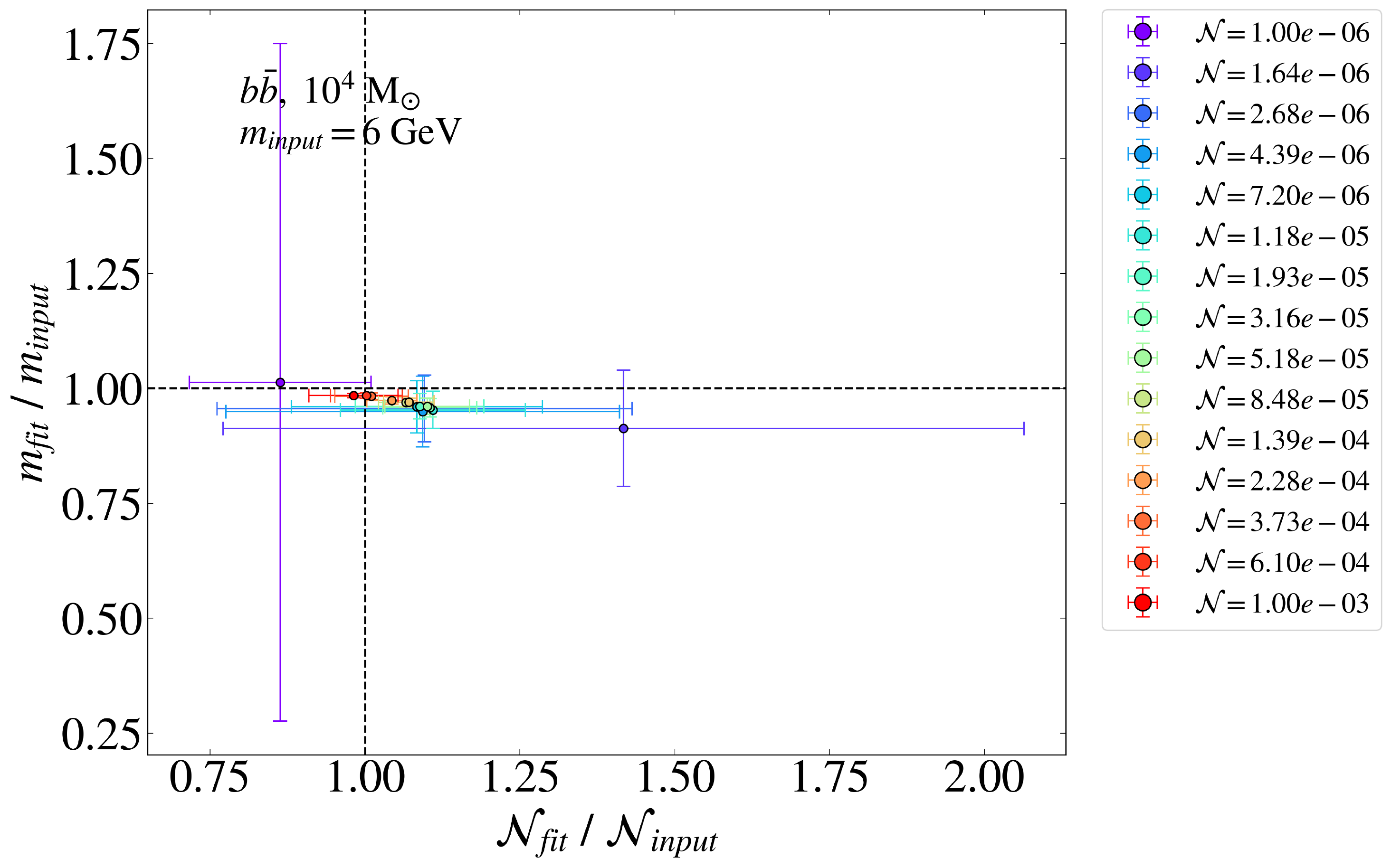}
\includegraphics[width=0.45\linewidth]{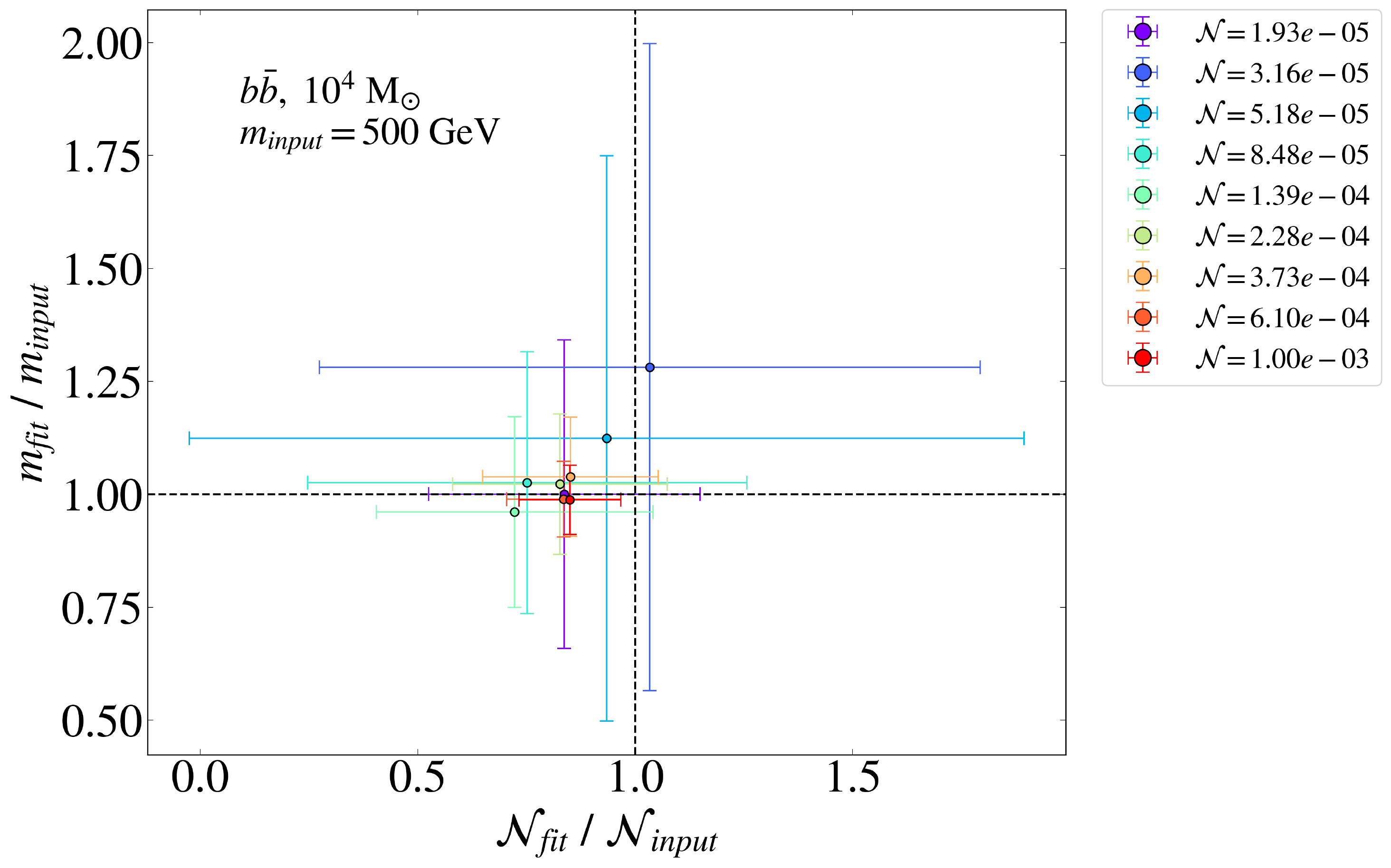}
\includegraphics[width=0.45\linewidth]{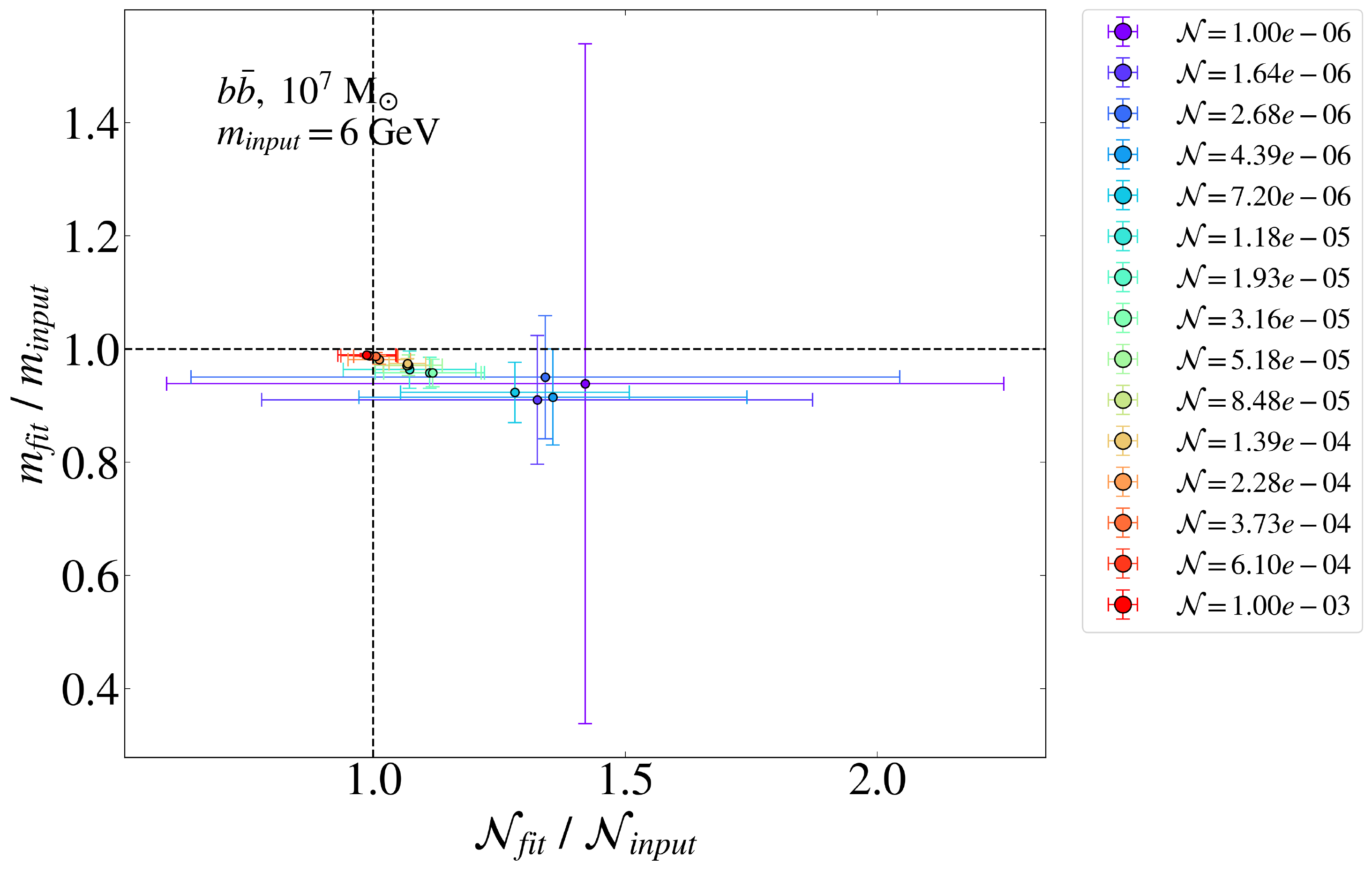}
\includegraphics[width=0.45\linewidth]{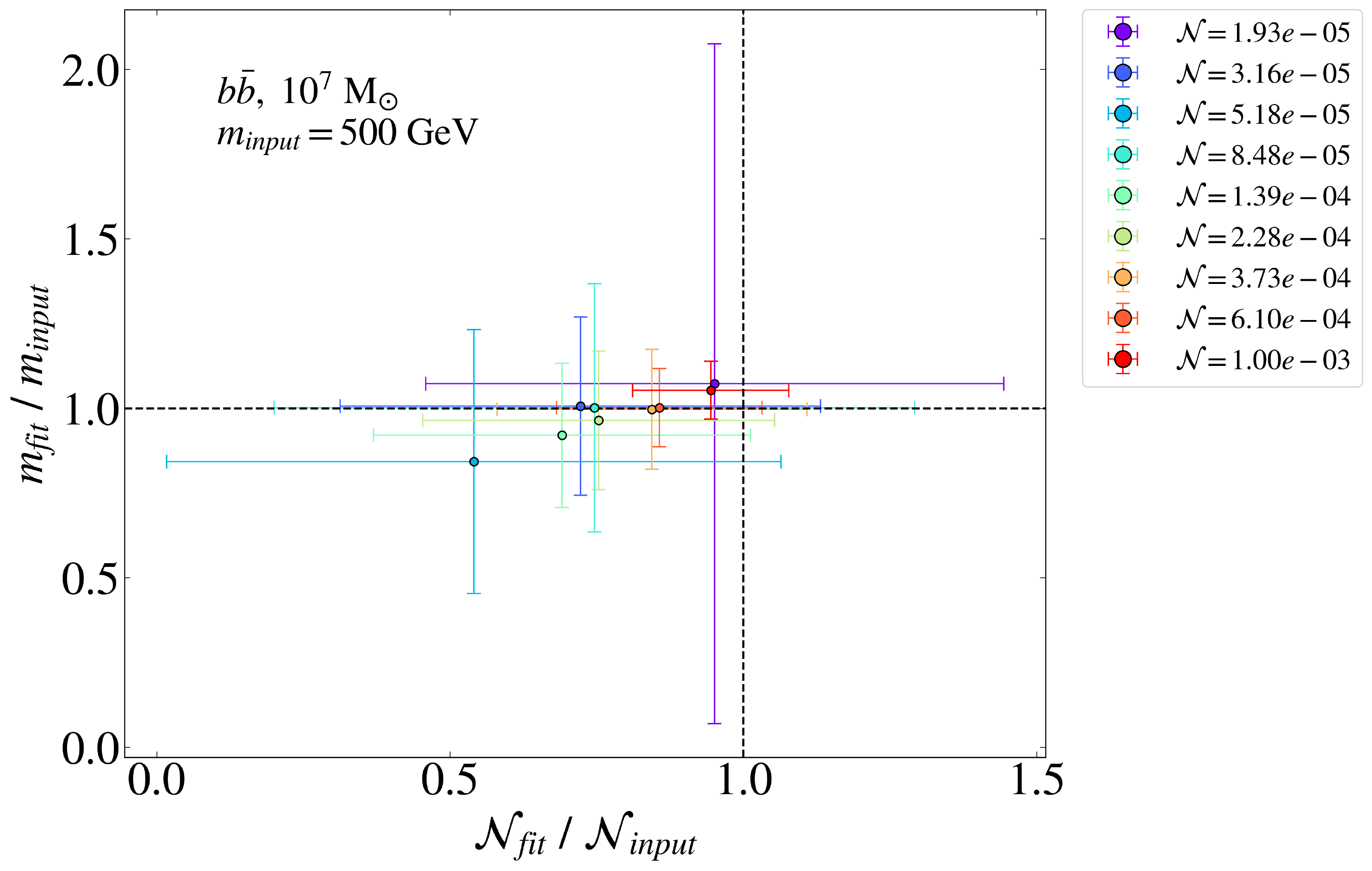}
\caption{Same as Figure \ref{fig:mass_norm_bb_model1}, but comparing the two subhalo models considered in our work. {Upper panels:} $10^4~\mathrm{M_\odot}$ subhalo. {Bottom panels:} $10^7~\mathrm{M_\odot}$ subhalo.}
\label{fig:mass_norm_bb_comparison}
\end{figure*}

As discussed, we study how well \textit{fermipy} can reconstruct the DM spectral parameters, this is, how precise the WIMP mass and normalization (J-factor $\times$ $\langle\sigma v\rangle$) fitted values are with respect to the injected values. We show the results in Figure \ref{fig:mass_norm_bb_model1} and Figure \ref{fig:mass_norm_bb_comparison}. For the sake of clarity, we only show results for the $b\bar{b}$ annihilation channel, although the same is found for the $\tau^+\tau^-$ channel. To ensure proper statistics, as in the rest of the paper, 10 realizations of the same WIMP mass/normalization setup are run, thus we report average values and $1-\sigma$ error bars corresponding to the standard deviation of these realizations.

From Figure \ref{fig:mass_norm_bb_model1} we find that, the lighter the WIMP, the better the accuracy of the recovered WIMP mass/normalization values. In addition, according to Figure \ref{fig:mass_norm_bb_comparison}, the two subhalo models provide similar results.

Therefore, as general conclusions, we find that:

\begin{itemize}
    \item \textit{fermipy} is generally able to recover the DM spectra with a few percent uncertainty in WIMP mass and normalization values, finding figures compatibles with the injected ones within 1-$\sigma$ uncertainties.
    
    \item The code is more precise when injecting signals with large normalizations, i.e., for large normalizations there are smaller uncertainties both in normalization and WIMP mass. In fact, since the overall normalization of the DM spectrum decreases when considering large DM masses, and our normalization grid is the same for all of them, the number of normalizations for which \textit{fermipy} detects the source is decreasing, not being able to detect a 10 TeV WIMP in any case.
    
    \item Both the WIMP mass and the normalization tend to be slightly underestimated -- a $\sim25\%$ for the mass and a $\sim50\%$ for the normalization at most, for the dimmest setups.
    
    \item There are no significant differences when considering different subhalo modelings and/or annihilation channels.
\end{itemize}

\subsection{Evolution of the signal with the normalization}
\label{app:signal_normalization}
\begin{figure*}[!ht]
\centering
\includegraphics[width=0.45\linewidth]{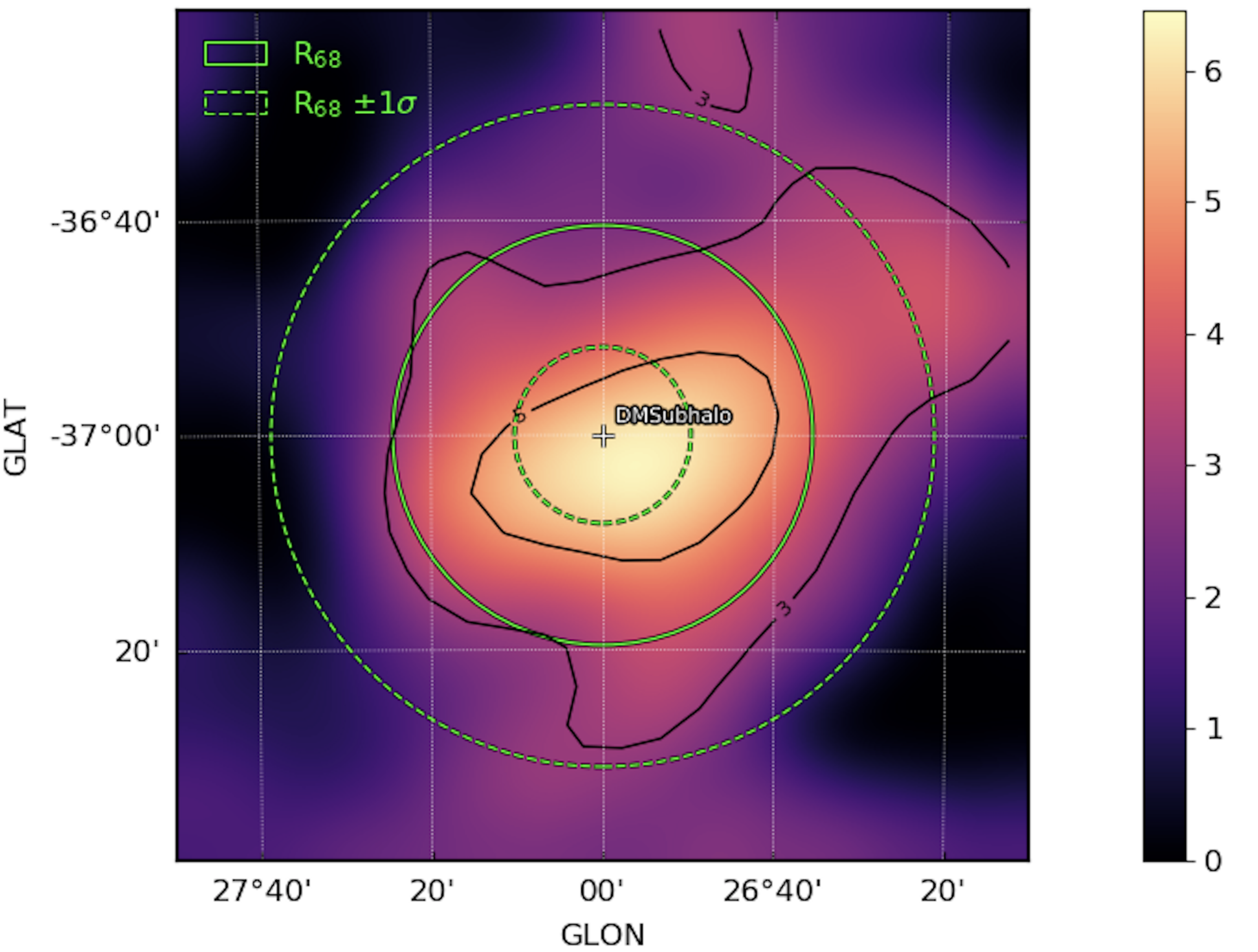}
\includegraphics[width=0.45\linewidth]{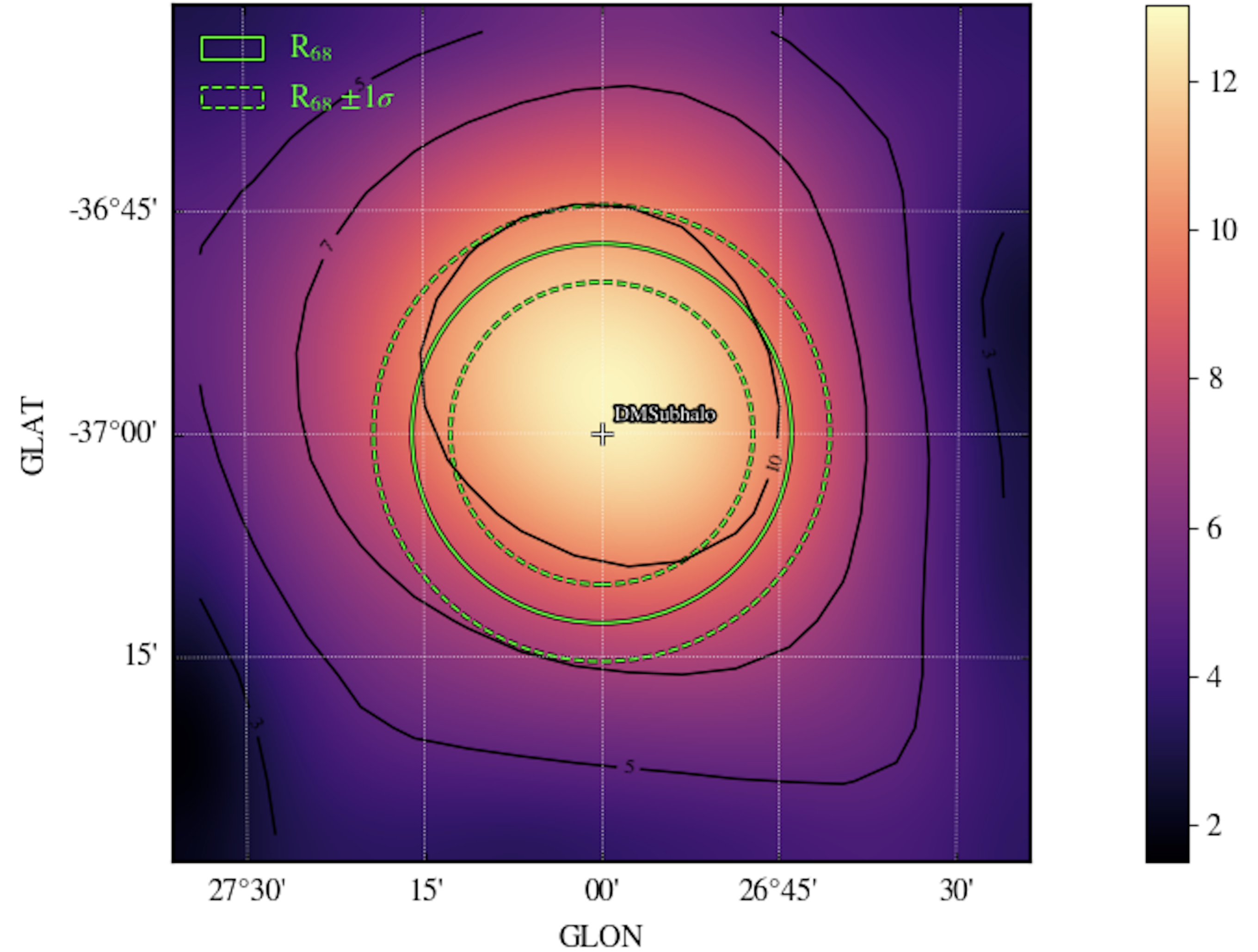}
\includegraphics[width=0.45\linewidth]{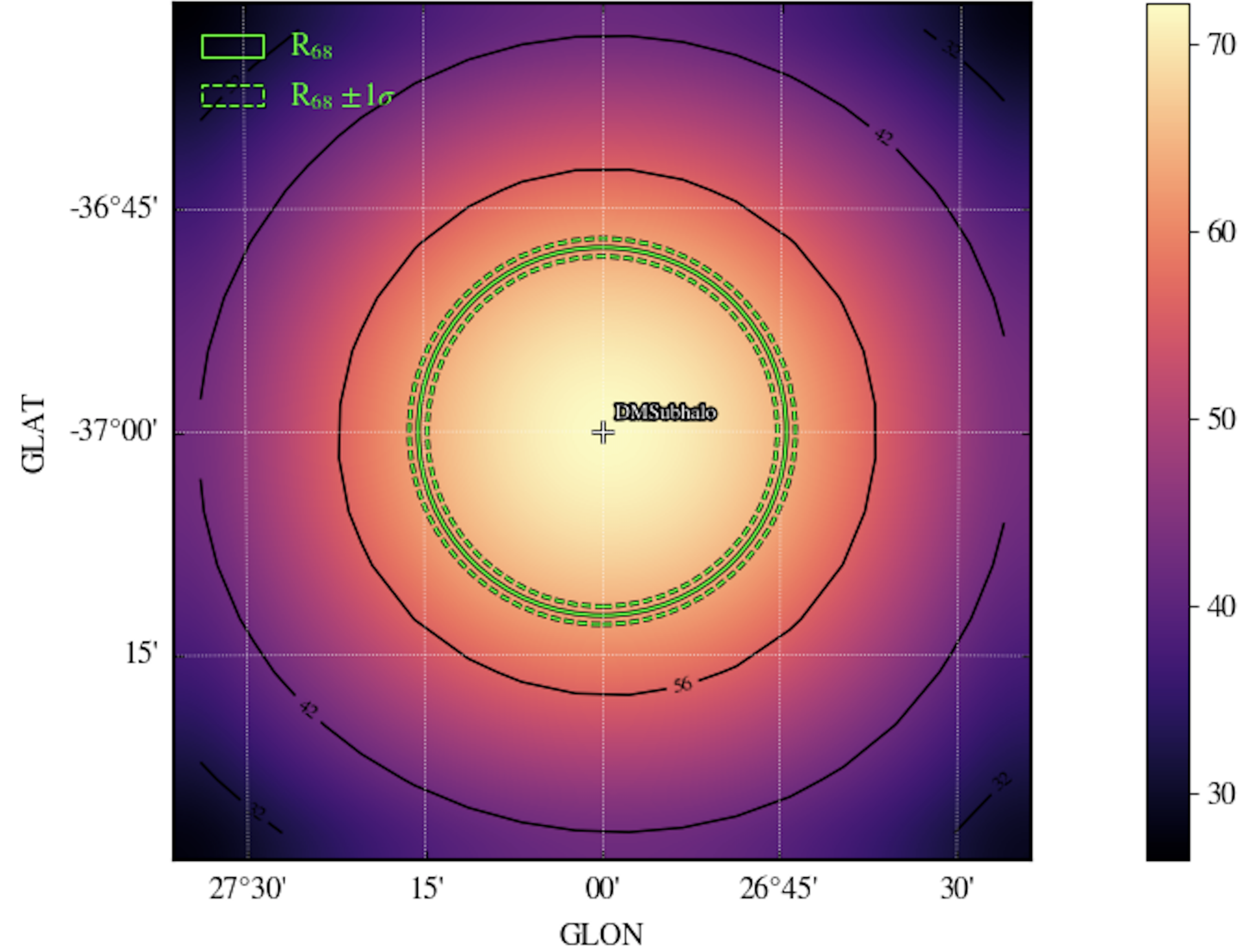}
\includegraphics[width=0.45\linewidth]{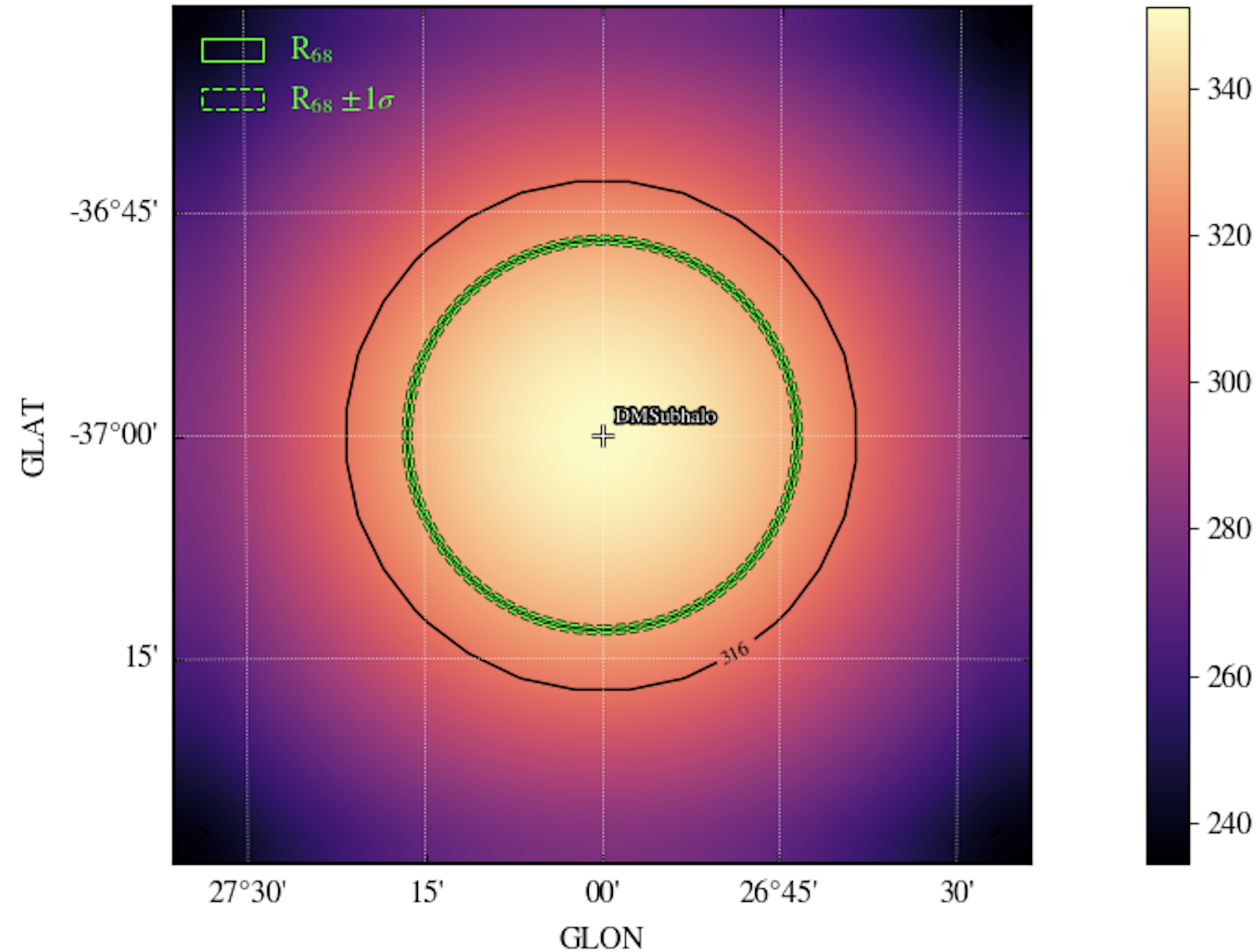}
\caption{Spatial morphology of the simulated DM subhalo for $\mathcal{N}=5\cdot10^{-6}$ (top left), $10^{-5}$ (top right), $10^{-4}$ (bottom left), and $10^{-3}$ (bottom right). Green, continuous circle is the best-fit extension, while dashed green circles are the $\pm1\sigma$ uncertainty. The color scale provides the delta log-likelihood values, the black curves being iso-contours for some particular values (note the different z-axis scale in each plot). See text for details of the analysis parameters.}
\label{fig:signal_morphology_comparison}
\end{figure*}

\begin{figure*}[!ht]
\centering
\includegraphics[width=0.45\linewidth]{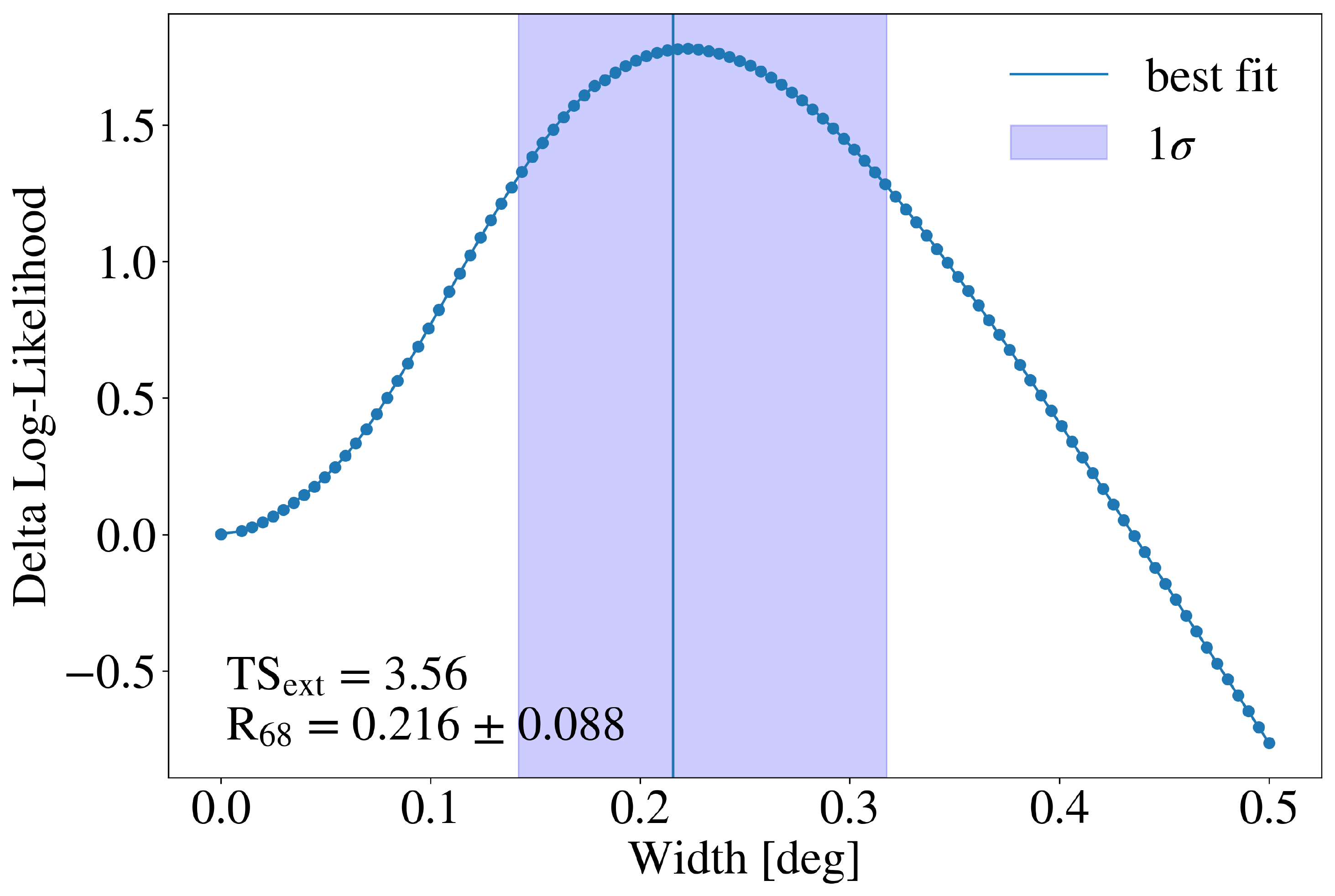}
\includegraphics[width=0.45\linewidth]{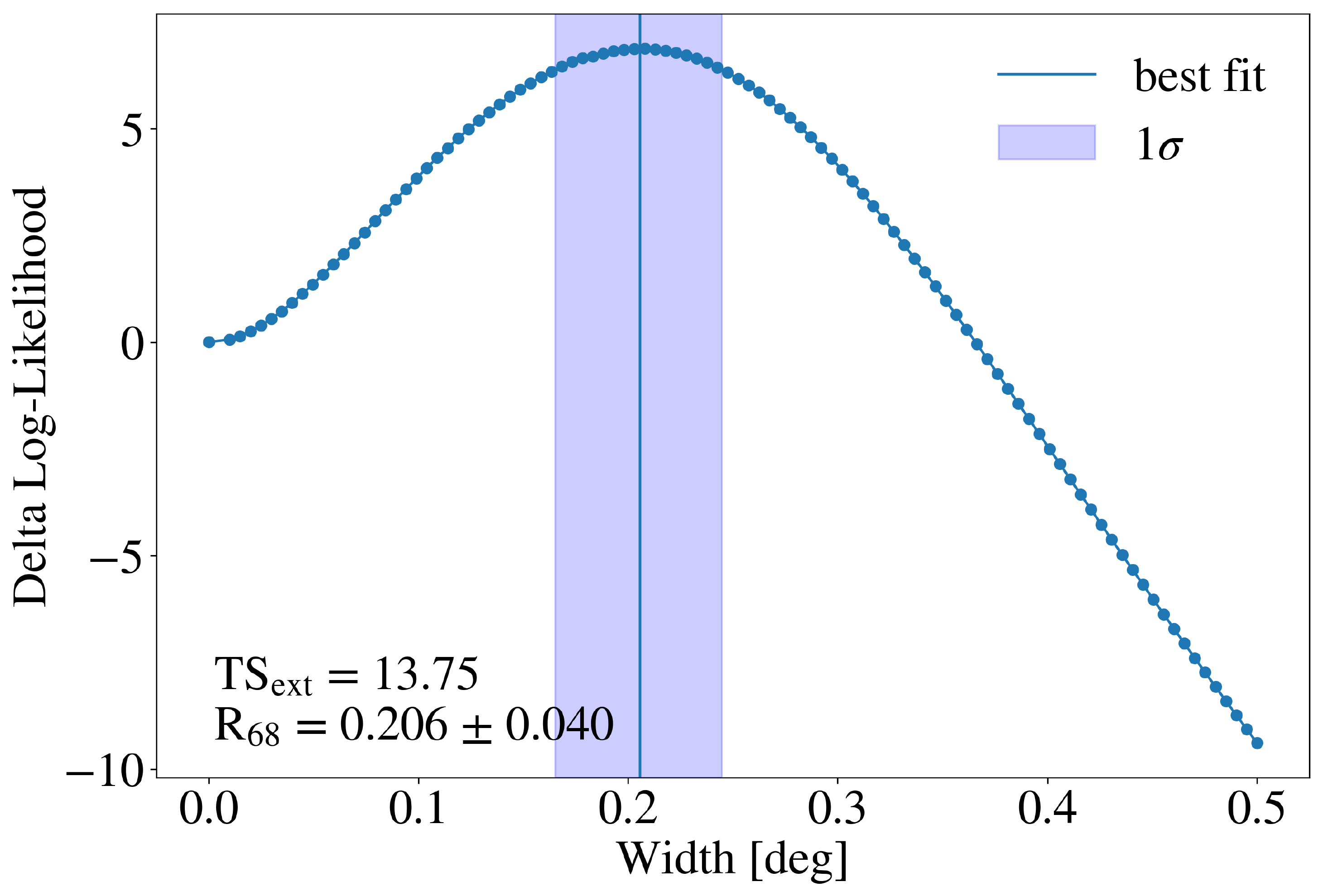}
\includegraphics[width=0.45\linewidth]{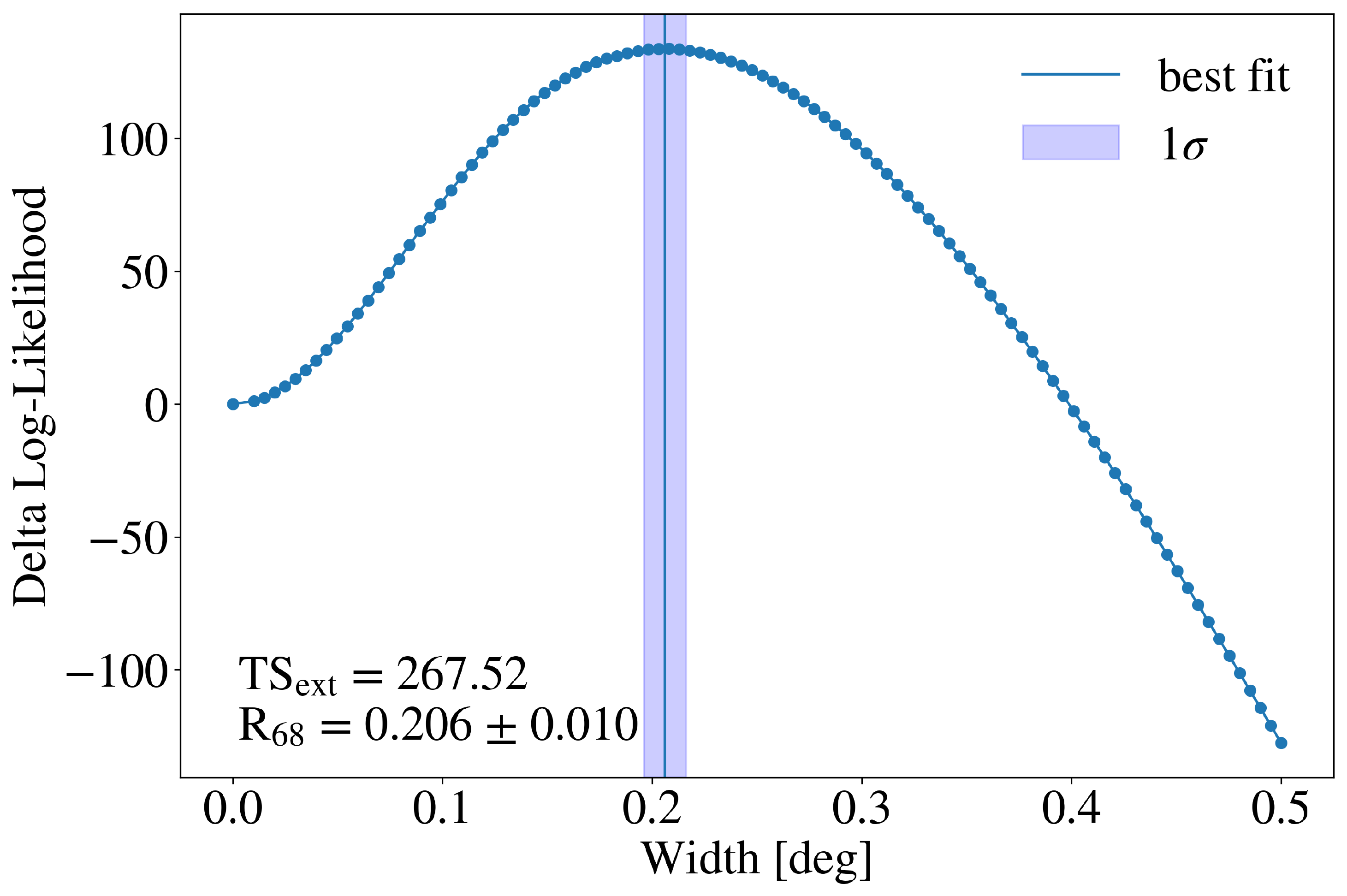}
\includegraphics[width=0.45\linewidth]{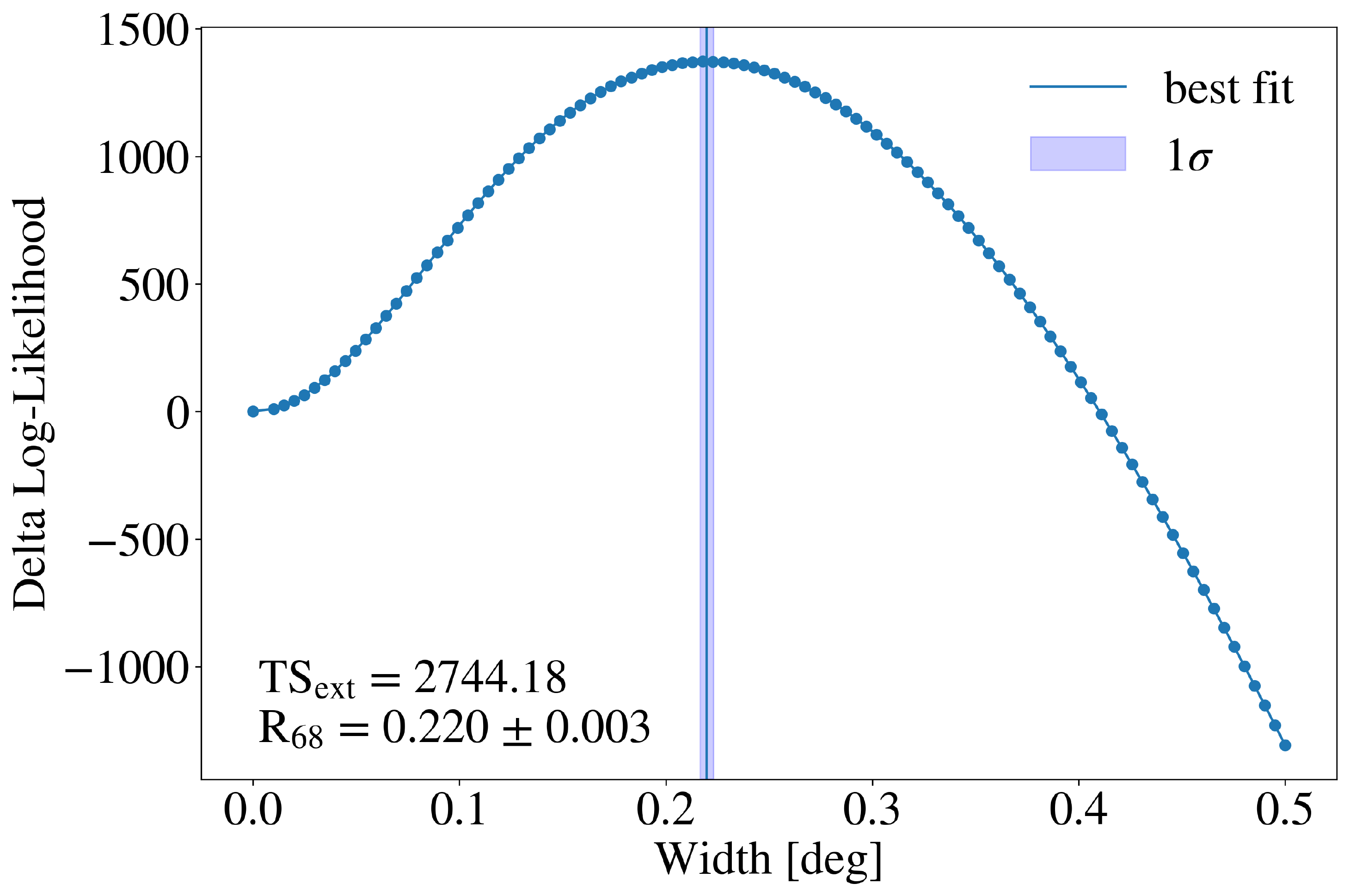}
\caption{Extension likelihood curves for $\mathcal{N}=5\cdot10^{-6}$ (top left), $10^{-5}$ (top right), $10^{-4}$ (bottom left), and $10^{-3}$ (bottom right). Blue, vertical line is the extension best-fit, while the violet band is the $1\sigma$ uncertainty. See text for details of the analysis parameters.}
\label{fig:likelihood_curves_comparison}
\end{figure*}

\begin{figure*}[!ht]
\centering
\includegraphics[width=0.45\linewidth]{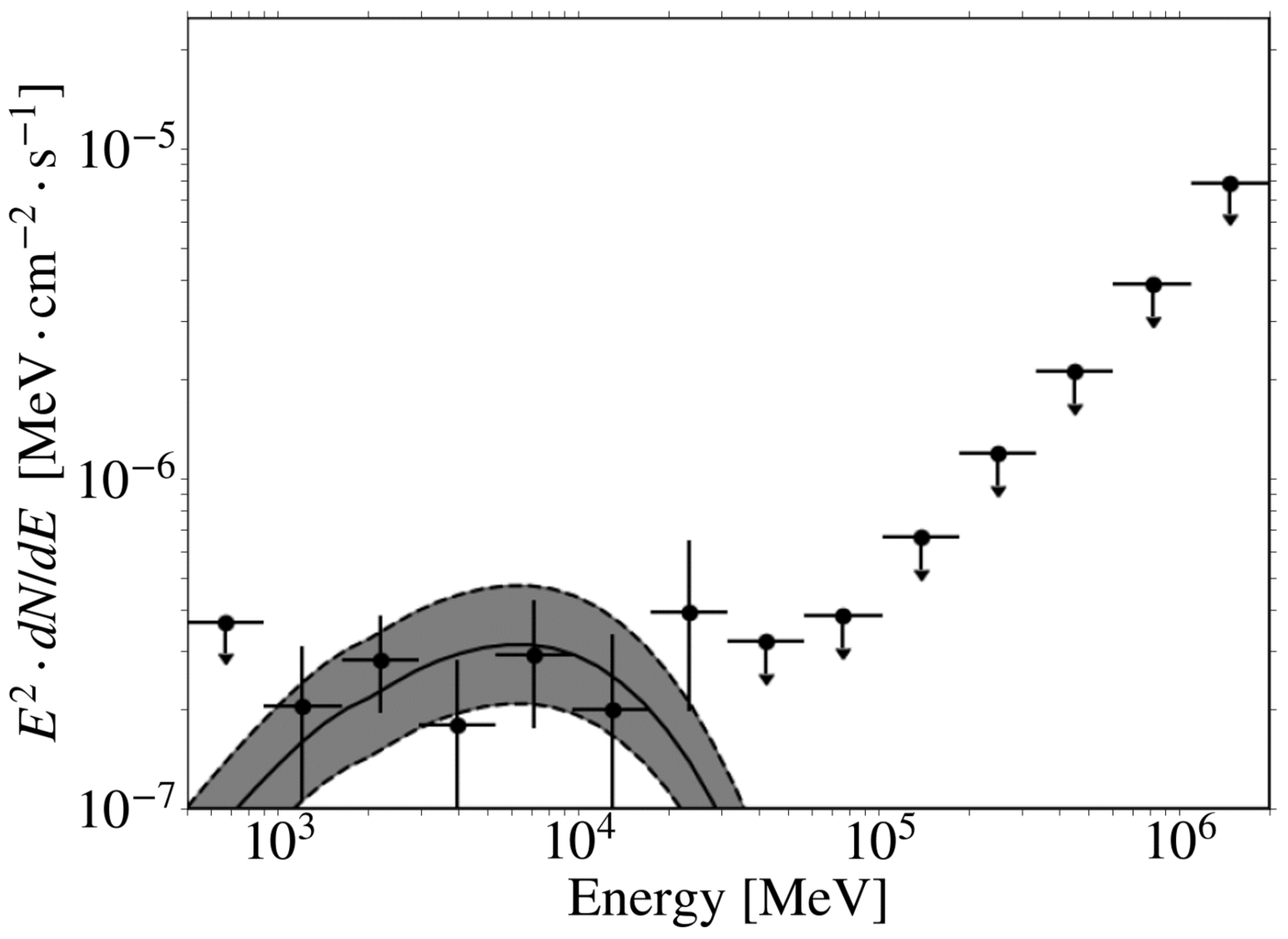}
\includegraphics[width=0.45\linewidth]{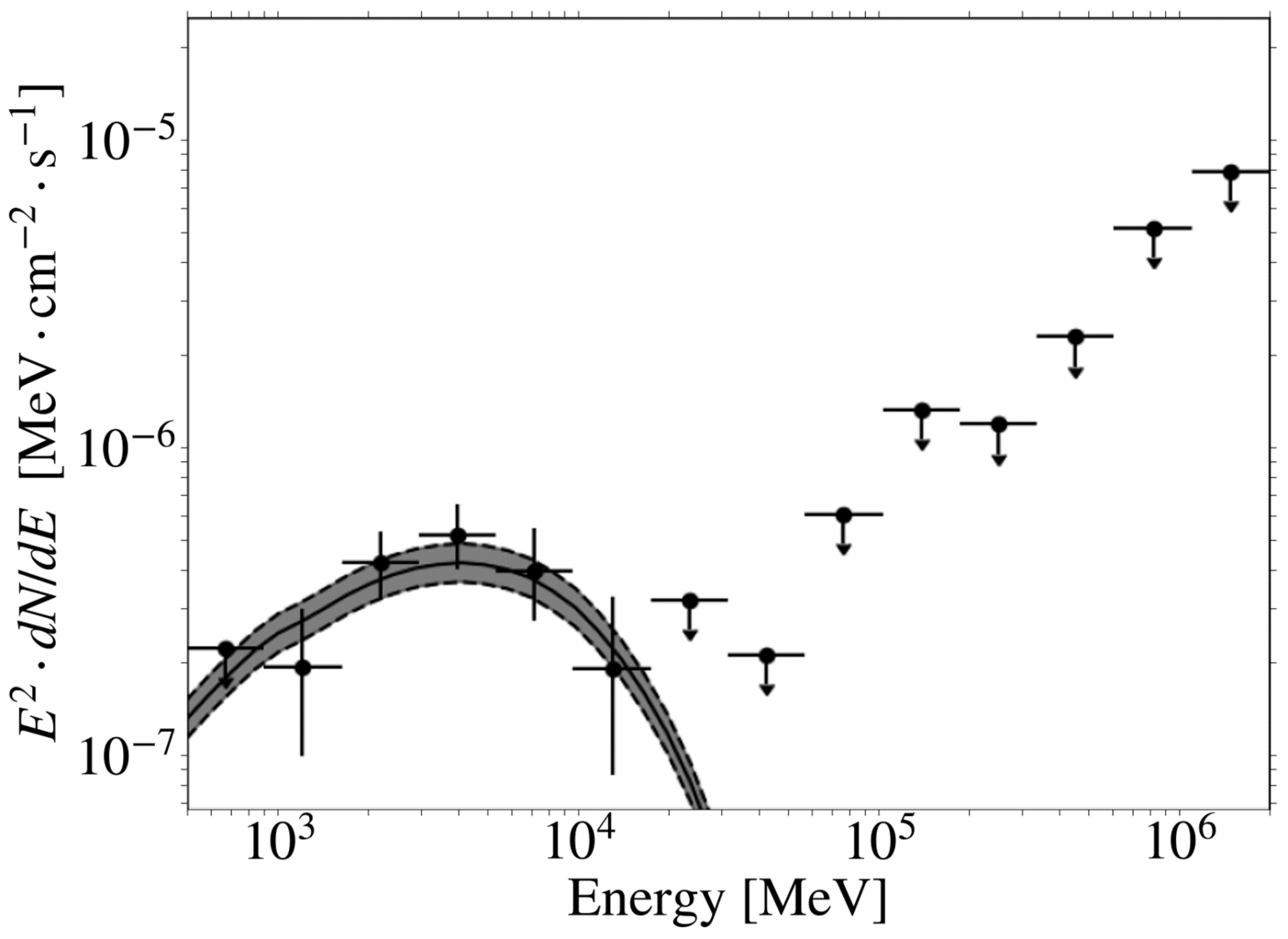}
\includegraphics[width=0.45\linewidth]{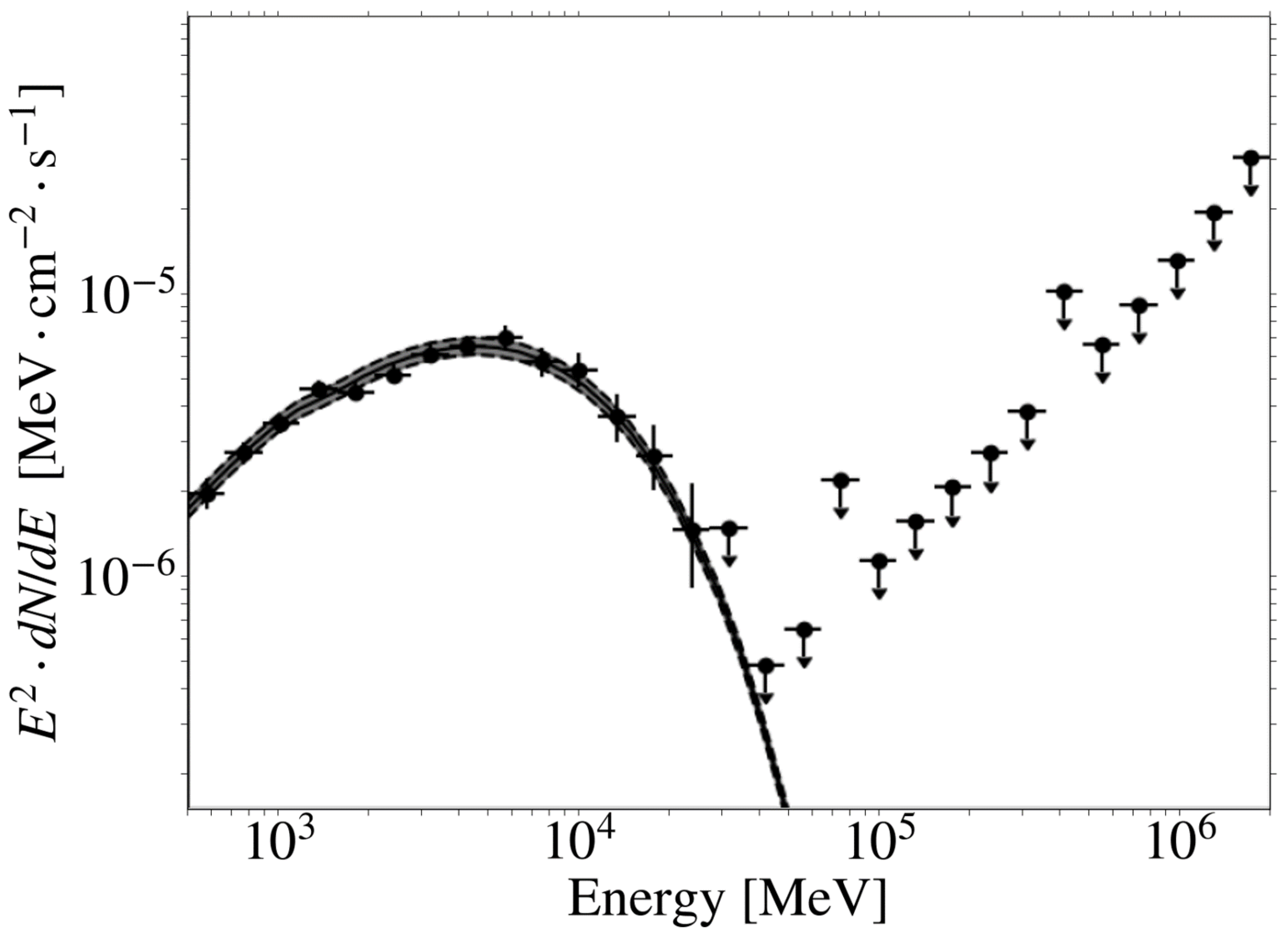}
\includegraphics[width=0.45\linewidth]{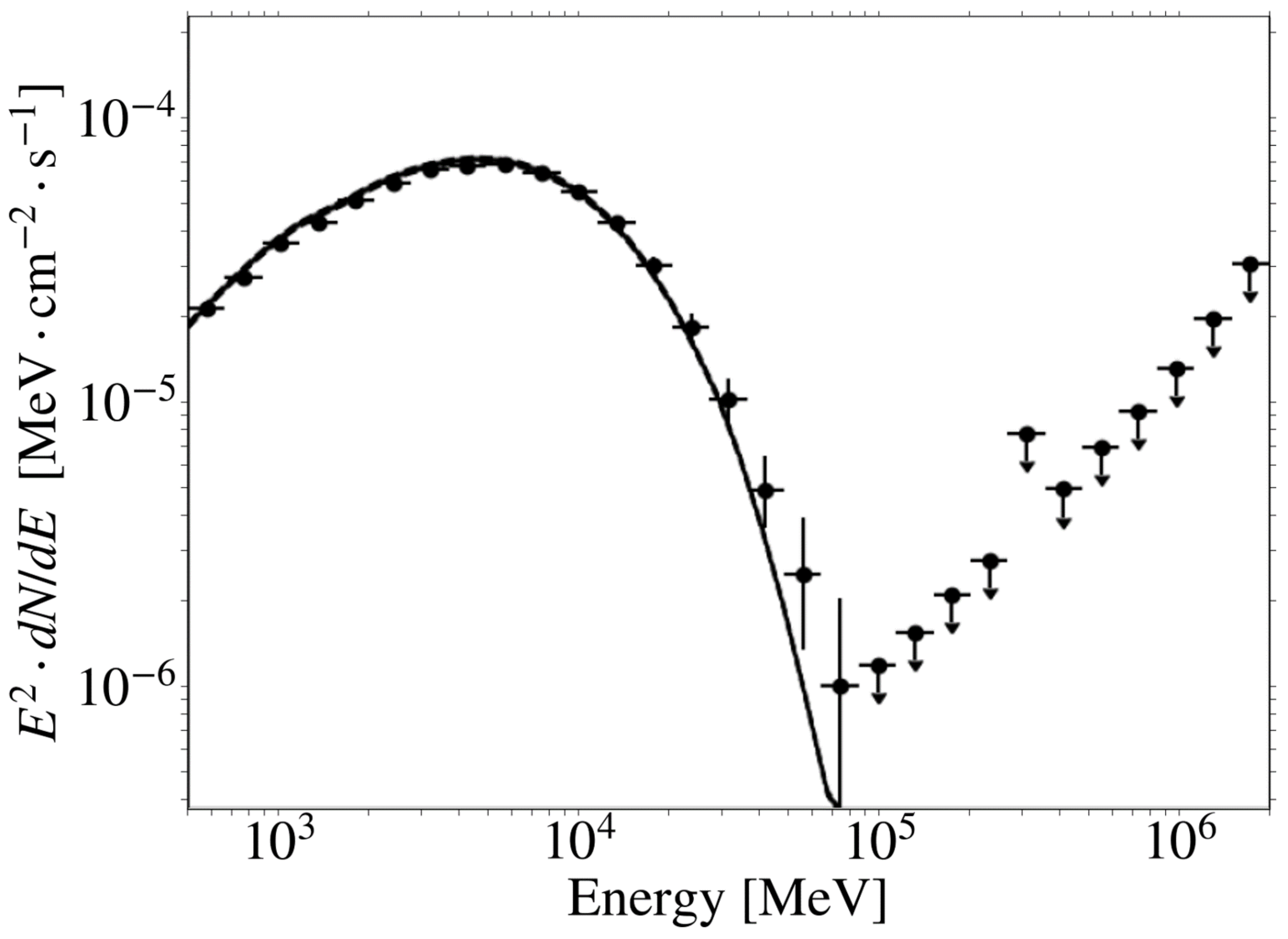}
\caption{Spectral energy distribution (SED) of a 100 GeV WIMP annihilating to $b\bar{b}$, for normalization values $\mathcal{N}=5\cdot10^{-6}$ (top left), $10^{-5}$ (top right), $10^{-4}$ (bottom left), and $10^{-3}$ (bottom right). The black band is the best-fit DM spectrum with $1\sigma$ uncertainty. The analyses in the two first cases are performed with 4 bins per energy decade, while the two latter 8, as the signal is bright enough to allow a finer binning. See text for details of the analysis parameters.}
\label{fig:DM_SED_comparison}
\end{figure*}

In this subsection, we study how the signal spatial morphology, DM spectral energy distribution (SED), and the likelihood curves of the extended analysis behave as a function of the input normalization. To do so, we consider a 100 GeV WIMP, annihilating into $b\bar{b}$, for the ($\mathrm{10^4~M_{\odot}}$, 0.5 kpc) subhalo model, although similar conclusions are found for the remaining configurations. In the next figures, these quantities are displayed for normalization values $\mathcal{N}=\{5\cdot10
^{-6},10^{-5},10^{-4},10^{-3}\}~\mathrm{GeV^2~cm^{-2}~s^{-1}}$. We start at $\mathcal{N}=5\cdot10^{-6}$ instead of $10^{-6}$ because there is no signal in the latter case. In Figure \ref{fig:signal_morphology_comparison} we show the evolution of the spatial morphology; in Figure \ref{fig:DM_SED_comparison} the DM SED; and, in Figure \ref{fig:likelihood_curves_comparison}, the likelihood curves of the spatial extension analysis.

There is a similar general trend in the three figures: as the normalization is increased, the recovery is more precise, i.e., the corresponding uncertainties are reduced. Indeed, Figures \ref{fig:signal_morphology_comparison} and \ref{fig:likelihood_curves_comparison} show that, although the best-fit $R_{68}$ is always stable, its uncertainty is reduced from almost 50\% to less than 2\%. Also, the preference for an extended source over the point-source model ($\mathrm{TS_{ext}}$, see Eq. \ref{eq:TS_ext}) increases accordingly. This is prominent in Figure \ref{fig:signal_morphology_comparison}, where the morphology of the signal evolves from an irregular
``blob'' to a perfect reconstruction of the input spatial template (see Figure \ref{fig:clumpy_maps}).

The same conclusions apply to the DM SEDs in Figure \ref{fig:DM_SED_comparison}: the SED starts as a poorly resolved spectrum, well fitted by either DM or a power law, and evolves to a refined, very well characterized spectrum with a much smaller uncertainty in most bins (indeed allowing for a finer analysis, with 8 bins per energy decade).

\bibliographystyle{apsrev4-2.bst}
\bibliography{references_ang_extension.bib}
\end{document}